\documentclass[sigconf, nonacm]{acmart}
\PassOptionsToPackage{table,xcdraw}{xcolor}

\usepackage[table,xcdraw]{xcolor}
\usepackage{tabularx, booktabs}

\usepackage{array}
\usepackage{macros}
\usepackage{float}
\AtBeginDocument{%
  \providecommand\BibTeX{{%
    \normalfont B\kern-0.5em{\scshape i\kern-0.25em b}\kern-0.8em\TeX}}}

\usepackage{hyperref}

\usepackage{xspace}
\usepackage{tcolorbox}
\tcbuselibrary{listings}
\usepackage{wrapfig}
\usepackage{graphicx}
\usepackage{hyperref}
\usepackage{multirow}

\def\markup{1}

\if\markup0
\usepackage[normalem]{ulem}
  \definecolor{myblue}{rgb}{0,0,0.75}
  \newcommand{\rv}[1]{{\leavevmode\color{myblue}#1}}
  \newcommand{\st}[1]{{\sout{#1}}}
\else
  \newcommand{\rv}[1]{#1}
  \newcommand{\st}[1]{}
\fi

\newcommand{\reportQuestion}[5]{%
(Q\textsubscript{#1}: 
$Mdn_C$=${#2}$ vs. $Mdn_B$=${#3}$, $p$=${#4}$, $r$=${#5}$)
}

\newtcblisting{shighlight}{
  listing only,
  nobeforeafter,
  after={\xspace},
  hbox,
  tcbox raise base,
  fontupper=\ttfamily\color{black},
  colback=pink,
  colframe=pink,
  size=fbox,
  boxrule=0pt,        
  left=-1pt,           
  right=-1pt,          
  top=-4.5pt,            
  bottom=-5pt,         
}

\newtcblisting{codesyntax}{
  listing only,
  nobeforeafter,
  after={\xspace},
  hbox,
  tcbox raise base,
  fontupper=\ttfamily\color{DarkGray},
  colback=LightGray,
  colframe=LightGray,
  size=fbox,
  boxrule=0pt,        
  left=-1pt,           
  right=-1pt,          
  top=-4pt,            
  bottom=-4pt,         
}

\newtcblisting{shortcut}{
  listing only,
  nobeforeafter,
  after={\xspace},
  hbox,
  tcbox raise base,
  fontupper=\ttfamily\color{Red500},
  colback=LightGray,
  colframe=LightGray,
  size=fbox,
  boxrule=0pt,        
  left=-1pt,           
  right=-1pt,          
  top=-4.5pt,            
  bottom=-5pt,         
}

\definecolor{Red500}{HTML}{EF4444}
\definecolor{LightGray}{HTML}{EBEBEB} 
\definecolor{DarkGray}{HTML}{4C4E52}
\definecolor{SH200}{HTML}{E3BDC5}

\usepackage{pgfplots}

\begin{document}

\title[CoLadder: Hierarchical Code Generation]{CoLadder: Supporting Programmers with Hierarchical Code Generation in Multi-Level Abstraction}

\author{Ryan Yen}
\orcid{0000-0001-8212-4100}

\affiliation{
  \institution{University of Waterloo}
  \streetaddress{200 University Ave W}
  \city{Waterloo}
  \state{Ontario}
  \country{Canada}
}
\email{r4yen@uwaterloo.ca}

\author{Jiawen Zhu}
\orcid{0009-0002-2652-7241}

\affiliation{
  \institution{University of Waterloo}
  \streetaddress{200 University Ave W}
  \city{Waterloo}
  \state{Ontario}
  \country{Canada}
}
\email{jiawen.zhu@uwaterloo.ca}

\author{Sangho Suh}
\orcid{0000-0003-4617-5116}

\affiliation{
  \institution{University of California San Diego}
  \streetaddress{9500 Gilman Dr}
  \city{La Jolla}
  \state{California}
  \country{USA}
}
\email{sanghosuh@ucsd.edu}

\author{Haijun Xia}
\orcid{0000-0002-9425-0881}

\affiliation{
  \institution{University of California San Diego}
  \streetaddress{9500 Gilman Dr}
  \city{La Jolla}
  \state{California}
  \country{USA}
}
\email{haijunxia@ucsd.edu}

\author{Jian Zhao}
\orcid{0000-0001-5008-4319}

\affiliation{%
  \institution{University of Waterloo}
  \streetaddress{200 University Ave W}
  \city{Waterloo}
  \state{Ontario}
  \country{Canada}
}
\email{jianzhao@uwaterloo.ca}

\newcommand{\sys}[0]{{{\it CoLadder}}}
\newcommand{\baseline}[0]{\textit{Baseline}}

\begin{abstract}
\rv{Programmers increasingly rely on Large Language Models (LLMs) for code generation. However, misalignment between programmers' goals and generated code complicates the code evaluation process and demands frequent switching between prompt authoring and code evaluation.
Yet, current LLM-driven code assistants lack sufficient scaffolding to help programmers format intentions from their overarching goals, a crucial step before translating these intentions into natural language prompts.}
To address this gap, we adopted an iterative design process to gain insights into programmers' strategies when using LLMs for programming. Building on our findings, we created \sys{}, a system that supports programmers by facilitating hierarchical task decomposition, direct code segment manipulation, and result evaluation during prompt authoring.
A user study with 12 experienced programmers showed that \sys{} is effective in helping programmers externalize their problem-solving intentions flexibly, improving their ability to evaluate and modify code across various abstraction levels, from goal to final code implementation.
\end{abstract}


\begin{teaserfigure}
  \includegraphics[width=\textwidth]{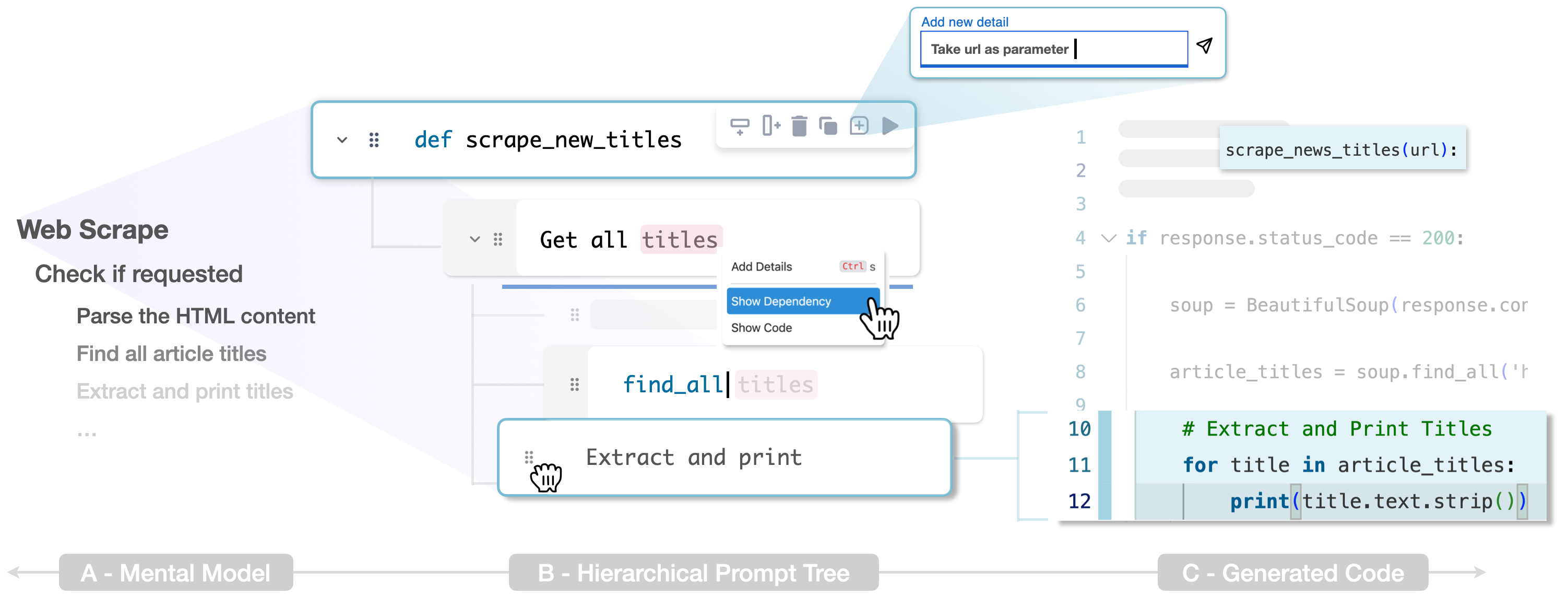}
  \caption{\sys{} enables programmers to flexibly decompose tasks, aligning with their mental models for solving programming tasks using LLM-driven code assistants (A). The system provides a tree-based editor that allows programmers to hierarchically express their intent through smaller, modular-based prompt blocks (B). This hierarchical prompt tree structure is then used to generate code, with each prompt block corresponding to a code segment (C). Programmers can directly manipulate the code based on the prompts using a series of block-based operations.}
  \Description{An overview of the \sys{} system. (A) illustrates how the system empowers programmers to flexibly decompose programming tasks, aligning with their mental models, and leveraging LLM-driven code assistance. (B) showcases the tree-based editor, enabling programmers to express their intent hierarchically through modular prompt blocks. These blocks facilitate the decomposition of tasks into manageable segments. (C) demonstrates the outcome of this hierarchical prompt tree structure: the generation of code, where each prompt block corresponds to a specific segment of code. Additionally, the figure highlights the capacity for programmers to directly manipulate the generated code based on the prompts using a series of block-based operations, enhancing their control and efficiency in code generation and verification.}
  \label{fig:teaser}
\end{teaserfigure}

\maketitle

\section{Introduction}

Recent advances in large language models (LLMs) have led to significant progress in AI-driven code assistants~\cite{chen2021evaluating, li2022competition, Friedman_2021} and brought changes to programmers workflows~\cite{mozannar2022reading, vaithilingam2022expectation, jiang2022discovering}.
These LLM-driven code assistants have extended their functionality beyond code completion to generate high-quality code suggestions in response to natural language (NL) prompts.
Programmers can now translate high-level goals into NL prompts without needing to deal with low-level code intricacies.
While this distinct capability can potentially enhance programming efficiency, recent research on programmers' interactions with LLM-driven code assistants has
revealed their \rv{challenges in evaluating the alignment between their intentions and the generated code~\cite{vaithilingam2022expectation, sarkar2022like, liu2023wants, mozannar2022reading}.}
This intention misalignment further necessitates programmers to iteratively refine prompts, adding to their workload and cognitive burden~\cite{strobelt2022interactive, liu2023pre, fiannaca2023programming, wu2022ai}.

\begin{figure*}[]
    \centering
    \includegraphics[width=0.85\linewidth]{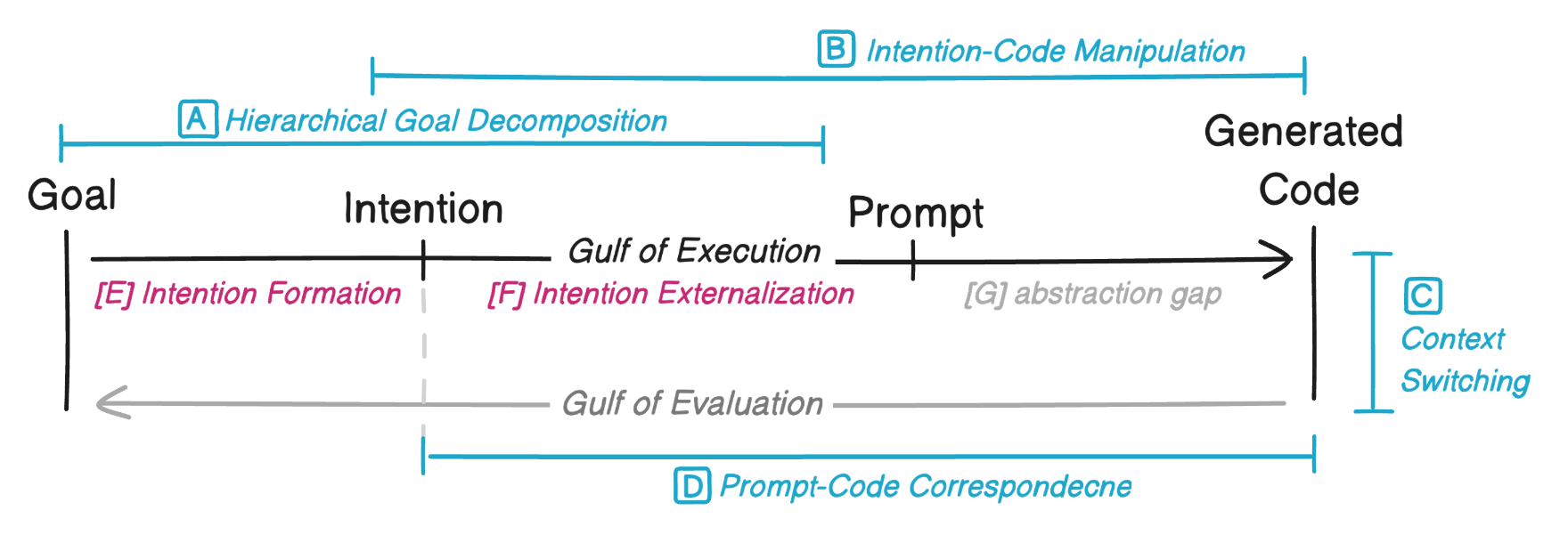}
    \caption{We have adapted Norman's seven stages of action to illustrate cognitive processes when interacting with LLM-driven systems. This model covers programmer intention formation (E) and externalization (F) within the \textit{gulf of execution}, while the \textit{gulf of evaluation} evaluates alignment between generated code and intentions. It underpins our system design, which addresses hierarchical goal decomposition (A), intention-code manipulation (B), context-switching challenges (C), and prompt-code correspondence (D).}
    \Description{The flow diagram represents stages of goal-oriented actions in the context of software interaction, specifically related to code generation. At the top, three overarching labels are presented: "A Hierarchical Goal Decomposition," "B Intention-Code Manipulation," and "C Context Switching." The diagram begins with the word "Goal" on the left, leading to a box labeled "Intention" with the sub-label "[E] Intention Formation." From the "Intention" box, a dashed arrow moves right to a dotted box labeled "Prompt," with the sub-label "[F] Intention Externalization," signifying the transition from internal thought to external expression. Beyond the "Prompt" box, another dashed arrow continues to the right, bridging the "Gulf of Execution," pointing to a box labeled "Generated Code" with a gap labeled "[G] abstraction gap," indicating a separation between the prompt and the code. Below these elements, there is a looping arrow that connects "Generated Code" back to "Goal," passing through a box labeled "Interpretation," representing the "Gulf of Evaluation." On the right side, a vertical line labeled "D Prompt-Code Correspondence" intersects the looping arrow.}
    \label{fig:cognitive}
\end{figure*}

\rv{
The evaluation challenge emerges due to the \textit{ambiguous} process of transforming a programmer's goal into generated code~(Fig.~\ref{fig:cognitive} Gulf of Execution). To overcome this challenge, it is necessary to bridge the gaps between the overarching goal and the specific intentions necessary to accomplish it (Fig.\ref{fig:cognitive}, Goal-Intention), as well as the gap between these intentions and the natural language prompts required for code generation (Fig.~\ref{fig:cognitive}, Intention-Code).
The first gap pertains to the \textit{intention formation} and \textit{intention externalization} processes. In this stage, programmers must articulate their intentions through planning and goal decomposition and subsequently translate these intentions into NL prompts.
The second gap is often termed the \textit{abstraction gap}~(Fig.~\ref{fig:cognitive}.G), leading to the challenge of \textit{abstraction matching}. Programmers must continually refine their prompts to ensure they contain the required level of detail for models to generate accurate code.
Because of the abstraction gap, programmers' prompts often lack the essential specificity and precision required for LLMs to accurately translate their intentions into generated code.
For instance, programmers may intend to validate an email address when the submit button is clicked, expressed as \qt{validate email when form submitted} in the prompt.
However, the generated code might implement a validation mechanism triggered when the API is called, deviating from the original intention of immediate email validation upon clicking the submit button in the user interface.
}

\rv{Researchers have proposed several techniques to scaffold the second gap---the abstraction gap between programmers' well-defined intentions and NL prompts.
For instance, Liu et al. introduced the technique of \textit{grounded abstraction matching}, which involves translating the code back into a predictable NL expression~\cite{liu2023wants}.}
Another major approach is to decompose complex prompts into sub-prompts of a pre-defined abstraction level~\cite{cai2023low, ritschel2022can, wu2022promptchainer, wu2022ai, huang2023anpl}. 
Specifically, the programming task is divided into smaller, more manageable sub-tasks, each at a set level of complexity or detail.

\rv{However, previous research has not addressed the first gap in the context of programmers using LLM-driven code assistants---the \textit{intention formation} and \textit{externalization} process~(Fig.~\ref{fig:cognitive}.E, F), which lies between the overarching goal and programmers' specific intentions---despite it being a pivotal factor in successfully tackling programming tasks~\cite{kim1995internal, hoc1977role}.
The \textit{intention formation process} involves the programmer's cognitive thinking process from a high-level goal to concrete intentions, which may entail determining what needs to be done or how to approach the goal.
The intention externalization process involves further operationalizing these intentions into executable NL prompts.
Continuing with the previous email validation example, programmers often begin with a higher-level goal, such as \qt{create a login page.} 
Programmers must further break down this goal into sub-goals that detail how or what actions to take, such as adding input fields or updating values when users type.
The process of goal decomposition and intention formation is a crucial step for both programmers in problem-solving and LLMs in generating code that aligns with programmers' intentions.
}

The goal of this research is thus to explore design opportunities for \rv{supporting programmers in the intention formation process and the subsequent externalization process.}
We conducted a formative study with six experienced programmers who regularly use LLM-driven code assistants. Findings from the study suggest that programmers are hindered due to the lack of informative prompt organization and the inability to directly control and manipulate generated code segments. 
Based on our findings, we propose \sys{}, a system that assists programmers with externalizing hierarchical prompt structures to generate code that aligns with their intentions.
\sys{} introduces the concept of \emph{hierarchical generation} which helps programmers decompose goals into subtasks (Fig.\ref{fig:teaser} A), create a task structure externalizing their intentions (Fig.\ref{fig:teaser} B), and generate corresponding code (Fig.~\ref{fig:teaser} C). Each prompt in \sys{} is a modular \emph{block} linked to code segments, allowing precise modification within the hierarchical task structure.
Furthermore, \sys{} enables evaluation during the prompt authoring process by providing scaffolding through multiple levels of abstraction, including goals, intentions, prompts, pseudocode, and generated code.

We further conducted a user study with 12 experienced programmers who frequently use LLM-driven code assistants to evaluate the usefulness of \sys{}.
The results validate that \sys{} helps programmers form and externalize their intentions.
The direct manipulation of prompt blocks at multiple levels of abstraction to modify the corresponding code provides programmers with control over the translation of their intentions into code.
With the supported scaffolding, \sys{} prevents programmers from disruptive cognitive switching between prompt authoring and code evaluation.
These findings imply the concept of \emph{hierarchical generation} as a design consideration for future developments in interactive LLM-code assistants.
In summary, our contribution is threefold:
\begin{itemize}
    \item A formative study identified programmers' strategies for forming and externalizing intentions to generate code.
    \item An interactive system, \sys{}, employs a hierarchical structure and block-based design to provide programmers with code generation capabilities across multiple levels of abstraction.
    \item A user study demonstrating improved usability by enhancing controllability with \textit{hierarchical generation} and enabling result evaluation during the prompt authoring process.
\end{itemize}

\section{Goal, Intention, Prompt, and Generated Code}
\rv{
In the pursuit of the \textit{goal} of a programming task, programmers must cultivate clear \textit{intentions}~\cite{heinonen2023synthesizing, kim1995internal, hoc1977role}. 
These intentions are articulated from the comprehension of what the program is intended to achieve (declarative knowledge) and the procedures involved in achieving it (procedural knowledge)~\cite{fix1993mental, clark2008cognitive}.
This \textit{intention formation process} involves the dissection of the overarching goal into smaller, manageable subgoals and then the \textit{externalization} of the program's elements with varying depths and extents~(Fig.~\ref{fig:cognitive} E)~\cite{hayes2013new, romero2001focal, fix1993mental, shargabi2015program, jeffries1982comparison}.
Externalizing layered intentions into actions is crucial~(Fig.~\ref{fig:cognitive} F), particularly in LLM-driven systems where the \textit{gulf of execution} can become \textit{fuzzy} due to LLMs' ability to generate results from various formats of NL prompts.
These prompts then carry the programmer's structured intentions into the LLM generation process.
Prior research strived to bridge the \textit{abstraction gap}~(Fig.~\ref{fig:cognitive} G)---the disparity between the human intention behind the prompt and the code generated by LLMs~\cite{liu2023wants, sarkar2022like}.
In contrast, our research emphasizes supporting intention formation (Fig. \ref{fig:cognitive}~A) and its subsequent externalization process, which empowers programmers to precisely guide the code generation with controllability (Fig. \ref{fig:cognitive}~B).
Moreover, our work incorporates features that reduce the \textit{gulf of evaluation}~\cite{norman1986cognitive} (Fig. \ref{fig:cognitive}~D), enhancing the programmer's ability to perceive, interpret, and evaluate the LLM's output without excessive context switching from prompt authoring and code evaluation (Fig. \ref{fig:cognitive}~C).
}

\section{Background and Related Work}
We reviewed related research on challenges in programmer-LLM interaction, existing solutions, and theories related to the intention formation process.

\subsection{Programmer-LLM Interaction}
While the interaction between programmers and AI has been explored in various contexts, our focus is on large language model-driven code generation tools. 
Recent advances in LLMs
mark a significant breakthrough in code generation compared to preceding deep learning models.
Previous research has conducted several user-centered studies to understand how programmers interact with LLMs-based code assistants and their perceptions of these tools~\cite{vaithilingam2022expectation, barke2023grounded, sarkar2022like, mozannar2022reading, nguyen2022empirical, tang2023empirical, pudari2023copilot, dakhel2023github, xu2022ide, liang2023understanding}.
Studies have shown that the accuracy of code assistants has improved significantly with the availability of state-of-the-art LLMs \cite{chen2021evaluating, pearce2023examining}, thereby increasing perceived productivity \cite{ziegler2022productivity, li2022competition}, especially in tasks that require writing simple code snippets repeatedly~\cite{sarkar2022like, barke2023grounded}.

\paragraph{\textbf{Challenges of \rv{Evaluation}}}
However, programmers now need to dedicate considerable time to evaluating AI-generated code suggestions \cite{mozannar2022reading, barke2023grounded}.
Excessive evaluation needs can lead to several issues~\cite{vasconcelos2023generation, weisz2021perfection, ross2023programmer}.
Programmers are often intimidated by the seemingly overwhelming effort required for code validation and bypass the evaluation step.
This causes problems like over-reliance on generated suggestions \cite{barke2023grounded, chen2021evaluating, xu2022ide}, and loss of control over their programs \cite{vaithilingam2022expectation}, which then introduces challenges during code modification \cite{cai2023low, al2022readable}. Programmers are also taxed with the extra cognitive load of switching between programming and debugging tasks \cite{vaithilingam2022expectation, ferdowsi2023live, bastola2023llm, ma2023demonstration}.

\paragraph{\textbf{Abstraction Matching Issue}}
Sarkar et al. \cite{sarkar2022like} observed that programmers often engage in iterative evaluation and prompt refinement to understand how well LLM-driven code assistants can interpret their prompts and generate the desired code, a process referred to as \textit{abstraction matching}.
Programmers are required to grasp the models' capabilities and limitations to understand the necessary naturalistic utterances to generate code that aligns with their intents. This issue is rooted in the notion of the \textit{gulf of execution} \cite{hutchins1985direct}.
The problem of matching abstractions became more noticeable with LLMs due to their capability to generate code at different \textit{levels of abstraction}, ranging from high-level, conceptual descriptions to low-level, detailed pseudo-code-like statements, which cover innumerable combinations of natural language expressions \cite{wu2022ai, liu2023wants}.
Our study extends the focus from abstraction matching from prompt-code to goal-code, considering the need for scaffolding intention formation process for programmers.

\subsection{Improving LLM-based Code Generation}
Compared to technical approaches like prompt engineering and few-shot learning, several design solutions and systems have been proposed to enhance interaction with LLM-driven code assistants.
These strategies encompass various techniques, such as introducing new programming languages \cite{huang2023anpl, beurer2023prompting}, automating prompt rewording~\cite{white2023prompt, fernando2023promptbreeder}, employing programming by demonstration techniques \cite{cypher1993watch, lieberman2001your}, and supporting task breakdowns \cite{ritschel2022can, wu2022ai}.
However, determining the `correct' level of abstraction remains a challenge, as overly detailed prompt decomposition can result in the programming process with LLMs resembling the use of a \qt{highly inefficient programming language}~\cite{sarkar2022like}.
Hence, prior research into natural language interfaces suggests the benefit of managing expectations and gradually revealing the capabilities of the system through user interaction and intervention \cite{luger2016like, liu2023wants, ross2023programmer, vasconcelos2023generation}.

Noticing this issue, prior research has proposed several approaches.
Liu et al. introduced \textit{Grounded Abstraction Matching}~\cite{liu2023wants} that provides a decomposed code example that users can modify and submit to the LLM as instructions, assisting programmers with unclear intentions and reducing abstraction matching problems.
AI Chains enhance programmer control and feedback by breaking problems into sub-tasks~\cite{wu2022ai}. Each sub-task corresponds to a specific step with an NL prompt, and results from previous steps inform prompts for subsequent tasks. This chaining method increases success rates when using the same model on multiple tasks~\cite{wu2022promptchainer, wu2022ai}.

While the previously mentioned approaches that rely on task breakdowns assist programmers in bridging their intent to code, they do not emphasize scaffolding programmers' intention formation process for solving programming tasks. Additionally, they primarily focus on local prompt-code correspondence without examining the overall structure matching, especially from task structure to code structure.
In contrast, \sys{} builds upon the concept of task decomposition and offers increased flexibility and controllability for programmers to not only craft effective prompts but also gain a deeper understanding of how their programming tasks can be logically structured.

\subsection{Programmers' \rv{Intention Formation Process}}
In the programming context, intentions encompass programmers' mental models of understanding and interpretation of the code, underlying programming tasks, and the overall structure of the programs they are working on~\cite{balijepally2012effect, dawson2013cognitive, ye1996expert}.
Several theories describe the formation of these intentions~\cite{von1995industrial, detienne2001software}. While some theories suggest a bottom-up approach, starting with understanding code syntax to derive semantic meanings, others advocate for a top-down strategy that begins with an initial hypothesis of code functionality and then evaluates it through syntax analysis~\cite{von1995program}.
Programmers must develop intentions at different levels of abstraction~\cite{ye1996expert, balz2010continuous}, encompassing specific code statements as well as larger program structures.
To support effective interaction and collaboration between programmers and LLMs in tackling programming tasks, it is crucial to provide scaffolding for these intentions~\cite{fiannaca2023programming, liu2023wants, sarkar2022like}.
In our work, our primary focus is on supporting programmers in the formation of intentions to tackle programming tasks and externalizing these intentions into prompts that generate code in alignment with their goals.
\section{CoLadder: Design Process \& Goals}
\label{sec:design}

\begin{table*}[]
\caption{The summary of challenges and strategies reported in Section~\ref{sec:challenges} and the resulting design goals (Sections~\ref{sec:design_goals}) and features in \sys{} (Sections~\ref{sec:system})} \label{tab:challenges}.
\begin{tabular}{p{0.22\textwidth}|p{0.24\textwidth}|p{0.24\textwidth}|p{0.2\textwidth}}
\toprule
\textbf{C}hallenges & \textbf{S}trategies & \textbf{D}esign \textbf{G}oals & Features \\
\toprule
\multirow{2}{0.22\textwidth}{\textbf{C1}: Unstructured Prompt to Externalize Intention}                        & \textbf{S1}: Task Decomposition                                      & \multirow{2}{0.22\textwidth}{\textbf{DG1}: Hierarchical Prompt Structure}              & \multirow{2}{*}{Prompt Tree Structure}  \\ \cline{2-2}
                                                                & \textbf{S2}: Hierarchical Structure                                  &                                                                 &                                         \\ \midrule
\multirow{4}{0.22\textwidth}{\textbf{C2}: Control Loss from Intents to Code}          & \textbf{S3}: Incremental Generation                                   & \multirow{4}{0.22\textwidth}{\textbf{DG2}:  Direct Manipulation of Prompts for Code Modification}                    & \multirow{2}{*}{Prompt Block}           \\ \cline{2-2}
                                                                & \textbf{S4}: In-Situ Generation                                 &                                                                 &                                         \\ \cline{2-2} \cline{4-4} 
                                                                & \textbf{S5}: Rearrange Code Segments                                  &                                                                 & \multirow{2}{0.22\textwidth}{Block-based Operations} \\ \cline{2-2}
                                                                & \textbf{S6}: Replace Code Segments                            &                                                                 &                                         \\ \midrule
\multirow{3}{0.22\textwidth}{\textbf{C3}: Disruptive Context Switching}             & \multirow{3}{0.22\textwidth}{\textbf{S7}: Evaluate Results during Prompt Authoring} & \multirow{3}{0.22\textwidth}{\textbf{DG3}: Enabling Code Evaluation during Prompt Authoring} & List Steps                              \\ \cline{4-4} 
                                                                &                                                             &                                                                 & Auto-Completion                         \\ \cline{4-4} 
                                                                &                                                             &                                                                 & Recommendation                          \\ \midrule
\multirow{2}{0.22\textwidth}{\textbf{C4}: Unclear Correspondence from Prompt to Code} & \textbf{S8}: Add Code to Prompt                                      & \multirow{2}{0.22\textwidth}{\textbf{DG4}: \rv{Enhancing Prompt-Code Correspondence}}            & Corresponding Code Highlight            \\ \cline{2-2} \cline{4-4} 
                                                                & \textbf{S9}: Evaluated by Comments                                    &                                                                 & Semantic Highlight                      \\ \bottomrule
\end{tabular}
\end{table*}

We conducted an iterative user-centered design to create \sys{}, an interface to help programmers decompose tasks based on their intentions and generate code accordingly. The design process consisted of three stages: 1) \textit{Understanding \& Ideation}---including an interview study with experienced programmers to discover the strategies they employ to address the challenges of programming with LLM-driven code assistants; 2) \textit{Prototype \& Walkthrough}---the design and development of \sys{} informed by established design goals and a cognitive walkthrough for feedback and iterative design (Section \ref{sec:system}); 3) \textit{Deploy \& Evaluate}---a user study to evaluate how programmers interact with \sys{} and their perceived usefulness (Section \ref{sec:study}).
In this section, we describe the first stage of our design process and report the obtained strategies and design goals that guided the design and development of \sys{} (Table.~\ref{tab:challenges}).

\subsection{Interview Study}
\label{sec:formative}
We recruited six participants ($5$ males, $1$ female; ages $25-27, M = 25.8, SD = 0.98$) through purposive sampling~\cite{etikan2016comparison}.
\rv{In our recruitment process, we sought participants experienced in both programming and the use of LLM-driven code generation tools.
Eligibility screening involved a pre-test survey that assessed participants' programming experience on a 5-point scale [1: very inexperienced; 5: very experienced], years of programming experience, and self-reported familiarity with LLM-driven code generation tools.} 
All recruited participants had more than five years of programming experience ($M = 6.67, SD = 1.75$ years) \rv{and were familiar with programming (score $M = 4.33, SD = 0.52$), familiar with LLM-code generation tools (score $M = 4.5, SD = 0.55$)}, and regularly used the LLM-code generation tools ($M = 8, SD = 2.56$ times/week). \rv{Detailed can be seen in Appendix~\ref{appendix:formative}}

Participants provided consent and were compensated with $20$ CAD for $45$-minute study sessions.
\rv{Before the study, we asked each participant to share at least three recent examples of their ChatGPT~\cite{openai2023chatgpt} usage for generating code to nudge them to reflect on how they use LLM-code generation tools.}
In the study, we interviewed participants to assess their challenges in forming and externalizing intentions, translating them into code, the strategies they employed to address these challenges and their needs.
All interviews were audio-recorded and subsequently transcribed into written text.
\rv{We analyzed the interviews using thematic analysis~\cite{braun2012thematic}, employing both inductive and deductive approaches. After interviewing six participants, the first two authors conducted the initial analysis collaboratively.}
We identified and categorized codes and themes related to the strategies participants used to address their challenges and their corresponding user needs. 
\rv{Any disagreements were resolved through discussion and ultimately leading to the final themes after the second iteration.}

\subsection{Interview Results: Strategies}
\rv{Overall, participants used LLM-code assistants across various programming languages (e.g., Python, JavaScript, and Bash) for a wide range of tasks, such as unfamiliar code generation, algorithm implementation, and code refactoring (Appendix~\ref{appendix:formative}).
}

\label{sec:challenges}
\subsubsection{\textbf{Structure Tasks and Prompts Hierarchically}}
We observed that the formation of intentions for participants to solve programming tasks when prompting LLM-code assistants involves two key facets. 
First, programmers need to form clear intentions for solving programming tasks, which is important to \pqt{verify if the generated code is correct or not}{P3}; 
Second, they must explore how to effectively construct the prompt to translate those intents to generated code. However, participants encountered difficulties with the linear representation of prompts (e.g., a sentence of comment), hindering their ability to externalize their intentions and understand their own prompts after composing them (\textbf{C1}).
For instance, P1 expressed, \qt{It [the prompt] becomes meaningless after a few iterations, as the prompt is not for humans but written for the LLM to understand me.}
Every participant adopted a similar strategy to alleviate the cognitive load when forming intentions --- by breaking down tasks into smaller subtasks (\textbf{S1}).
Most (5 out of 6 participants, 5/6 henceforth) participants took things a step further by structuring their tasks hierarchically to externalize their intentions (\textbf{S2}).
Most (4/6) participants used bullet points to structure the prompt, \pqt{I will use indent when writing the prompts; this hierarchy structure helps me think about the detailed steps.}{P6}

\subsubsection{\textbf{Generate and Edit Code Segments by Segments.} }
While participants made efforts to structure their prompts to become more comprehensive, the LLMs exacerbated the difficulties of evaluation by generating an entire codebase based on the whole prompt.
This issue resulted in a sense of control loss and \textit{`fear'} over the code generation process (\textbf{C2}), where participants expressed the desire to \pqt{generate the program bit by bit.}{P2}
Some (3/6) participants mentioned the difficulties in the long-term maintenance of the program, \pqt{I want to be able to return to my code months later and still be able to debug it.}{P6}
As a result, participants generally preferred to accept the generated code line by line, similar to the use of auto-completion features (\textbf{S3}), rather than generating the entire code base at once.
Participants frequently used an additional strategy (\textbf{S4}) to control the length of the generated code. This strategy involved inserting line breaks between prompts, allowing them to generate code selectively between specific prompts rather than generating lengthy code encompassing all prompts. However, implementing this strategy introduced a misalignment issue between the structured prompts and the code structure, which subsequently made it challenging for participants to determine \pqt{where to insert newly generated code.}{P2}
Participants (4/6) thus reported an alternative strategy, where they generated self-contained code segments independently, and then merged them into the existing codebase (\textbf{S5}). P5 explained the reason, \qt{I will combine it [generated code] by myself as I didn't know what it would look like beforehand.}
Another common strategy is to select a segment of existing code and replace it with the newly generated code (\textbf{S6}), enabling manipulation of targeted code segments without impacting other sections.
Yet, P3 outlined this tedious process of preserving and combining both existing and newly generated code, \qt{sometimes if the generated code is partially accurate, I need to initially accept it, make a copy, undo the changes, and finally paste the copied code below my original code.}

\begin{figure*}
    \centering
    \includegraphics[width=1\linewidth]{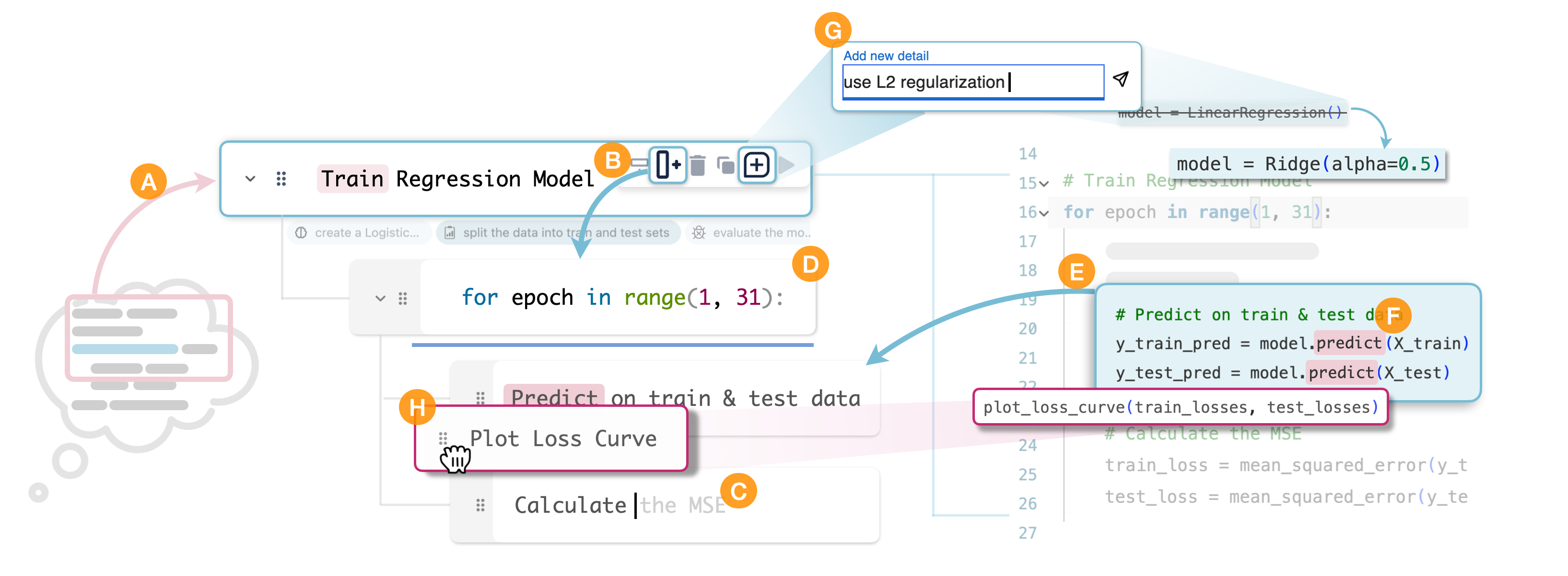}
    \caption{An example workflow of using \sys{}. A user creates a high-level prompt (A) based on their intentions of the programming task. Subsequently, they add a sub-task indented underneath (B), incorporating code syntax within the prompt (D). The \textit{List Steps} feature is employed to summarize the generated code into prompts (E). Following evaluation, the user modifies the prompt, accepting \textit{auto-completed} suggestions (C). To ensure code accuracy, the user employs the \textit{semantic highlight} feature (F). When additional details are needed, they use the \textit{supplement} feature to add detail to the prompt (G). Finally, the user rearranges the prompt structure using the \textit{Drag and Drop} feature (H).}
    \Description{This figure demonstrates a typical workflow in \sys{}. Initially, a user creates a high-level prompt (A) based on their intentions of the programming task. Subsequently, they add a sub-task indented underneath (B), incorporating code syntax within the prompt (D). The 'List Steps' feature is employed to summarize the generated code into prompts (E). Following evaluation, the user modifies the prompt, accepting auto-completed suggestions (C). To ensure code accuracy, the user employs the semantic highlight feature (F). When additional details are needed, they transition to a lower level (L2) and supplement the prompt (G). Finally, the user rearranges the prompt structure using the Drag and Drop (DnD) feature (H).}
    \label{fig:usage-scenario}
\end{figure*}

\subsubsection{\textbf{Evaluate Generated Code through Prompts}}
All participants expressed frustration with current tools that constantly suggest results, leading to continuous switching between prompt authoring and code evaluation (\textbf{C3}).
They noted that although this context switching may seem trivial, it significantly disrupts their overall programming flow. P6 mentioned, \qt{Sometimes I know it [the generated code] would not be correct, but I will still look into it,} and P4 explained that \qt{probably this time the generated code is correct, who knows?}
However, we observed that participants more frequently accepted auto-completed prompts, using them to quickly \pqt{verify if the system had accurately captured my [their] intents.}{P3} (\textbf{S7}) Some participants (P1, P2, P5), deliberately waited for the code assistants to auto-complete their prompts, which further \pqt{reduced the time needed to verify the generated code.}{P2}

\subsubsection{\textbf{Adding Cues to Navigate between Prompt-Code}}
Participants usually need to go through several iterations of the prompt, which makes it challenging to locate the phrases to modify as the prompt grows increasingly lengthy. All participants thus reported difficulties in navigating back and forth between prompt and generated code segments, finding it \pqt{difficult to find which part [of the prompt] is causing the error that needs to be modified.}{P1} (\textbf{C4})
Several programmers (4/6) incorporate code syntax into their prompts to facilitate code evaluation by making it easier to locate keywords (\textbf{S8}). For instance, P2 mentioned using pseudo-code-like prompts as a strategy \qt{to control the generation} and {reduce the cognitive load during the evaluation process.} 
Some participants (2/6) also mentioned that they would rather rewrite the whole segment of code, rather than attempt to edit and debug it. \pqt{After generating code, I need to spend a lot of time finding the code causing the error and also finding the prompt that resorted to this result, I would just start all over again.}{P1}
Participants usually valued the generated comments (in NL) and used them as anchor points to match segments of prompts to code (\textbf{S9}). P4 mentioned the reason that \qt{the generated comments are useful to check if the generated code matches my instructions step-by-step.}

\subsection{Design Guidelines for Supporting Programmers' Strategies}
\label{sec:design_goals}

To support the reported strategies (\textbf{S1-S9}) programmers leveraged to overcome challenges they encountered (\textbf{C1-C4}), we formulated four design guidelines (\textbf{DGs}).
The main design goal is to offer \textit{hierarchical generation}, enabling programmers to form and externalize abstract intentions into generated code while ensuring alignment.

\paragraph{\textbf{DG1: Offering Hierarchical Prompt Structure}}

The lack of structured prompts hinders programmers from forming and externalizing the intentions for solving programming tasks (\textbf{C1}). The system should support prompt decomposition (\textbf{S1}) with a hierarchical representation of the prompt structure that externalizes the task structure programmers possess in mind (\textbf{S2}).
Previous research adopted prompt decomposition by pre-defining the permitted abstraction levels \cite{wu2022ai, ritschel2022can, huang2023anpl, beurer2023prompting}, which may not necessarily reflect programmers' own intuition of how they would have decomposed the task \cite{xie2017influential, carroll1988mental}. Participants adopt a more flexible prompt structure that varies based on the task's nature and complexity. Hence, the system should enable participants to define abstraction levels for externalizing their intentions.

\paragraph{\textbf{DG2: Direct Manipulation of Prompt for Code Modification}}
Programmers often experience control loss when evaluating and modifying large amounts of generated code (\textbf{C2}). This issue highlights the need to provide programmers with control over the prompt authoring process. 
Such support should include the ability to easily identify, and select the range of modifications (\textbf{S6}), insert new prompts (\textbf{S4}), and reorganize the structure of prompts (\textbf{S5}). The system must facilitate modifications at different levels of abstraction that allow programmers to make the necessary changes without inadvertently modifying unrelated code segments. The system should also generate results incrementally (\textbf{S3}) instead of generating entire code simultaneously and overwhelming the programmer.

\begin{figure*}
    \centering
    \includegraphics[width=\linewidth]{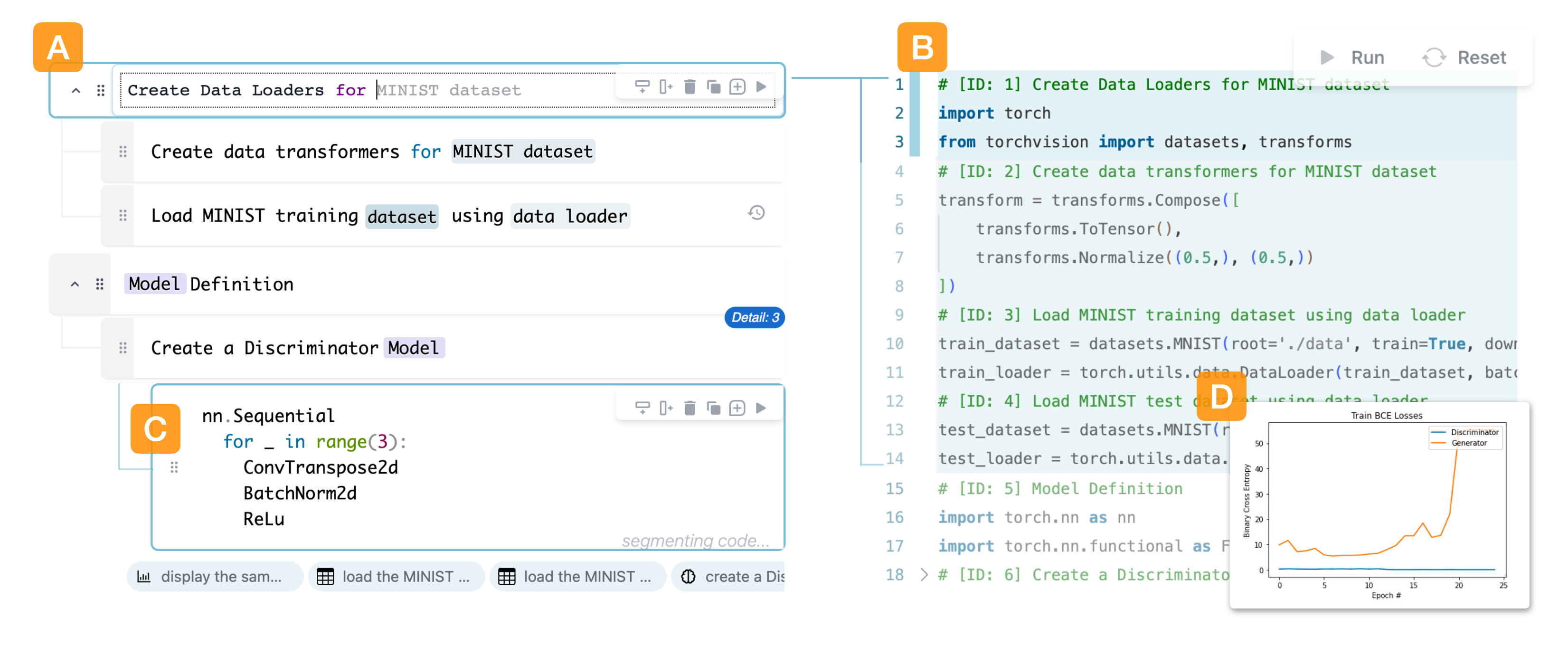}
    \caption{\sys{} comprises four key components: (A) the prompt tree editor, allowing programmers to decompose their intent into smaller prompt blocks; (B) the code editor, facilitating code evaluation and editing; (C) the prompt block, enabling programmers to compose prompts in \textit{mixed mode}, incorporating both code and natural language; and (D) the execution result panel, which displays the execution result and any associated error messages.}
    \label{fig:sys_overview}
    \Description{The \sys{} system overview is presented in this image, showcasing its four main components: Prompt Tree Editor (Left): This component allows programmers to break down their intent into smaller prompt blocks. It is a crucial tool for organizing and structuring their programming tasks hierarchically. Code Editor (Right): The code editor enables programmers to evaluate and edit the generated code. It serves as a workspace for fine-tuning and modifying the code to align with their intent. Prompt Block (Within the Prompt Tree): In this component, programmers can write prompts in `mixed mode.' This means they can seamlessly integrate both code and natural language within the same prompt block, providing flexibility in communication with the system. Execution Result Panel (Bottom Right): The execution result panel is designed to display the outcomes of code execution, including the generated results and any error messages that may occur during the coding process. This image offers a comprehensive visual representation of the \sys{} system, illustrating how its various components work together to assist programmers in generating and evaluating code.}
\end{figure*}

\paragraph{\textbf{DG3: Enabling Code \rv{Evaluation} during Prompt Authoring}}
During the iterative refinement of prompts, programmers often experience cognitive overload due to the disruptive context switching between code evaluation and prompt authoring (\textbf{C3}). 
Findings highlight the possibility of assisting programmers in evaluating generated results during the prompt authoring process without necessitating additional context switching.
The system should deliver feedback in a non-intrusive manner and provide context to reflect the system's understanding of the task. This approach helps programmers understand whether the system accurately captures their intent (\textbf{S7}).

\paragraph{\textbf{DG4: Enhancing Prompt-Code \rv{Correspondence for Evaluation}}}
To facilitate navigation and modification across various levels of abstraction, from user intents to the final generated code (\textbf{C4}), the system must ensure correspondence between prompts and code, including a matching between the overall task and code structure.
The system should also provide visual cues to highlight the segments of the generated code according to the prompt structure (\textbf{S9}), for programmers to navigate and modify the desired code segments. In addition, the system should enable programmers to write prompts containing code syntax (\textbf{S8}) to help them efficiently pinpoint the corresponding code.

\subsection{Usage Scenario}
Casey is a data scientist who wants to build a regression model on a wine-quality dataset. 
While being experienced in Python, she aims to leverage LLM-driven code assistant to speed up her development process, and thus she launches \sys{}.
Casey starts by considering the main steps to approach this task by outlining primary objectives, such as partitioning the dataset, building and evaluating the regression model, and plotting the results.

{\small\textbf{Prompt Authoring with Hierarchical Decomposition. }}
Casey translates her intent into a set of high-level prompts, such as \qt{Train Regression Model} (Fig.~\ref{fig:usage-scenario}~A), to externalize the code structure she envisioned.
Next, she goes deeper and adds some sub-tasks under the high-level prompts with the \textbf{[Add Child]} \includegraphics[width=0.028\linewidth]{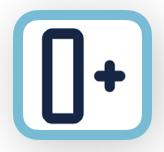} button, adjusting the level of detail as needed (Fig.~\ref{fig:usage-scenario}~B). For example, under \qt{Train Regression Model}, she adds sub-tasks such as \qt{Partition the Dataset.} 
Casey maintains this breadth-first approach, gradually detailing each high-level task with the \textbf{[Add Siblings]} \includegraphics[width=0.028\linewidth]{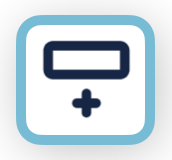} button.
As Casey added each prompt block, the code was updated in real time according to the overall task structure. However, the code editor displayed only the code for existing prompt blocks, with the rest of the code segments remaining folded.
When Casey crafts these prompts, she leverages the prompt \textbf{[Auto-Complete]} feature to quickly formulate more detailed prompts (Fig.~\ref{fig:usage-scenario}~C). 
In some cases, she uses code syntax expressions such as 
\begin{codesyntax}
for epoch in range(1, 31):
\end{codesyntax}
or 
\begin{codesyntax}
load_boston()
\end{codesyntax}
without the need to translate the code statements to NL (Fig.~\ref{fig:usage-scenario}~D).
To define finer-grained steps under each sub-task, Casey sometimes adds sub-prompt blocks manually and sometimes utilizes the \textbf{[List Steps]} feature, which automatically suggests step-by-step guidance for the code to generate.
For example, under \qt{for epoch in range(1, 31)}, the listed steps recommend actions like \qt{Predict on Train and Test data} that is summarized from the relevant code snippets concerning the model prediction (Fig.~\ref{fig:usage-scenario}~E)
assisting her in evaluating the alignment of the intent-code.

{\small\textbf{Navigating and Evaluating through Multi-level Prompts.}}
Casey then navigates through the hierarchical structure with up/down arrow keys and evaluates highlighted code segments corresponding to the specific prompt block. 
The \textbf{[Semantic Highlight]} feature helps her correlate phrases in her prompts with the code segments (Fig.~\ref{fig:usage-scenario}~F). For instance, her prompt mentions the ``
\begin{shighlight}
Predict
\end{shighlight}
\textit{on train \& test data}''
results in highlighted 
\begin{shighlight}
.predict()
\end{shighlight}
in the code segments and phrases in the prompt block (e.g., ``
\begin{shighlight}
Train
\end{shighlight}
\textit{Regression Model}'') at the parent level with lower opacity representing lower correlation.
Casey now affirms that the LLM accurately interpreted her intent based on the tree structure.

{\small\textbf{Modification and Block-based Operations.}}
As Casey navigates through prompts, she identifies code segments that require adjustments. 
In the block \qt{Train Regression Model}, she modifies the prompt by the \textbf{[Supplement]} \includegraphics[width=0.028\linewidth]{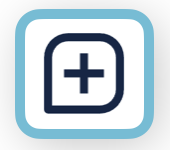} feature to specify using L2 regularization and resulted in changing the code from 
\begin{codesyntax}
LinearRegression()
\end{codesyntax}
to
\begin{codesyntax}
Ridge(alpha=0.5)
\end{codesyntax}
(Fig.~\ref{fig:usage-scenario}~G). 
Further, she uses the \textbf{[DnD]} \includegraphics[width=0.028\linewidth]{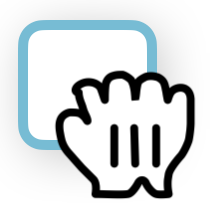} feature to relocate the block \qt{Plot Loss Curve} under the block \qt{for epoch in range(1, 31)}, resulting in the generation of a plot for each training epoch (Fig.~\ref{fig:usage-scenario}~H).
In the end, Casey compiles and runs her code, observing that the system successfully outputs 30 graphs displaying loss curves that match her intents.
\section{\textbf{CoLadder}}
\label{sec:system}
\begin{figure*}
    \centering  
    \includegraphics[width=\linewidth]{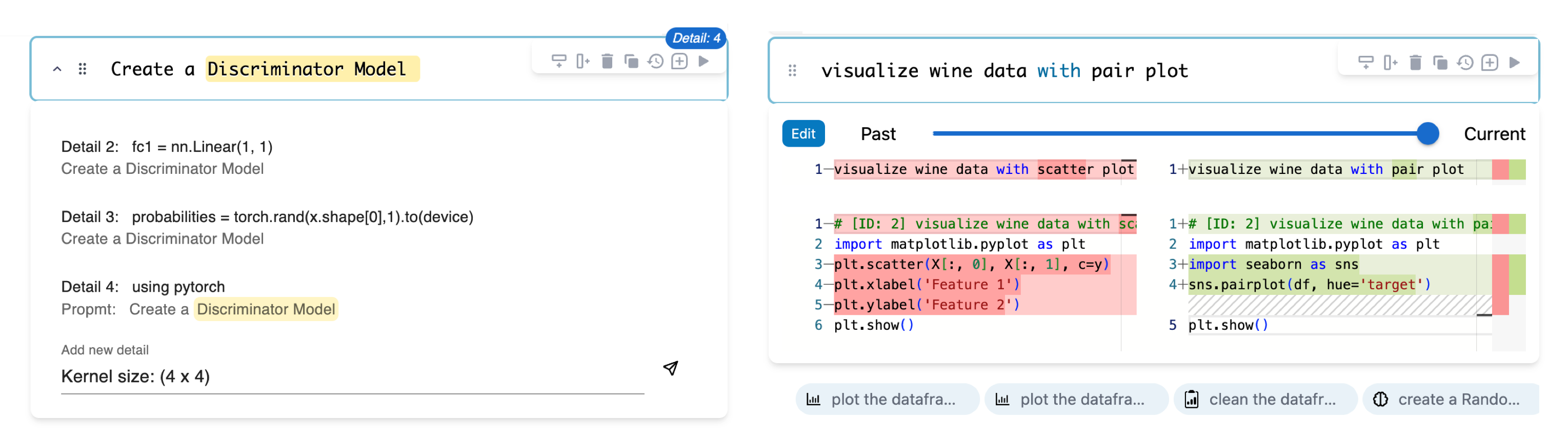}
    \caption{Supplement View (Left): Programmers can add additional details to their prompts via a conversational UI. It also shows the history of supplements and expand or hide by clicking the top right icon; History's Diff View (Right): programmers can observe various types of changes (such as edits, additions, supplements), iterations of prompts, and their corresponding generated code in the diff view.}
    \label{fig:supplement_diff}
    \Description{Supplement View (Left): Programmers can use this view to add additional details to their prompts via a conversational user interface. It also allows programmers to access the history of supplements, which they can expand or hide by clicking the top right icon displaying the number of supplements; History's Diff View (Right): programmers can observe various types of changes (such as edits, additions, supplements), iterations of prompts, and their corresponding generated code. These iterations of changes are visually presented in the diff view. There is a slider to locate the specific iterations.}
\end{figure*}

\sys{} consists of two main UI components: 1) The \textbf{prompt tree editor} (Fig.~\ref{fig:sys_overview} A) allows the programmer to externalize their intentions by decomposing the programming task into smaller prompt blocks (Fig.~\ref{fig:sys_overview} C); 2) The \textbf{code editor} (Fig.~\ref{fig:sys_overview} B) allows the programmer to evaluate the generated code and directly edit the program. The programmer can also compile and run the current code to see the results or errors below the code editor (Fig.~\ref{fig:sys_overview} D).

In the following, we discuss \sys{}'s functionalities and design in detail based on \textbf{DG1-4}.
We also articulated insights from the \rv{same six experienced programmers who participated in our interview studies, inviting them to take part in cognitive walkthrough experiments during the iterative design process.
During the walkthrough, an experimenter presented the Figma prototype, verbally explained the interaction flow, and instructed participants to perform specific actions. 
Following the system walkthrough, we conducted semi-structured interviews to collect feedback on the detailed design of features, their effectiveness, and possibilities for future improvements.}

\subsection{From Task Structure to Code Structure}
To assist programmers in structuring their prompts hierarchically to externalize their intentions (\textbf{DG1}), we offer a tree-based prompt editor that enables programmers to construct prompts that reflect both the task and code structure. Furthermore, we decomposed tasks into smaller sub-tasks at multiple levels of abstraction. This approach enables programmers to directly manipulate their intent to code based on the hierarchical structure (\textbf{DG2}).

\subsubsection{\textbf{Prompt Tree Editor.}}
The tree-based visualization helps programmers organize tasks in line with the top-down mental programming model. This tree editor allows programmers to convey task structure through the horizontal indentation of sub-tasks while still maintaining the program structure vertically. For instance, if a programmer has added a task, \qt{Extracting quotes from a web page,} they can represent the hierarchical task structure and execution order of the code by adding a sub-task, \qt{Find all class=quote,} indented underneath.
Rather than automatically decomposing tasks, \sys{} allows programmers to construct prompts flexibly that align with their intentions. Based on expert walkthrough suggestions, we implemented a foldable prompt tree, aligning with the foldable code editor. This is useful for longer programs, eliminating the need for constant scrolling.

\subsubsection{\textbf{Prompt Block.}} 
Each decomposed task in the tree nodes is referred to as a \textit{prompt block}, where programmers can write the prompt in \textit{mixed mode} (Fig.~\ref{fig:sys_overview} C). \rv{Programmers have the flexibility to input both NL and code syntax, aligning with their preference for occasionally using code syntax in the prompt to express their thoughts.
For instance, participants often opt for statements like \qt{for i in...} rather than NL expressions such as \qt{iterate through the...}.}
Each block functions as a miniature code editor, with semantic highlighting of code syntax and prompts.
Several participants (4/6) wanted prompt block revision history to aid in recalling the reasoning behind specific prompts. 
\sys{} documents prompt block iterations, visualizing differences in prompts and generated code, enabling programmers to efficiently navigate and recover specific iterations as required~(Fig.~\ref{fig:supplement_diff} Right).

\begin{figure*}
    \centering
    \includegraphics[width=\textwidth]{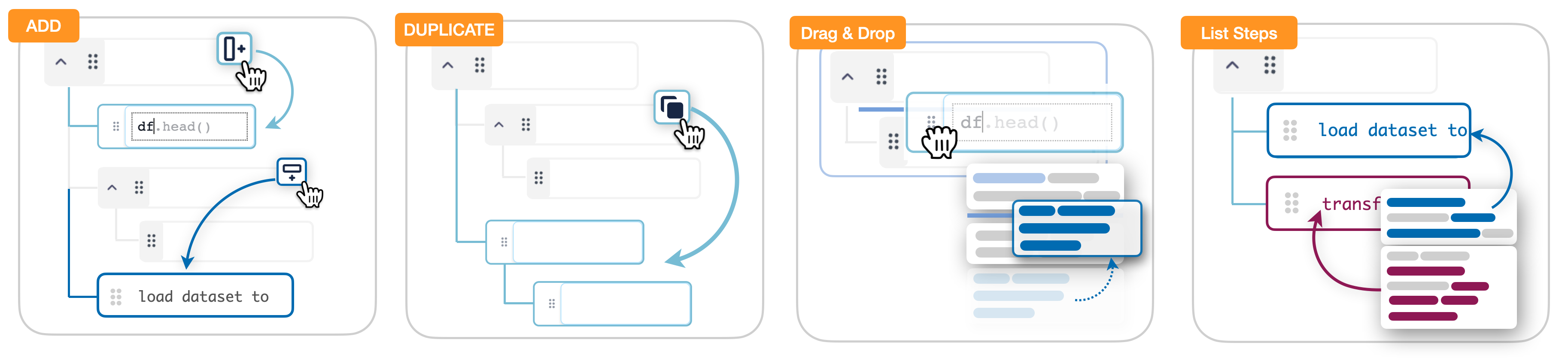}
    \caption{\textit{ADD} Operation enables the addition of sibling blocks or child blocks to the existing prompt structure; \textit{DUPLICATE} operation allows for the cloning of sub-trees; \textit{DnD} operation empowers programmers to reorganize prompt blocks; \textit{List Steps} provides high-level, summarized steps from the generated code.}
    \label{fig:operations}
    \Description{ADD Operation enables the addition of sibling blocks or child blocks to the existing prompt structure; DUPLICATE operation allows for the cloning of sub-trees; DnD operation empowers programmers to reorganize prompt blocks; List Steps provides high-level, summarized steps from the generated code.}
\end{figure*}

\subsubsection{\textbf{Block-based Operations.}} 
\sys{} supports direct manipulation of the prompt structure by providing several prompt block operations that are activated through either buttons or shortcuts. Each block-based operation will update the corresponding code and propagate changes to the rest of the code as necessary.

\begin{enumerate}
    \item \textbf{[Add]} either a block as a sibling (same level) or child (sub-level) based on their intent (Fig.~\ref{fig:operations} ADD).
    After adding a block, the programmer can start entering their prompt to guide the system to generate code based on the current task structure in the prompt tree editor; 

    \item \textbf{[Edit]} allows programmers to refine prompts when they want to modify specific code segments. This modification adheres to a hierarchical structure, so when programmers edit a parent block containing multiple child blocks, the changes apply uniformly to all code segments within those child blocks, ensuring consistency across related sections.

    \item \textbf{[Delete]} unneeded prompt blocks (e.g., parent blocks and all their children). Similar to the \textbf{[Edit]}, this operation will only affect a segment of the code and propagate the changes across the rest of the code to prevent errors.

    \item \textbf{[Duplicate]} copies prompt blocks, and if the code block is a parent block with sub-blocks, it duplicates all its child blocks, creating an identical structure (Fig.~\ref{fig:operations} DUPLICATE);

    \item \textbf{Drag and Drop ([DnD])} restructure prompt blocks or elevate certain code segments to a higher-level scope, allowing them to create reusable functions or methods accessible by other parts of the program (Fig.~\ref{fig:operations} Drag \& Drop);

    \item \textbf{[Supplement]} adds extra details either to the entire prompt block or specific phrases within the NL prompt. This feature addresses the need for additional information required by the LLM but not necessarily by the programmer to understand the program.
    Once a programmer submits supplementary information, it will appear as a badge in the top-right corner of the prompt block, accessible by expanding it. 
\end{enumerate}

\subsection{Evaluate Results from Prompt}
\sys{} offers diverse informative feedback to assist programmers in assessing whether the system accurately comprehends their intent (\textbf{DG3}). This capability enables programmers to concentrate on crafting prompts without experiencing disruptive cognitive shifts between prompt authoring and code evaluation.

\begin{figure*}
    \centering
    \includegraphics[width=1\linewidth]{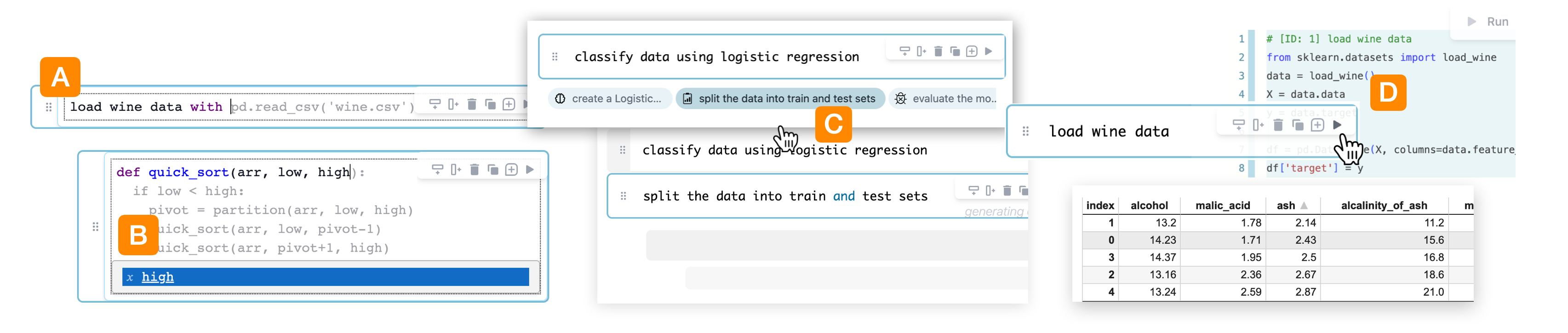}
    \caption{(A) Auto-completion that could be completed in the format of natural language and code syntax; (B) Auto-completion based on the variable name used before; (C) Recommendation feature that suggests the next step based on the prompt tree structure; (D) Live execution showing the interim results of the current block.}
    \label{fig:complete_recommend_exc}
    \Description{(A) Auto-completion that could be completed in the format of natural language and code syntax; (B) Auto-completion based on the variable name used before; (C) Recommendation feature that suggests the next step based on the prompt tree structure; (D) Live execution showing the intermediate results of the current block.}
\end{figure*}

\subsubsection{\textbf{List Steps at Lower-Level.}} To help programmers evaluate the generated code effectively, we support the \textit{List Steps} operation, which allows programmers to understand how the model generated the code for the current prompt block (Fig.~\ref{fig:operations} List Steps).
\rv{This feature semantically segments the generated code and provides a step-by-step summarization of these segments, which is then added as new sub-prompt blocks.} 
It serves as scaffolding for programmers to comprehend the lower-level details of generated code.

\subsubsection{\textbf{Auto-completion. }} Participants in the formative study value the in-line auto-completion and view it as a step to evaluate the result. \sys{} support programmers with two types of auto-complete while editing prompt blocks: 1) Word-level auto-completion based on variables or semantically related naturalistic utterance (Fig.~\ref{fig:complete_recommend_exc} B), and 2) Sentence-level prompt auto-completion from LLM suggests in-line auto-completed prompts based on the context of the current tree structure (Fig.~\ref{fig:complete_recommend_exc} B). Programmers can press the tab button to accept these suggestions.

\subsubsection{\textbf{Recommendation}} After adding a new prompt block, \sys{} provides multiple possible next steps to the programmer (Fig.~\ref{fig:complete_recommend_exc} C). These recommended prompts are displayed below the current prompt block in order of relevance scores suggested by the model. Programmers can select one of them to add below. This feature assists programmers in accomplishing their goals in a step-by-step manner and helps them evaluate if \sys{} correctly understands their intent by successfully recommending the appropriate next steps.

\subsubsection{\textbf{Interim Results. }} The \textbf{[Compile]} operation allows programmers to independently execute each block to display the interim results (Fig.~\ref{fig:complete_recommend_exc} D). \sys{} wraps the code from the prompt tree and compiles the code to display the results. Programmers can evaluate the interim results to determine the next step.

\subsection{Facilitating Prompt-Code Correspondence}
\sys{} offers features for programmers to navigate various abstraction levels to locate and modify targeted prompt blocks (\textbf{DG4}).

\subsubsection{\textbf{Showing Corresponding Code.} } 
Participants found that evaluating individual code segments was often sufficient.
In response, \sys{}'s code editor view highlights only relevant code when programmers select the corresponding prompt block, folding other code segments.
Different code editor glyph's decorations illustrate the current location of the prompt in the prompt tree. \rv{For finer adjustments, programmers can directly modify the code in the code editor, where changes will be propagated back to the corresponding prompt block.}

\subsubsection{\textbf{Semantic Highlight and Dependency.} }
\begin{figure*}
    \centering
    \includegraphics[width=0.8\linewidth]{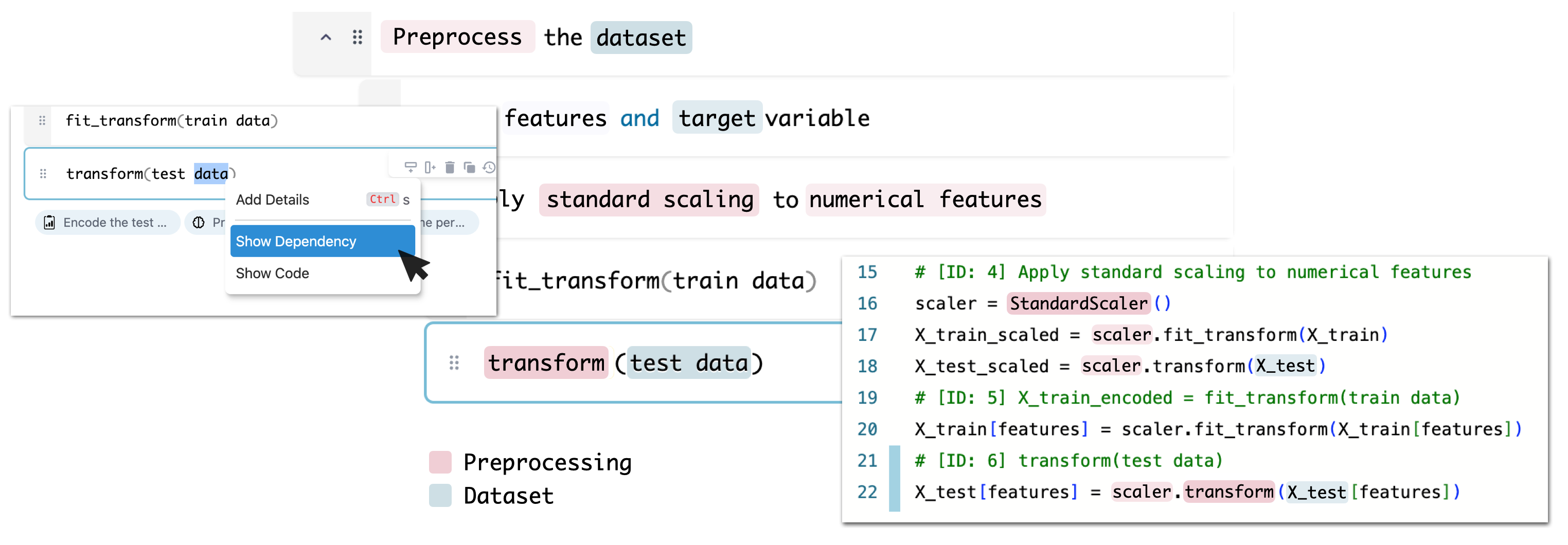}
    \caption{Programmers can select a phrase within the prompt, and the system will highlight correlated phrases throughout the tree structure and code segments. Colours are used to represent the entity type, while opacity indicates the correlation score.}
    \label{fig:dependency}
    \Description{The system establishes a direct correspondence between prompts and the generated code while providing semantic highlighting. Programmers can select a phrase within the prompt, and the system will highlight correlated phrases throughout the tree structure and code segments. Colours are used to represent the entity type, while opacity indicates the correlation score, offering a visual aid for better understanding and alignment between prompts and code.}
\end{figure*}

\sys{} offers two types of syntax highlighting for prompt blocks: semantic highlighting to differentiate between code and NL in mixed-mode editing.
Secondly, highlight NL phrases to help understand dependencies across prompt blocks at different levels. This is beneficial when the programmer refers to the same variable in the code with different terms (e.g., df, data, table that all refer to the same dataframe). The programmer can select phrases within the prompt, and \sys{} will display semantically related phrases throughout the tree, with different entity types shown in distinct colours, and opacity determined by the relevance score (Fig.\ref{fig:dependency}).
The relevant code segment is highlighted in the code editor for easy identification by the programmer via corresponding phrases from the prompt.

\subsubsection{\textbf{Keyboard Navigation.}}
\sys{} offers keyboard shortcuts that enable programmers to access features and effectively navigate across blocks. The arrow keys ($\downarrow$$ / \uparrow$) can move through prompt blocks at various levels with the highlighted corresponding code segments in the code editor. Programmers can also use 
\begin{shortcut}
Enter
\end{shortcut}
to start editing, 
\begin{shortcut}
Esc
\end{shortcut}
to record the editing, 
\begin{shortcut}
Alt
\end{shortcut}
+ $\downarrow / \uparrow $ to create siblings/ children/ and 
\begin{shortcut}
Alt
\end{shortcut}
+
\begin{shortcut}
Enter
\end{shortcut}
to activate the \textit{List Steps} feature.

\subsection{System Implementation}
\sys{} is built on the Next.js framework, enabling server-side rendering for API calls, including the OpenAI GPT-4 API~\cite{openai2023gpt4} for hierarchical code generation.
The user interface incorporates the Monaco Editor~\cite{microsoft2023monaco}, providing an intuitive coding experience in both prompt blocks and the main code editor view.
\rv{To execute and compile results from LLMs, \sys{} utilizes Pyodide~\cite{pyodide2023}, a potent Python web compiler.
Logging is managed through Firebase's Real-Time Database, categorizing interactions and responses by unique user and condition IDs.
The entire \sys{} platform is built and deployed on Vercel, accessible through a public domain URL.}

\label{sec:tree-prompt}
\subsubsection{Prompting Techniques}
\rv{
For all code generation features, we structured the prompt tree into a text-based tree structure with indices and depth specifying the location of each prompt block~\cite{khot2022decomposed}.
We incorporated a set of few-shot examples derived from hierarchical prompt use cases identified in the interview study to facilitate the model's in-context learning of the tree structure~\cite{brown2020language}. We further developed an output parser to organize LLM responses into a tree format, where nodes contain unique indexes, prompts, and code.
In addition, we adopted the Chain-of-Thought prompting technique~\cite{wei2023chainofthought, li2023chain} by using LangChain~\cite{langchain2023}. This involves guiding the LLM to first generate intermediate reasoning steps in NL, forming a logical sequence that leads to the final code output. This approach is beneficial for decomposing complex tasks into manageable steps, ensuring logical consistency and adherence to programming practices.
For all block-based operations, recommendations, or auto-completions, the parsed prompt tree serves as the context, combined with specific prompt templates. These templates and examples vary based on the operation, allowing for targeted code generation that aligns with the intended action (Appendix~\ref{appendix:prompt}). We developed and tested the prompt template using OpenAI's GPT-4, the most advanced and publicly available LLM to date.

\subsubsection{Features and Block-based Operations}
In the [Add] operation, we instruct the LLM to adopt a bottom-up approach, starting from the lowest indentation levels and progressively integrating the child nodes' code with their parent nodes.
For [Edit] operations, the LLM generates code corresponding to the specific prompt block, taking into account the history of each prompt iteration as part of the context. This history acts as a buffer memory, aiding in semantic searches for related results to regenerate by capturing the difference between variants and discerning the programmer's intent. Note that \sys{} automatically records the prompt iteration history when a prompt block is edited.

Error prevention and correction mechanisms are implemented post-code generation to ensure any necessary changes are applied throughout the codebase (e.g., variable adjustments).
After performing prompt block operations, a sequential chain is established to use preceding operation outputs as inputs for subsequent actions. If changes are needed, the system parses the generated text with the Myers diff algorithm, updating only the segments with applicable changes.
For scenarios where programmers edit the code, the corresponding prompt blocks, including all blocks in the subtree containing that segment of code, are updated to ensure subsequent operations consider the manually modified code.

In addition to block-based operations, the Semantic Highlighting feature follows a process where it initially segments utterances and pairs them with corresponding code segments.
Subsequently, it prompts LLMs to assess the similarity between selected utterances and code segments in comparison to other pairs of utterances and code (Appendix~\ref{appendix:prompt2}). A relevance score is determined based on the cosine similarity of text embeddings, and it is further categorized using named entity recognition with a predefined set of entities (e.g., dataset, preprocessing, and variable).
}

\section{Evaluation}
\label{sec:study}
We conducted a study involving 12 experienced and frequent LLM-based code assistants programmers to assess the efficacy of \sys{} in addressing the following \textbf{R}esearch \textbf{Q}uestions (aligned with the A-D in Fig.~\ref{fig:cognitive}):
\begin{itemize}
    \item \textbf{RQ1:} Whether and how \sys{} supports programmers in their \rv{intention formation and externalization process}?
    \item \textbf{RQ2:} Whether \sys{} provides programmers with \rv{control when translating intentions into generated code}?
    \item \textbf{RQ3:} Whether and how \sys{} reduces \rv{context switching between prompt authoring and results evaluation}?
    \item \textbf{RQ4:} Whether \sys{} enhances \rv{prompt-code correspondence} for programmers to \rv{evaluate generated code}?
\end{itemize}

\begin{figure*}
    \centering
    \includegraphics[width=\linewidth]{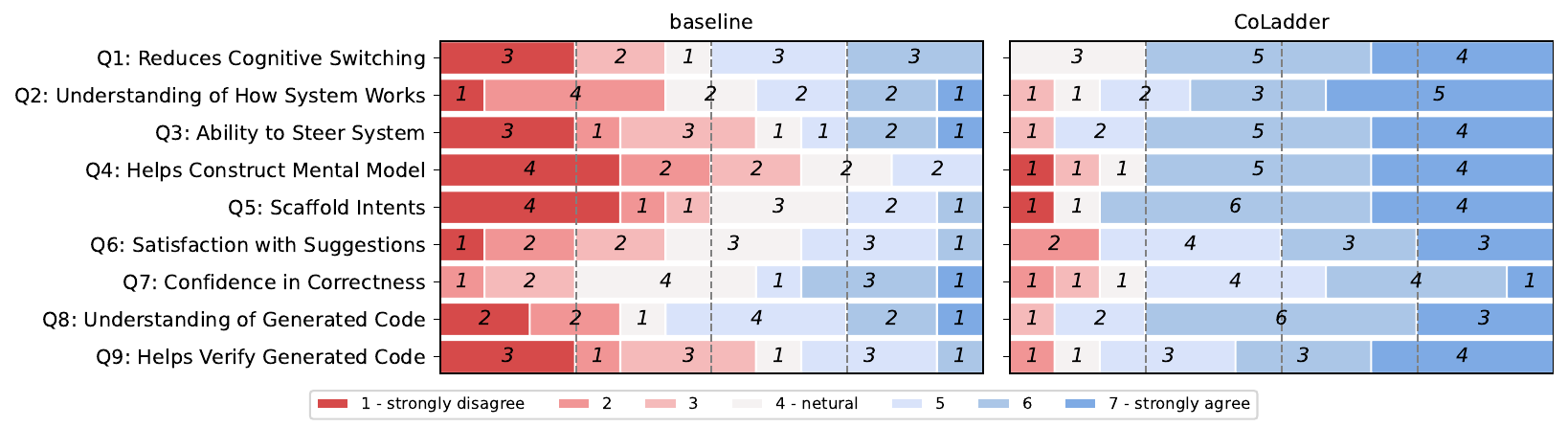}
    \caption{User perception of utility of \baseline{} and \sys{}, measured on self-defined $7$-point Likert scales (Appendix~\ref{appendix:questionnaire}).}
    \Description{
    User perception of utility of \baseline{} and \sys{}, measured on self-defined $7$-point Likert scales. Overall, \sys{} performed better than \baseline{} for all items, and the difference was significant for most statements. For a complete list of all Likert scale items evaluated, see the appendix.
    }
    \label{fig:self-likert}
\end{figure*}

\subsection{Participants}
We recruited 12 participants ($7$ males, $5$ females; ages $23-36, M = 26, SD = 3.54$) through \rv{purposive} sampling~\cite{etikan2016comparison} via the university mailing list. 
\rv{We selected experienced programmers with Python proficiency scores of 4 or higher on a 1 to 5 scale~\cite{xu2022ide, vaithilingam2022expectation}. This choice is because experienced programmers can better construct mental representations of programming solutions compared to novices, who often struggle to connect code segments with their intended goals~\cite{romero2001focal, hoc1977role, wiedenbeck1993characteristics, fix1993mental}. Novices may find it challenging to fully utilize \sys{}, which requires externalizing and matching intentions with code when their intentions are not well-formed.
Furthermore, we screened participants based on their familiarity with LLM-code assistants, particularly with GPT-4, using a self-assessed rating on a 5-point Likert scale. This step was taken to ensure that our participants have a solid understanding of the capabilities of the language model we utilized and are experienced in crafting effective prompts.} 
The final 12 participants we recruited for the study are experienced programmers ($M = 7.88, SD = 4.34$ years) and confident in Python programming (score $M = 4.42, SD = 0.51$).
They also regularly use LLM-code generation tools ($M = 8, SD = 2.56$ times/week) and self-reported being familiar with them based on a $5$-point Likert scale (score $M = 4, SD = 0.74$).

\subsection{Programming Tasks}
\rv{
To select tasks for our study, we applied three criteria: 1) Time: We aimed for tasks that could be completed within 12-15 minutes to minimize participant fatigue; 2) Question Type: We focused on tasks that required participants to perform specific actions, excluding queries about language features or package installations; 3) Complexity: Tasks needed to be complex enough to prevent GPT-4 from generating complete solutions, forcing participants to form intentions for problem-solving and evaluation.
We drew inspiration from programming tasks used by Xu et al.~\cite{xu2022ide} and Vaithilingam et al.~\cite{vaithilingam2022expectation}, sourced from Stack Overflow and categorized into seven common programming task types~(Table~\ref{tab:program-tasks}). We selected medium to high-difficulty categories and asked participants to self-rate their expertise in these categories on a 5-point Likert scale during screening, with a threshold of a rating above 3 for category selection.
Based on participants' ratings, we chose Machine Learning
(score $Mdn$=$4$, $\sigma$=$1.19$) and Data Visualization (score $Mdn$=$4$, $\sigma$=$0.75$) as study categories. We adapted tasks to meet our complexity criteria, refining them through discussions and testing. After a pilot study with two participants, we finalized two tasks in each category~(Appendix~\ref{appendix:tasks}). These tasks featured indirect, verbose descriptions~\cite{liu2023wants}, presented in picture format to prevent direct copying, and re-ordered requirements to encourage independent planning.
}

\subsection{Baseline and Apparatus}
In addition to \sys{}, we implemented a \baseline{} with which to compare \sys{}.
\baseline{} is a web-based code editor that generates code based on inline comments, similar to GitHub Copilot~\cite{Github_2023}.
\rv{We designed the \baseline{} to closely resemble the code editor that participants are familiar with, allowing them to craft prompts based on their own experiences freely. 
Participants have the freedom to write, edit, and compile code within the \baseline{} environment. 
In \baseline{}, code suggestions are automatically generated and appear as grey text after the cursor position when users pause typing. Users can choose to accept these suggestions by pressing the tab key or ignore them and keep typing.
It is important to note that both \baseline{} and \sys{} are powered by GPT-4 to maintain a fair comparison.
Instead of creating a new system, this \baseline{} allows for a meaningful comparison between their individual experienced workflow and \sys{}.}
Both systems are web applications accessible on all operating systems and web browsers. Participants had the option to join either in-person or remotely via Zoom, the video conferencing platform.

\subsection{Procedure}
We chose a within-subjects design to compare \sys{} and \baseline{}.
Each study session lasted about $60$-$75$ minutes, and participants were compensated with CAD$\$30$. The study was approved by the university’s ethics review board.
We began the study by giving the participant an overview of the study procedure. Participants completed a consent form and a pre-study questionnaire that collected their demographic information and education level. 
Each participant was allocated to one of two distinct task categories, which included machine learning and data visualization. 
\rv{To prevent participants from simply copying and pasting the questions into the editor, we presented the assigned tasks as screenshots rather than text.}
Every participant completed two Python programming tasks based on their assigned category, one using \sys{} and the other using \baseline{}.
\rv{
Participants were assigned to tasks and categories using a Latin square design, ensuring an equal distribution of each condition (baseline or system) across participants and counterbalancing all task-category combinations.}

Before each task, participants received a tutorial on both \sys{} and \baseline{}, followed by a $5-10$ minute practice session for each condition.
Participants then had up to $12$ minutes to attempt each task within the assigned category. 
\rv{Participants determined the completion of the task based on their satisfaction and judgment. If they finished a task early, they would notify the experimenter to proceed to the next task.
Participants received incentives (CAD$\$5$) if they correctly met all the criteria specified for the tasks.
They were also asked to think aloud while completing the task~\cite{fonteyn1993description}.}
After each task, participants completed a post-task questionnaire evaluating the usability and utility of \sys{} and \baseline{}. Usability was measured using the UMUX-LITE scale, which is directly related to the SUS score \cite{lewis2013umuxlite}, and the NASA-TLX scale for perceived cognitive load~\cite{hart1988development}~(Appendix~\ref{appendix:umux}). Utility was measured using self-defined Likert scale items~(Appendix~\ref{appendix:umux}). 
In the end, we conducted a semi-structured interview to gain participants' insights about both \sys{} and \baseline{} and to understand their behaviour during the task.
\rv{Additionally, we recorded participants' screen activity and later played it back to assist them in recalling and explaining their observed behaviours during the task after the recall test.
Both \sys{} and \baseline{} logged various types of events based on participants' interactions across the timeline, including code editing, prompt authoring, and prompt editing.}

\subsection{Data Analysis}
We transcribed the think-aloud data and post-study interviews for all participants. Subsequently, we analyzed these transcriptions using \rv{reflexive thematic analysis~\cite{braun2019reflecting}. Our approach combined inductive and deductive methods to identify codes and themes, with a particular emphasis on participants' intention formation processes and editing experiences. The initial analysis involved the first two authors collaborating to group codes into broader categories related to prompt and code editing behaviors, and any disagreements were resolved by revisiting the interview transcripts.

We conducted statistical analysis on the comparative survey data by comparing responses between the \baseline{} and \sys{} conditions using the Wilcoxon signed-rank test, given the ordinal nature of Likert-scale responses and the small sample size. 
In the upcoming sections, we will present the data in the following format: (Q\textsubscript{question number if come from questionnaire}:
Median\textsubscript{\sys{}} vs. Median\textsubscript{\baseline{}}, $p$-value, $r$=effect size).
Furthermore, prompts collected from participants' logs were classified into \textit{procedural}, \textit{declarative}, and \textit{mixed} block types across multiple layers. 
Two researchers coded the data collaboratively, achieving an initial inter-coder agreement of $97\%$, which was iteratively refined to $100\%$. 
}

\section{Findings}
\begin{figure*}[!tbp]
  \centering
  \begin{minipage}[b]{0.42\textwidth}
    \includegraphics[width=\textwidth]{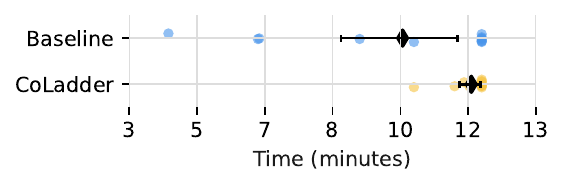}
    \caption{Distribution of time spent on tasks for each participant in both conditions.}
    \label{fig:time-dist}
  \end{minipage}
  \hfill
  \begin{minipage}[b]{0.54\textwidth}
    \includegraphics[width=\textwidth]{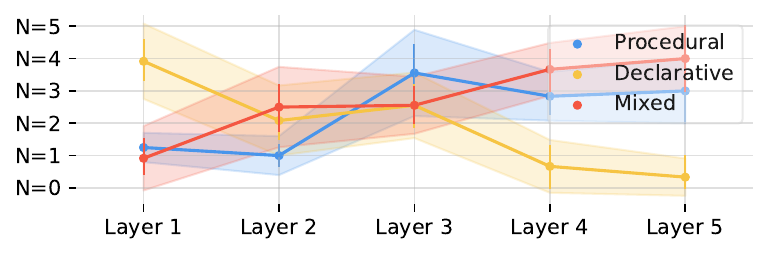}
    \caption{Frequency distribution of distinct block types across five layers in the prompt tree structure, with shaded areas representing standard deviation and error bars indicating 95\% confidence intervals.}
    \label{fig:level-count}
    \Description{The image is a line plot with five discrete x-axis categories labelled from 'level1' to 'level5', representing different levels of analysis. Each line—blue for 'how', orange for 'what', and red for 'mixed'—plots the count of each block type at the corresponding level. The shaded areas around each line represent the standard deviation, providing a visual indication of data spread and variability. Point markers at each level show the mean count for each category, and the error bars extending from these points illustrate the 95\% confidence intervals.}
  \end{minipage}
\end{figure*}

Our results highlight the usefulness of \sys{} in facilitating programmers to flexibly decompose their tasks, and scaffold their intentions into code across multiple levels of abstraction through hierarchical generation.
We present a detailed qualitative analysis and system log data corresponding to the four \textbf{D}esign \textbf{G}uidelines and \textbf{R}esearch \textbf{Q}uestions.

\subsection{General Impression}
\subsubsection{Self-Perceived Task Completion and Completion Time}
\label{sec:time}
\rv{Based on participants' self-evaluation, similar numbers of participants completed the tasks in two conditions ($7/12$ for \sys{} and $6/12$ for \baseline{}).}
However, on average, participants took significantly more ($p=.040$) time to complete tasks with \sys{} ($M=11.74, SD=0.50$ min) compared to \baseline{} ($M=10.05, SD=2.79$ min). \rv{The detailed time distribution in Figure~\ref{fig:time-dist} reveals that two participants spent less than 7 minutes in \baseline{} condition. This result does not come as a surprise to us, as our goal was to encourage programmers to allocate more time to planning and articulating their intentions and prompts.
The qualitative insights from the recall test will discuss the reasons for this outcome and explain why this is not a disadvantage for \baseline{}~(Section~\ref{sec:recall}).}

\subsubsection{Task Correctness}
\rv{We compiled participants' code after the study to validate if they correctly met all the specified criteria.
With \baseline{}, a higher proportion of tasks ($50.0\%$) remained incomplete, while with \sys{}, this percentage was slightly lower ($41.67\%$). 
The \sys{} condition showed a higher rate of tasks that were both completed and correct ($50.0\%$) compared to the \baseline{} condition ($25.0\%$).
Notably, more tasks were completed but found to be incorrect using \baseline{} ($25.0\%$) compared to the \sys{} condition ($8.33\%$).}
\subsubsection{Satisfaction and Confidence}
Compared to \baseline{}, participants found that when using \sys{}, they were more satisfied with the suggestions of the system~\reportQuestion{6}{6}{4}{.028}{.35}.
However, while there was an increase in the median confidence level regarding the correctness of the system-generated code, \rv{this increase was not statistically significant compared to the baseline condition~\reportQuestion{7}{5}{4}{.58}{.16}. 
These results suggest that \sys{} had no significant impact on participants' perception of the model's accuracy, but it did lead to generated results that were more closely aligned with their intentions.}

\subsubsection{Usability (UMUX-LITE)}
To measure the usability of \sys{}, we computed the SUS scores based on the UMUX-LITE.
The average SUS scores were significantly greater ({\small $p = .02$}) for \sys{} ({\small $\text{Mdn} = 90.61$}), compared to \baseline{} ({\small $\text{Mdn} = 68.94$}).
Typically, a SUS score above $70$ is considered ``acceptable'' and one above $85$ ``excellent'' \cite{aaron2008sus}. This indicates that \sys{} has good usability and is much more usable than \baseline{}.

\subsubsection{Perceived Cognitive Load (NASA-TLX)}
We also used NASA-TLX to measure the perceived workload associated with each system.
Compared to \baseline{}, \sys{} had lower mental ($\text{Mdn} = 3.0 < 4.5, p = .10$), physical ($\text{Mdn} = 1.0 < 2.0, p = .08$), and temporal ($\text{Mdn} = 3.0 < 4.5, p = .16$) demand, required less effort ($\text{Mdn} = 3.0 < 4.5, p = .39$), and led to better performance ($\text{Mdn} = 6.0 > 5.0, p = .11$) and statistically significantly less frustration ($\text{Mdn} = 2.0 < 5.0, p = .04$). The overall perceived workload, obtained by averaging all six raw NASA-TLX scores (with the ``Performance'' measure inverted), was also lower for \sys{} than \baseline{} ($\text{Mdn} = 2.33 < 3.75, p = .11$). Thus participants found \sys{} to be less taxing to use compared to \baseline{}, though this difference was not statistically significant.

\begin{figure*}
    \centering \includegraphics[width=.94\linewidth]{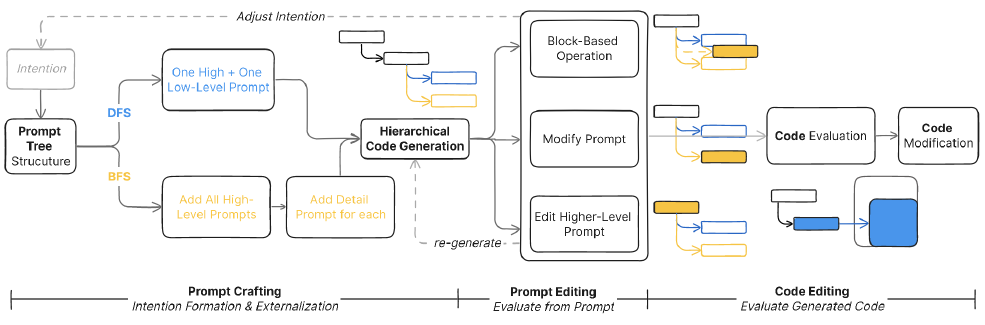}
    \caption{Participants typically initiate their workflow by externalizing their intentions through either a Depth-First Search (DFS) or Breadth-First Search (BFS) approach. Subsequently, \sys{} generates code structured in alignment with their carefully crafted task structure. Following code generation, participants proceed with block-based operations, supplement additions, and, in some cases, modify higher-level prompts to regenerate the code. 
    Participants using \sys{} typically evaluate and compile the code at the final stage, whereas those in the \baseline{} condition often need to evaluate and compile code more frequently throughout the process. }
    \Description{In the workflow diagram, participants typically begin by externalizing their intentions, utilizing either a Depth-First Search (DFS) or Breadth-First Search (BFS) approach. Next, \sys{} generates code that aligns with their meticulously designed task structure. After code generation, participants engage in various block-based operations, add supplements, and, occasionally, modify higher-level prompts to regenerate the code. Finally, participants using \sys{} typically evaluate and compile the code at the final stage. In contrast, those in the \baseline{} condition often find it necessary to evaluate and compile code more frequently throughout the process.}
    \label{fig:scaffold-result}
\end{figure*}

\subsection{\rv{Intention Formation and Externalization} (RQ1)}
Participants felt the prompt tree structure of \sys{} significantly helped construct their intentions for solving the programming task compared to \baseline{}~\reportQuestion{4}{6}{2.5}{.008}{.76}. 

\subsubsection{Various Prompt Types are Used in Different Layers.}
Every participant constructed a prompt tree with at least $2$ layers ($Mdn=4, SD=1.17$ layers), leveraging the prompt tree as an externalization of their mental task structure. As P2 mentioned, \qt{the tree structure clarifies my approach both in solving the task and in guiding the model to generate the desired code.}
\rv{
Figure~\ref{fig:level-count} shows that the \textit{declarative} type of prompt blocks (i.e., what tasks to be done) decrease as the number of layers increases. This suggests participants' preference for specifying declarative knowledge in the first three layers.
Conversely, the \textit{procedural} type (i.e., the how-to of the task) appears less frequent at first but exhibits an increase in the third layer, indicating an increase in procedure-oriented prompts as the layers progress. 
}

\subsubsection{Strategies for Structuring Prompt Tree}
Participants mostly (11/12) structured their prompts to horizontally map to the task structure through indentation (e.g., tasks to sub-tasks), while also aligning the order of prompts vertically with the structure of the generated code.
P6 elaborated: \qt{I prioritize defining the overall task structure, using child blocks to differentiate sub-tasks [...] and will adjust the code structure later on based on its execution sequence.}
While the flexibility of \sys{} allows free-form prompt structuring, our observations revealed two major workflows~(Fig.~\ref{fig:scaffold-result}):
\begin{enumerate}
    \item \textbf{Breadth-First Structure}: Participants (4/12) outlined all primary tasks first and subsequently delved into the finer details by adding sub-tasks (or \textit{Supplements}). Before moving to sub-tasks, they utilized the \textit{List Steps} feature to evaluate how the code is being implemented. \rv{This result is similar to the top-down decomposition method leverage for code comprehension~\cite{fix1993mental}.}
    \item \textbf{Depth-First Structure}: Participants (7/12) addressed all sub-tasks within a main task before moving on to the next primary task. Participants with a well-defined intention beforehand were more inclined to adopt this approach, leveraging the \textit{List Steps} feature as a \pqt{cross-validation mechanism}{P8} to ensure the generated code aligned with their specified sub-tasks. \rv{This approach is similar to the stepwise refinement in program design~\cite{fix1993mental}.}
\end{enumerate}

Without the prompt tree structure, participants using \baseline{} typically faced two challenges. First, some participants (4/12) spent a significant amount of time mapping out the task to code, often constructing a lengthy section of comments that included all the steps required to approach the task. This approach required them to \pqt{think about the code in very detail first.}{P2}
On the other hand, most participants (8/12) developed their intention while authoring the prompt by adding more context to the prompt to adjust the output. Participants reported that using the linear representation of prompts \pqt{could not fully express their thoughts.}{P4} 
\rv{While two participants constructed a layered prompt as discovered in the formative study, they did not use it as the prompt for code generation. 
Instead, they utilized it to establish the overall context and then proceeded to refine their prompts in a more detailed manner during the code-writing process.}

In summary, participants using \sys{} with both approaches found that the tree structure encouraged them to \pqt{contemplate the implementation of the code layer by layer,}{P5} alleviating the \pqt{cognitive burden of thinking about the entire code structure beforehand.}{P11}

\begin{figure*}
    \includegraphics[width=\linewidth]{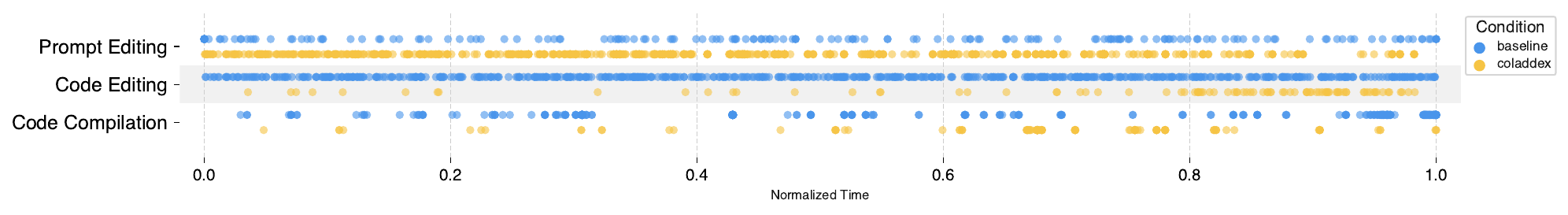}
    \caption{A scatter plot displays events occurring throughout normalized time, synthesized from all participants. In \baseline{}, code editing is spread throughout the workflow, while with \sys{}, participants tend to edit code in the later stages.}
    \Description{The plot visualizes data points along a horizontal timeline representing different stages of work, categorized as 'Prompt Editing', 'Code Editing', and 'Code Compilation'. The data points are arranged linearly from left to right, signifying the progression of normalized time. Each point is colored according to the condition it pertains to, with blue for the 'baseline' and yellow for 'colladex'.}
    \label{fig:timeline}
\end{figure*}

\subsection{Controlled Scaffolding from Intention to Code (RQ2)}
\rv{After externalizing their intentions through the prompt tree, participants proceeded to evaluate the results and made edits to prompts and code as needed.}
Overall, participants found \sys{} to be more helpful in scaffolding their intentions to generate the desired code~\reportQuestion{5}{6}{3.5}{.007}{.77} compared to the \baseline{}.

\subsubsection{Editing Prompts before Code.}
\begin{figure*}
    \centering
    \includegraphics[width=1\textwidth]{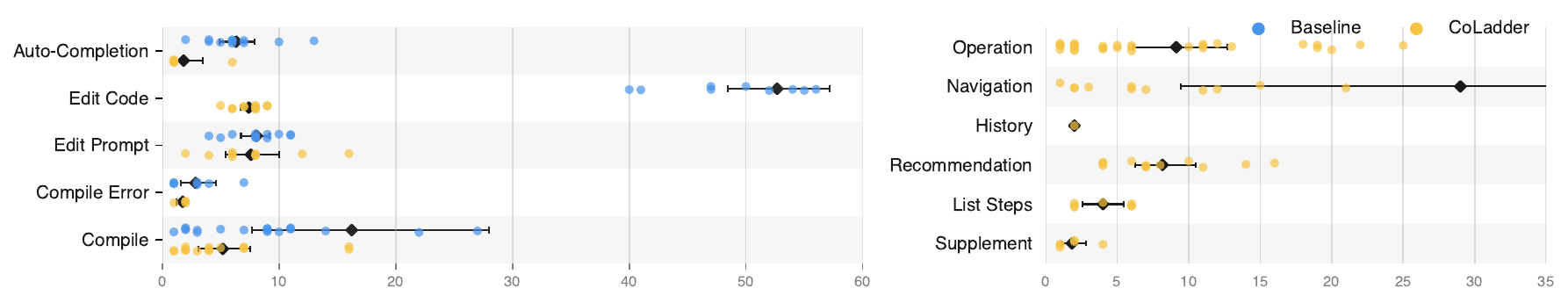}
    \caption{Comparison of event counts between \baseline{} and \sys{}.}
    \Description{The plot displays event counts for 10 types of actions: "Auto-Completion", "Edit Code", "Edit Prompt", "Compile Error", "Compile", "Operation", "Navigation", "History", "Recommendation", "List Steps", "Supplement". Dots are coloured according to condition, where blue is for baseline and yellow is for CoLadder. Error bars represention 95\% confidence intervals are included as well.}
    \label{fig:event-count}
\end{figure*}

We observed that participants typically began code evaluation after drafting the prompt tree structure. P6 explained, \qt{The generated code will not be accurate unless I provide details [e.g., by adding child blocks, \textit{Supplement} operation].}
\rv{Participants typically edit prompts first instead of modifying the generated code when using \sys{}~(Fig.~\ref{fig:timeline}). 
P9 explained that the reason for adjusting prompts in \sys{} is \qt{not because it generates the wrong code, but to adjust my approach for solving the task [intention].} 
Analysis of the log data~(Fig.~\ref{fig:timeline}) revealed that participants made significant code edits in the latter third of their work when using \sys{}. 
Participants mentioned that they felt it was more like \pqt{maintaining a document}{P4} and \pqt{fully noting my thought process}{P1} rather than engaging in prompt engineering, as was the case in the \baseline{} condition.

\subsubsection{Block-based Operations Help in Prompt Editing.}
As depicted in Figure~\ref{fig:event-count} Left, participants in the \baseline{} condition tended to manually edit the code significantly more than in the \sys{} condition ($Mdn$=$8.0$ vs. $Mdn$=$53.0$, $p$=$.002$, $r$=$0.9$).
This finding aligns with the observation that they only edited the code towards the end when using \sys{}. 
However, there is no significant difference in the amount of prompt editing ($Mdn$=$7.0$ vs. $Mdn$=$8.0$, $p$=$.72$, $r$=$0.08$), suggesting that block-based operations (as shown in Fig.~\ref{fig:event-count} Right) may reduce the necessity for code editing. 
Participants found these block-based operations to be more \qt{intuitive} and \qt{direct} ways to modify the generated code compared to directly editing through text.
}
Participants highlighted the block-based operations (e.g., \textit{DnD}) could help them focus more on structuring prompts, \pqt{I love the drag and drop feature, which allows me to structure the code freely based on my mental model without concerns about the code's structure.}{P7}

\subsubsection{Modular-based Design Enhance Controllability}
Participants reported that they could steer \sys{} more controllably towards the task goal~\reportQuestion{3}{6}{3.0}{.007}{.77} with the \pqt{step-by-step approach}{P6}.
When comparing the authoring processes in both systems, participants (11/12) felt that \sys{} provided them with more controllability while \rv{modifying the generated code from prompts.} 
Participants found that the modular-based design (i.e., prompt blocks) allowed them to focus on one segment of code at a time, which prevented them from \pqt{getting lost while verifying the generated code.}{P2} Participants could also easily identify where to add prompt blocks because they \pqt{only need to ensure that the high-level task structure is correctly ordered, without having to search through the code to find the exact position.}{P1}
They found it useful for modifying the targeted code segments, \pqt{without worrying about affecting other sections.}{P4}
Participants also found that the tree structure assisted them in identifying where to modify the code more easily compared to \baseline{}. \pqt{It's simpler to make changes to the code [with \sys{}] when there's an error. [...] I can modify the parent level and the changes will be reflected in all child blocks as well.}{P8}

\subsection{Results \rv{Evaluation} during Prompt Authoring (RQ3)}
Compared to \baseline{}, participants found that \sys{} significantly reduced the need for cognitive switching between prompt authoring and code evaluation~\reportQuestion{1}{6.0}{4.5}{.01}{.67}. 

\subsubsection{List Steps, Auto-Complete, Recommendation Enhancing Intention Alignment Evaluation}
All participants experimented with the \textit{List Steps} feature in \sys{} (Fig.~\ref{fig:event-count} Right), primarily to \pqt{assess generated code alignment with [their] intents.}{P9} P2 explained, \qt{If the steps are correct, I am confident the code will be too.} 
This approach was similarly adopted with the \textit{Recommendation} feature and \textit{Auto-Complete} features~(Fig.~\ref{fig:event-count} Right); participants mostly utilized them \pqt{as a cue to see if the system captured my intent}{P8}. 
Specifically, participants leverage recommendations to evaluate the alignment between their intents and the system's comprehension, \rv{rather than accepting the recommendation as the next prompt.}
Without these features in \baseline{}, participants have to manually identify specific changes in the code segments corresponding to their prompt modifications by \pqt{comparing the previous and current generated code.}{P7}
Overall, \rv{the purpose of evaluating the code for participants using \sys{} was to modify their prompts, but not to evaluate the correctness of the generated code.}

\subsubsection{Improved Recall Performance after Participants Using \sys{}}
\label{sec:recall}
Despite the reduced need for frequent code evaluation, participants using \sys{} demonstrated a significantly better understanding of their programs compared to using \baseline{}~\reportQuestion{8}{6.0}{5.0}{.007}{.77}.
A recall test was conducted to investigate participant code comprehension~\cite{dunsmore2000comparative}. Participants could recall code implementation systematically after using \sys{}, from higher-level (e.g., the purpose of the task) to lower-level code implementations with details in each step. 
P11 explained the reason, \qt{the system [\sys{}] helped me think through the programming task already when I was drafting prompts.} P9 added, \qt{I do not need to spend time on comprehending code, as I have already verified segments of code corresponded to each prompt before.} 
In contrast, in \baseline{}, participants started with detailed codes and gradually summarized and mapped the task steps.
\rv{This result can also be attributed to \sys{}'s support in intention formation and externalization compared to the \baseline{}. This trade-off in terms of time efficiency aligns with our previous finding that using \sys{} was slower than \baseline{} (Section~\ref{sec:time}). Participants tended to invest more time in structuring clear intentions in their minds, which aids them in evaluating the code with greater ease and reduces cognitive load.}

\subsubsection{Transitioning from Opportunistic Programming to Comprehensive Code Understanding}
We observed from Figure~\ref{fig:event-count} that participants compiled the code (i.e., \textit{compile error} and \textit{compile}) significantly more in \baseline{} compared to the \sys{} \rv{($Mdn$=$24.0$ vs. $Mdn$=$18.5$, $p$=$.012$, $r$=$0.56$)}, which was used as an alternative approach for \pqt{verifying the generated code.}{P12} 
P1's strategy in using \baseline{} was to \qt{try to compile the code to see if it works,} without the need to evaluate the generated code.
\rv{Participants in the \baseline{} condition compiled the code throughout the session, whereas participants in the \sys{} condition tended to compile the code at a later stage (Fig.~\ref{fig:timeline}). 
While this approach might increase the overall completion time,
} 
participants using this \textit{opportunistic} approach in \baseline{} faced challenges to modify the code when compiled results were incorrect, where they \pqt{had to check by cross-referencing task descriptions and generated code.}{P1}

\subsection{\rv{Enhancing Prompt-Code Correspondence} (RQ4)}
Participants navigated across various layers of prompt blocks and noted that \sys{} significantly enhanced their ability to evaluate generated code~\reportQuestion{9}{6.0}{3.0}{.006}{.80} in comparison to \baseline{}. 

\subsubsection{Enhancing Navigation and Precise Modification through Corresponding Code Highlight in Modular-Based Approach}
Participants found it more effective to navigate and make precise modifications with \sys{} as they could adopt a modular approach. 
They could evaluate code in segments with the code highlight feature, which \pqt{reduced cognitive load.}{P4}
Folding and presenting different segments of code depending on the structure of the task allows programmers to \pqt{more easily control the depth of verification required.}{P6} 
Some suggested that having a structural intention formed beforehand allowed them to \pqt{swiftly locate the segments needing changes.}{P8}
We observed that most participants (10/12) do not evaluate the entire program but evaluate them segments by segments, as they do not worry about the \pqt{possibility of the LLM incorrectly concatenating my [their] code or using different variable names.}{P11} 
\rv{
Participants in the \baseline{} condition, on the other hand, encountered difficulties in identifying differences from the integrated prompt. 
The majority of participants (9/12) expressed frustration with the time-consuming process of \pqt{constantly verifying the same code,}{P7} which at times led them to opt for directly modifying the code. 
P5 further reported the challenges of not knowing whether all their intentions were effectively conveyed to \baseline{} and not being able to discern \qt{which prompts were attributed to incorrect results.}
}

\subsubsection{Facilitating Detailed Code Evaluation with Semantic Highlight and Mixed Methods Writing}
As the prompt blocks increased, participants valued the \textit{Semantic Highlighting} feature that simplified the linkage between the task and its corresponding code. The majority (10/12) felt that the \textit{Mixed Methods} writing feature accelerated the evaluation process by checking \pqt{if the keyword is showing in the right place as [they] thought.}{P2} 
While some participants (4/12) added code syntax to the prompt when using \baseline{}. They did so primarily as an intervention method when the generated code consistently produced errors, rather than as a means to facilitate the evaluation process.
We also observed that participants preferred the use of inline keywords (e.g., \textit{read\_csv()}, \textit{sns.pairplot()}) rather than multi-line code when drafting prompts, where they could \pqt{easily identify and track changes}{P11} across multi-level abstractions.
\rv{
Figure~\ref{fig:level-count} displays the count of prompt blocks written with mixed methods, ranging from the first to the maximum layer. Interestingly, participants tended to embrace mixed methods in the later layers, considering them to be the form closest to the generated code.
While most participants used mixed methods to reflect their intentions, one participant explained the advantage of employing mixed methods in the final layer because \pqt{it is the nearest [in distance] to the generated code [on the right-hand side].}{P7}
}
\section{Limitations}
Our primary limitation is associated with the diversity of programming tasks evaluated and the constraints of prompt programming. Some open-ended programming tasks, particularly those focused on rapid iteration (e.g., exploratory programming, exploratory data analysis)~\cite{kery2017exploring, tukey1977exploratory}, tend to prioritize quick idea iteration over code quality. Programmers often need to make swift, small changes, such as adjusting parameters and variables~\cite{snoek2012practical, yoon2014longitudinal}. NL-based programming may not be as effective in these cases, as programmers can quickly modify specific parts of the code without waiting for code generation.
Despite the option to make direct code changes within \sys{}, participants mainly focus on prompt authoring and incorrectly perceive that they \pqt{always have to refine the prompt to change the code}{P6}.
Some participants (3/12) \rv{who had forgotten about the bi-directional editing feature expressed a desire to be able to directly modify the code} within \sys{}, as they wanted to \pqt{adjust some low-level code details directly [in \sys{}]}{P10}. 
\rv{This finding suggests that while \sys{} supports bi-directional editing, the current design lacks clarity in indicating whether bi-directional editing is functioning. Future design improvements should incorporate visual cues to inform participants when changes have been recorded.
}

Another limitation of \sys{} is that the prompt tree structure may not always align with the actual code structure.
For instance, while in the programmer's intention, \qt{plotting the loss curve} may be considered a sub-task under \qt{model evaluation}, it might exist within the global function scope in the generated code.
While we initially designed the separation of the user interface for the prompt editor and code editor with consideration of this issue, two participants mentioned that our system may be less effective in certain programming languages or scenarios (e.g., Object-Oriented programming~\cite{dawson2013cognitive} or \rv{complex projects with multiple files}) where the task and code structures can deviate significantly. 
\rv{
We argue that while the correspondence between prompts and code might be less obvious, the highlight of the corresponding (DG4) and hierarchical generation could still be beneficial.
This is in contrast to the \baseline{}, where programmers would need to locate the generated code and manually map it to the prompts.

We acknowledge the limitation of testing \sys{} exclusively with scripting languages and recognize that certain programming languages, particularly compiled languages, may face challenges with existing features, such as compilation errors. However, our design of \sys{} primarily emphasizes support for the intention formation and externalization stages, arguing that the hierarchical mental representation (DG1) is broadly applicable across various programming scenarios and languages~\cite{hoc1977role, fix1993mental, kim1995internal}. 
The need for scaffolding intentions with controllability (DG2) and reducing cognitive switching (DG3) are also common challenges raised by prior works in different programming settings~\cite{liang2023large}.
Overall, while \sys{} was primarily designed to tackle challenges in general NL-based programming, we acknowledge the future need for designs more tailored to specific programming tasks and languages.
For example, exploring the integration of visualizations such as class diagrams to better convey program structures and relationships between classes or components~\cite{gutwenger2003new}.
}

\section{Discussion and Future Work}
We discussed how \sys{} assists programmers in forming intentions for programming tasks, differentiates controllability in program and interaction with LLM, mitigates over-reliance issues, and its potential applicability in both familiar and unfamiliar programming tasks.

\subsection{Intentions Formation \& Development}
The findings from our study demonstrate that \sys{} effectively supports code generation at multiple levels of abstraction, thereby assisting programmers in scaffolding their intent. This aligns with existing literature on the challenges of understanding the capabilities and limitations of language model-driven code generation systems, as well as the need for clear and naturalistic input to generate specific code that matches the programmer's intent~\cite{wu2022ai, liu2023wants, fiannaca2023programming}

Additionally, our findings resonate with prior research highlighting the importance of programmers forming intentions of code at different levels of abstraction~\cite{sarkar2022like}, from specific code statements to larger program structures~\cite{ye1996expert, balz2010continuous}. This underscores the need for scaffolding useful intentions to facilitate interactions and collaboration between programmers and LLMs in solving programming tasks. In our work, we place a strong emphasis on assisting programmers in forming these intentions to craft effective prompts that generate code aligning with their intentions~\cite{xie2017influential, carroll1988mental}.

Our study highlights the role of \sys{} in aiding programmers as they evolve their intentions throughout the problem-solving process~\cite{maalej2014comprehension}. By offering hierarchical prompt structures and block-based operations, \sys{} enables programmers to easily adapt and refine their task representations as they gain deeper insights into their programming tasks. This aligns with agile development principles, fostering dynamic adjustments in problem-solving approaches~\cite{heinonen2023synthesizing}.

\subsection{Controlling Program or AI Interactions}
\rv{
In the context of programming with AI assistants, we discerned a subtle distinction in \textit{controllability} over the resulting program versus the process of scaffolding programmers' intentions.
The former pertains to the ability to directly and manually edit the code, while the latter involves translating decomposed intentions into prompts and generated code.
Three participants found it less intuitive to edit the code in \sys{} without visual cues. This finding suggests that, although \sys{} effectively enhances control over the intention scaffolding process (\textbf{RQ2}), it may not necessarily improve control over the program itself.
However, we observed that participants required less code editing when using \sys{}, especially in the early stages. This reduction was attributed to participants having alternative methods for easily modifying the generated code segments through block-based operations. While these operations may not necessarily enhance control over the program, they do aid in externalizing intentions and intuitive modifications.
}
Future research should explore two facets of controllability that are essential for programmers: controllability in their interactions with AI~\cite{hook2000steps, amershi2019guidelines, norman1994might} and controllability over the program itself~\cite{navarro2001visual, green1996usability}.

\subsection{Over-Reliance and Program Comprehension}
In our findings, we observed a trend where programmers often accepted the suggestions from the baseline code assistant and subsequently modified the generated code. This observation aligns with prior studies that have suggested the possibility of programmers developing an over-reliance on LLM-driven code assistants~\cite{barke2023grounded, chen2021evaluating, xu2022ide}. However, in the context of \sys{}, participants meticulously crafted their programs step by step, demonstrating a deeper understanding of the overall programming structure. This approach appeared to yield benefits in the recall test, as participants showed a greater ability to comprehend the program they had constructed~\cite{maalej2014comprehension}.

This finding raises intriguing questions about the potential implications for future maintenance tasks. It prompts further inquiry into whether the careful, step-by-step program crafting facilitated by \sys{} might result in more maintainable codebases or offer advantages in scenarios where long-term code comprehension and modification are required~\cite{al2017source}. Exploring these aspects in future research could shed light on the effectiveness and sustainability of \sys{} in addressing the over-reliance issue commonly encountered in human-AI interaction~\cite{amershi2019guidelines}.

\subsection{Experiences and Task Familiarity}
We designed and studied \sys{} with experienced programmers because they can form more well-defined intentions when addressing programming tasks compared to novices~\cite{ye1996expert, heinonen2023synthesizing, corritore1991novices, fix1993mental}. However, we are also interested in how task familiarity might cause differences among experienced programmers. 
According to our pre-study questionnaire, all participants reported being familiar with basic machine learning ($M = 4.17, SD = 1.19$) and basic data visualization ($M = 4.25, SD = 0.75$) tasks in Python.
We calculated the Spearman correlation between self-perceived familiarity (on a 5-point Likert scale) and the NASA-TLX and self-defined questionnaire responses.
The correlation was weak\footnote{As a general rule of thumb, a Spearman correlation below $0.3$ is considered weak and one between $0.3-0.6$ is moderately strong.} for all items except \qt{Performance} ($r = -0.348$) from NASA-TLX, and \qt{Understanding of How System Works} ($r = -0.447$) and \qt{Satisfaction with Suggestions} ($r = -0.327$) from the self-defined questionnaire, which had moderate correlations.

\rv{
These findings imply that while participants' prior familiarity with specific programming tasks may have had some influence on their perceptions, it was not a dominant factor.
Potential future research could investigate the impact of varying degrees of task familiarity on the workflow when using \sys{}.
In this work, we focus on the significance of the overall programming experience, as the ability to formulate effective intentions is crucial in the translation process to prompts with controllability and in determining what aspects to evaluate.
}
\section{Conclusion}
In this paper, we present \sys{}, an interactive system that aids programmers in code generation and evaluation. It achieves this by providing hierarchical task decomposition, modular-based code generation, and result evaluation during prompt authoring. Our iterative design process uncovered strategies and programmers' needs for externalizing intentions and translating them into NL prompts for code generation. A user study with 12 experienced programmers further validated \sys{}, demonstrating its capacity to enhance programmers' ability to navigate and edit code across various abstraction levels, from initial goal to final code implementation. Our work provides valuable design insights into the concept of \textit{hierarchical generation} for future LLM-driven systems.

\bibliographystyle{ACM-Reference-Format}
\bibliography{main}


\begin{thebibliography}{91}


\ifx \showCODEN    \undefined \def \showCODEN     #1{\unskip}     \fi
\ifx \showDOI      \undefined \def \showDOI       #1{#1}\fi
\ifx \showISBNx    \undefined \def \showISBNx     #1{\unskip}     \fi
\ifx \showISBNxiii \undefined \def \showISBNxiii  #1{\unskip}     \fi
\ifx \showISSN     \undefined \def \showISSN      #1{\unskip}     \fi
\ifx \showLCCN     \undefined \def \showLCCN      #1{\unskip}     \fi
\ifx \shownote     \undefined \def \shownote      #1{#1}          \fi
\ifx \showarticletitle \undefined \def \showarticletitle #1{#1}   \fi
\ifx \showURL      \undefined \def \showURL       {\relax}        \fi
\providecommand\bibfield[2]{#2}
\providecommand\bibinfo[2]{#2}
\providecommand\natexlab[1]{#1}
\providecommand\showeprint[2][]{arXiv:#2}

\bibitem[Al~Madi(2022)]%
        {al2022readable}
\bibfield{author}{\bibinfo{person}{Naser Al~Madi}.} \bibinfo{year}{2022}\natexlab{}.
\newblock \showarticletitle{How readable is model-generated code? examining readability and visual inspection of GitHub copilot}. In \bibinfo{booktitle}{\emph{Proceedings of the 37th IEEE/ACM International Conference on Automated Software Engineering}}. \bibinfo{pages}{1--5}.
\newblock


\bibitem[Al-Saiyd(2017)]%
        {al2017source}
\bibfield{author}{\bibinfo{person}{Nedhal~A Al-Saiyd}.} \bibinfo{year}{2017}\natexlab{}.
\newblock \showarticletitle{Source code comprehension analysis in software maintenance}. In \bibinfo{booktitle}{\emph{2017 2nd International Conference on Computer and Communication Systems (ICCCS)}}. IEEE, \bibinfo{pages}{1--5}.
\newblock


\bibitem[Amershi et~al\mbox{.}(2019)]%
        {amershi2019guidelines}
\bibfield{author}{\bibinfo{person}{Saleema Amershi}, \bibinfo{person}{Dan Weld}, \bibinfo{person}{Mihaela Vorvoreanu}, \bibinfo{person}{Adam Fourney}, \bibinfo{person}{Besmira Nushi}, \bibinfo{person}{Penny Collisson}, \bibinfo{person}{Jina Suh}, \bibinfo{person}{Shamsi Iqbal}, \bibinfo{person}{Paul~N Bennett}, \bibinfo{person}{Kori Inkpen}, {et~al\mbox{.}}} \bibinfo{year}{2019}\natexlab{}.
\newblock \showarticletitle{Guidelines for human-AI interaction}. In \bibinfo{booktitle}{\emph{Proceedings of the 2019 chi conference on human factors in computing systems}}. \bibinfo{pages}{1--13}.
\newblock


\bibitem[Balijepally et~al\mbox{.}(2012)]%
        {balijepally2012effect}
\bibfield{author}{\bibinfo{person}{VenuGopal Balijepally}, \bibinfo{person}{Sridhar Nerur}, {and} \bibinfo{person}{RadhaKanta Mahapatra}.} \bibinfo{year}{2012}\natexlab{}.
\newblock \showarticletitle{Effect of task mental models on software developer's performance: An experimental investigation}. In \bibinfo{booktitle}{\emph{2012 45th Hawaii International Conference on System Sciences}}. IEEE, \bibinfo{pages}{5442--5451}.
\newblock


\bibitem[Balz et~al\mbox{.}(2010)]%
        {balz2010continuous}
\bibfield{author}{\bibinfo{person}{Moritz Balz}, \bibinfo{person}{Michael Striewe}, {and} \bibinfo{person}{Michael Goedicke}.} \bibinfo{year}{2010}\natexlab{}.
\newblock \showarticletitle{Continuous maintenance of multiple abstraction levels in program code}. In \bibinfo{booktitle}{\emph{International Workshop on Future Trends of Model-Driven Development}}, Vol.~\bibinfo{volume}{2}. SCITEPRESS, \bibinfo{pages}{68--79}.
\newblock


\bibitem[Bangor et~al\mbox{.}(2008)]%
        {aaron2008sus}
\bibfield{author}{\bibinfo{person}{Aaron Bangor}, \bibinfo{person}{Philip~T. Kortum}, {and} \bibinfo{person}{James~T. Miller}.} \bibinfo{year}{2008}\natexlab{}.
\newblock \showarticletitle{An Empirical Evaluation of the System Usability Scale}.
\newblock \bibinfo{journal}{\emph{International Journal of Human–Computer Interaction}} \bibinfo{volume}{24}, \bibinfo{number}{6} (\bibinfo{year}{2008}), \bibinfo{pages}{574--594}.
\newblock
\urldef\tempurl%
\url{https://doi.org/10.1080/10447310802205776}
\showDOI{\tempurl}
\showeprint{https://doi.org/10.1080/10447310802205776}


\bibitem[Barke et~al\mbox{.}(2023)]%
        {barke2023grounded}
\bibfield{author}{\bibinfo{person}{Shraddha Barke}, \bibinfo{person}{Michael~B James}, {and} \bibinfo{person}{Nadia Polikarpova}.} \bibinfo{year}{2023}\natexlab{}.
\newblock \showarticletitle{Grounded copilot: How programmers interact with code-generating models}.
\newblock \bibinfo{journal}{\emph{Proceedings of the ACM on Programming Languages}} \bibinfo{volume}{7}, \bibinfo{number}{OOPSLA1} (\bibinfo{year}{2023}), \bibinfo{pages}{85--111}.
\newblock


\bibitem[Bastola et~al\mbox{.}(2023)]%
        {bastola2023llm}
\bibfield{author}{\bibinfo{person}{Ashish Bastola}, \bibinfo{person}{Hao Wang}, \bibinfo{person}{Judsen Hembree}, \bibinfo{person}{Pooja Yadav}, \bibinfo{person}{Nathan McNeese}, {and} \bibinfo{person}{Abolfazl Razi}.} \bibinfo{year}{2023}\natexlab{}.
\newblock \showarticletitle{LLM-based Smart Reply (LSR): Enhancing Collaborative Performance with ChatGPT-mediated Smart Reply System (ACM)(Draft) LLM-based Smart Reply (LSR): Enhancing Collaborative Performance with ChatGPT-mediated Smart Reply System}.
\newblock \bibinfo{journal}{\emph{arXiv preprint arXiv:2306.11980}} (\bibinfo{year}{2023}).
\newblock


\bibitem[Beurer-Kellner et~al\mbox{.}(2023)]%
        {beurer2023prompting}
\bibfield{author}{\bibinfo{person}{Luca Beurer-Kellner}, \bibinfo{person}{Marc Fischer}, {and} \bibinfo{person}{Martin Vechev}.} \bibinfo{year}{2023}\natexlab{}.
\newblock \showarticletitle{Prompting is programming: A query language for large language models}.
\newblock \bibinfo{journal}{\emph{Proceedings of the ACM on Programming Languages}} \bibinfo{volume}{7}, \bibinfo{number}{PLDI} (\bibinfo{year}{2023}), \bibinfo{pages}{1946--1969}.
\newblock


\bibitem[Braun and Clarke(2012)]%
        {braun2012thematic}
\bibfield{author}{\bibinfo{person}{Virginia Braun} {and} \bibinfo{person}{Victoria Clarke}.} \bibinfo{year}{2012}\natexlab{}.
\newblock \bibinfo{booktitle}{\emph{Thematic analysis.}}
\newblock \bibinfo{publisher}{American Psychological Association}.
\newblock


\bibitem[Braun and Clarke(2019)]%
        {braun2019reflecting}
\bibfield{author}{\bibinfo{person}{Virginia Braun} {and} \bibinfo{person}{Victoria Clarke}.} \bibinfo{year}{2019}\natexlab{}.
\newblock \showarticletitle{Reflecting on reflexive thematic analysis}.
\newblock \bibinfo{journal}{\emph{Qualitative research in sport, exercise and health}} \bibinfo{volume}{11}, \bibinfo{number}{4} (\bibinfo{year}{2019}), \bibinfo{pages}{589--597}.
\newblock


\bibitem[Brown et~al\mbox{.}(2020)]%
        {brown2020language}
\bibfield{author}{\bibinfo{person}{Tom~B. Brown}, \bibinfo{person}{Benjamin Mann}, \bibinfo{person}{Nick Ryder}, \bibinfo{person}{Melanie Subbiah}, \bibinfo{person}{Jared Kaplan}, \bibinfo{person}{Prafulla Dhariwal}, \bibinfo{person}{Arvind Neelakantan}, \bibinfo{person}{Pranav Shyam}, \bibinfo{person}{Girish Sastry}, \bibinfo{person}{Amanda Askell}, \bibinfo{person}{Sandhini Agarwal}, \bibinfo{person}{Ariel Herbert-Voss}, \bibinfo{person}{Gretchen Krueger}, \bibinfo{person}{Tom Henighan}, \bibinfo{person}{Rewon Child}, \bibinfo{person}{Aditya Ramesh}, \bibinfo{person}{Daniel~M. Ziegler}, \bibinfo{person}{Jeffrey Wu}, \bibinfo{person}{Clemens Winter}, \bibinfo{person}{Christopher Hesse}, \bibinfo{person}{Mark Chen}, \bibinfo{person}{Eric Sigler}, \bibinfo{person}{Mateusz Litwin}, \bibinfo{person}{Scott Gray}, \bibinfo{person}{Benjamin Chess}, \bibinfo{person}{Jack Clark}, \bibinfo{person}{Christopher Berner}, \bibinfo{person}{Sam McCandlish}, \bibinfo{person}{Alec Radford}, \bibinfo{person}{Ilya Sutskever},
  {and} \bibinfo{person}{Dario Amodei}.} \bibinfo{year}{2020}\natexlab{}.
\newblock \bibinfo{title}{Language Models are Few-Shot Learners}.
\newblock
\newblock
\showeprint[arxiv]{2005.14165}~[cs.CL]


\bibitem[Cai et~al\mbox{.}(2023)]%
        {cai2023low}
\bibfield{author}{\bibinfo{person}{Yuzhe Cai}, \bibinfo{person}{Shaoguang Mao}, \bibinfo{person}{Wenshan Wu}, \bibinfo{person}{Zehua Wang}, \bibinfo{person}{Yaobo Liang}, \bibinfo{person}{Tao Ge}, \bibinfo{person}{Chenfei Wu}, \bibinfo{person}{Wang You}, \bibinfo{person}{Ting Song}, \bibinfo{person}{Yan Xia}, {et~al\mbox{.}}} \bibinfo{year}{2023}\natexlab{}.
\newblock \showarticletitle{Low-code LLM: Visual Programming over LLMs}.
\newblock \bibinfo{journal}{\emph{arXiv preprint arXiv:2304.08103}} (\bibinfo{year}{2023}).
\newblock


\bibitem[Carroll and Olson(1988)]%
        {carroll1988mental}
\bibfield{author}{\bibinfo{person}{John~M Carroll} {and} \bibinfo{person}{Judith~Reitman Olson}.} \bibinfo{year}{1988}\natexlab{}.
\newblock \showarticletitle{Mental models in human-computer interaction}.
\newblock \bibinfo{journal}{\emph{Handbook of human-computer interaction}} (\bibinfo{year}{1988}), \bibinfo{pages}{45--65}.
\newblock


\bibitem[Chen et~al\mbox{.}(2021)]%
        {chen2021evaluating}
\bibfield{author}{\bibinfo{person}{Mark Chen}, \bibinfo{person}{Jerry Tworek}, \bibinfo{person}{Heewoo Jun}, \bibinfo{person}{Qiming Yuan}, \bibinfo{person}{Henrique Ponde de~Oliveira Pinto}, \bibinfo{person}{Jared Kaplan}, \bibinfo{person}{Harri Edwards}, \bibinfo{person}{Yuri Burda}, \bibinfo{person}{Nicholas Joseph}, \bibinfo{person}{Greg Brockman}, {et~al\mbox{.}}} \bibinfo{year}{2021}\natexlab{}.
\newblock \showarticletitle{Evaluating large language models trained on code}.
\newblock \bibinfo{journal}{\emph{arXiv preprint arXiv:2107.03374}} (\bibinfo{year}{2021}).
\newblock


\bibitem[Clark et~al\mbox{.}(2008)]%
        {clark2008cognitive}
\bibfield{author}{\bibinfo{person}{Richard~E Clark}, \bibinfo{person}{David~F Feldon}, \bibinfo{person}{Jeroen~JG Van~Merrienboer}, \bibinfo{person}{Kenneth~A Yates}, {and} \bibinfo{person}{Sean Early}.} \bibinfo{year}{2008}\natexlab{}.
\newblock \showarticletitle{Cognitive task analysis}.
\newblock In \bibinfo{booktitle}{\emph{Handbook of research on educational communications and technology}}. \bibinfo{publisher}{Routledge}, \bibinfo{pages}{577--593}.
\newblock


\bibitem[Corritore and Wiedenbeck(1991)]%
        {corritore1991novices}
\bibfield{author}{\bibinfo{person}{Cynthia~L Corritore} {and} \bibinfo{person}{Susan Wiedenbeck}.} \bibinfo{year}{1991}\natexlab{}.
\newblock \showarticletitle{What do novices learn during program comprehension?}
\newblock \bibinfo{journal}{\emph{International Journal of Human-Computer Interaction}} \bibinfo{volume}{3}, \bibinfo{number}{2} (\bibinfo{year}{1991}), \bibinfo{pages}{199--222}.
\newblock


\bibitem[Cypher and Halbert(1993)]%
        {cypher1993watch}
\bibfield{author}{\bibinfo{person}{Allen Cypher} {and} \bibinfo{person}{Daniel~Conrad Halbert}.} \bibinfo{year}{1993}\natexlab{}.
\newblock \bibinfo{booktitle}{\emph{Watch what I do: programming by demonstration}}.
\newblock \bibinfo{publisher}{MIT press}.
\newblock


\bibitem[Dakhel et~al\mbox{.}(2023)]%
        {dakhel2023github}
\bibfield{author}{\bibinfo{person}{Arghavan~Moradi Dakhel}, \bibinfo{person}{Vahid Majdinasab}, \bibinfo{person}{Amin Nikanjam}, \bibinfo{person}{Foutse Khomh}, \bibinfo{person}{Michel~C Desmarais}, {and} \bibinfo{person}{Zhen Ming~Jack Jiang}.} \bibinfo{year}{2023}\natexlab{}.
\newblock \showarticletitle{Github copilot ai pair programmer: Asset or liability?}
\newblock \bibinfo{journal}{\emph{Journal of Systems and Software}}  \bibinfo{volume}{203} (\bibinfo{year}{2023}), \bibinfo{pages}{111734}.
\newblock


\bibitem[Dawson(2013)]%
        {dawson2013cognitive}
\bibfield{author}{\bibinfo{person}{Linda Dawson}.} \bibinfo{year}{2013}\natexlab{}.
\newblock \showarticletitle{Cognitive processes in object-oriented requirements engineering practice: analogical reasoning and mental modelling}.
\newblock In \bibinfo{booktitle}{\emph{Information Systems Development: Reflections, Challenges and New Directions}}. \bibinfo{publisher}{Springer}, \bibinfo{pages}{115--128}.
\newblock


\bibitem[D{\'e}tienne(2001)]%
        {detienne2001software}
\bibfield{author}{\bibinfo{person}{Fran{\c{c}}oise D{\'e}tienne}.} \bibinfo{year}{2001}\natexlab{}.
\newblock \bibinfo{booktitle}{\emph{Software design--cognitive aspect}}.
\newblock \bibinfo{publisher}{Springer Science \& Business Media}.
\newblock


\bibitem[Dunsmore and Roper(2000)]%
        {dunsmore2000comparative}
\bibfield{author}{\bibinfo{person}{Alastair Dunsmore} {and} \bibinfo{person}{Marc Roper}.} \bibinfo{year}{2000}\natexlab{}.
\newblock \showarticletitle{A comparative evaluation of program comprehension measures}.
\newblock \bibinfo{journal}{\emph{The Journal of Systems and Software}} \bibinfo{volume}{52}, \bibinfo{number}{3} (\bibinfo{year}{2000}), \bibinfo{pages}{121--129}.
\newblock


\bibitem[Etikan et~al\mbox{.}(2016)]%
        {etikan2016comparison}
\bibfield{author}{\bibinfo{person}{Ilker Etikan}, \bibinfo{person}{Sulaiman~Abubakar Musa}, \bibinfo{person}{Rukayya~Sunusi Alkassim}, {et~al\mbox{.}}} \bibinfo{year}{2016}\natexlab{}.
\newblock \showarticletitle{Comparison of convenience sampling and purposive sampling}.
\newblock \bibinfo{journal}{\emph{American journal of theoretical and applied statistics}} \bibinfo{volume}{5}, \bibinfo{number}{1} (\bibinfo{year}{2016}), \bibinfo{pages}{1--4}.
\newblock


\bibitem[Ferdowsi et~al\mbox{.}(2023)]%
        {ferdowsi2023live}
\bibfield{author}{\bibinfo{person}{Kasra Ferdowsi}, \bibinfo{person}{Michael~B James}, \bibinfo{person}{Nadia Polikarpova}, \bibinfo{person}{Sorin Lerner}, {et~al\mbox{.}}} \bibinfo{year}{2023}\natexlab{}.
\newblock \showarticletitle{Live Exploration of AI-Generated Programs}.
\newblock \bibinfo{journal}{\emph{arXiv preprint arXiv:2306.09541}} (\bibinfo{year}{2023}).
\newblock


\bibitem[Fernando et~al\mbox{.}(2023)]%
        {fernando2023promptbreeder}
\bibfield{author}{\bibinfo{person}{Chrisantha Fernando}, \bibinfo{person}{Dylan Banarse}, \bibinfo{person}{Henryk Michalewski}, \bibinfo{person}{Simon Osindero}, {and} \bibinfo{person}{Tim Rockt{\"a}schel}.} \bibinfo{year}{2023}\natexlab{}.
\newblock \showarticletitle{Promptbreeder: Self-referential self-improvement via prompt evolution}.
\newblock \bibinfo{journal}{\emph{arXiv preprint arXiv:2309.16797}} (\bibinfo{year}{2023}).
\newblock


\bibitem[Fiannaca et~al\mbox{.}(2023)]%
        {fiannaca2023programming}
\bibfield{author}{\bibinfo{person}{Alexander~J Fiannaca}, \bibinfo{person}{Chinmay Kulkarni}, \bibinfo{person}{Carrie~J Cai}, {and} \bibinfo{person}{Michael Terry}.} \bibinfo{year}{2023}\natexlab{}.
\newblock \showarticletitle{Programming without a Programming Language: Challenges and Opportunities for Designing Developer Tools for Prompt Programming}. In \bibinfo{booktitle}{\emph{Extended Abstracts of the 2023 CHI Conference on Human Factors in Computing Systems}}. \bibinfo{pages}{1--7}.
\newblock


\bibitem[Fix et~al\mbox{.}(1993)]%
        {fix1993mental}
\bibfield{author}{\bibinfo{person}{Vikki Fix}, \bibinfo{person}{Susan Wiedenbeck}, {and} \bibinfo{person}{Jean Scholtz}.} \bibinfo{year}{1993}\natexlab{}.
\newblock \showarticletitle{Mental representations of programs by novices and experts}. In \bibinfo{booktitle}{\emph{Proceedings of the INTERACT'93 and CHI'93 conference on Human factors in computing systems}}. \bibinfo{pages}{74--79}.
\newblock


\bibitem[Fonteyn et~al\mbox{.}(1993)]%
        {fonteyn1993description}
\bibfield{author}{\bibinfo{person}{Marsha~E Fonteyn}, \bibinfo{person}{Benjamin Kuipers}, {and} \bibinfo{person}{Susan~J Grobe}.} \bibinfo{year}{1993}\natexlab{}.
\newblock \showarticletitle{A description of think aloud method and protocol analysis}.
\newblock \bibinfo{journal}{\emph{Qualitative health research}} \bibinfo{volume}{3}, \bibinfo{number}{4} (\bibinfo{year}{1993}), \bibinfo{pages}{430--441}.
\newblock


\bibitem[Friedman(2021)]%
        {Friedman_2021}
\bibfield{author}{\bibinfo{person}{Nat Friedman}.} \bibinfo{year}{2021}\natexlab{}.
\newblock \bibinfo{title}{Introducing GitHub Copilot: your AI pair programmer}.
\newblock
\newblock
\urldef\tempurl%
\url{https://github.blog/2021-06-29-introducing-github-copilot-ai-pair-programmer/}
\showURL{%
\tempurl}


\bibitem[Github(2023)]%
        {Github_2023}
\bibfield{author}{\bibinfo{person}{Github}.} \bibinfo{year}{2023}\natexlab{}.
\newblock \bibinfo{title}{Github Copilot, Your AI pair programmer}.
\newblock
\newblock
\urldef\tempurl%
\url{https://github.com/features/copilot}
\showURL{%
\tempurl}


\bibitem[Green and Petre(1996)]%
        {green1996usability}
\bibfield{author}{\bibinfo{person}{Thomas R.~G. Green} {and} \bibinfo{person}{Marian Petre}.} \bibinfo{year}{1996}\natexlab{}.
\newblock \showarticletitle{Usability analysis of visual programming environments: a ‘cognitive dimensions’ framework}.
\newblock \bibinfo{journal}{\emph{Journal of Visual Languages \& Computing}} \bibinfo{volume}{7}, \bibinfo{number}{2} (\bibinfo{year}{1996}), \bibinfo{pages}{131--174}.
\newblock


\bibitem[Gutwenger et~al\mbox{.}(2003)]%
        {gutwenger2003new}
\bibfield{author}{\bibinfo{person}{Carsten Gutwenger}, \bibinfo{person}{Michael J{\"u}nger}, \bibinfo{person}{Karsten Klein}, \bibinfo{person}{Joachim Kupke}, \bibinfo{person}{Sebastian Leipert}, {and} \bibinfo{person}{Petra Mutzel}.} \bibinfo{year}{2003}\natexlab{}.
\newblock \showarticletitle{A new approach for visualizing UML class diagrams}. In \bibinfo{booktitle}{\emph{Proceedings of the 2003 ACM symposium on Software visualization}}. \bibinfo{pages}{179--188}.
\newblock


\bibitem[Hart and Staveland(1988)]%
        {hart1988development}
\bibfield{author}{\bibinfo{person}{Sandra~G Hart} {and} \bibinfo{person}{Lowell~E Staveland}.} \bibinfo{year}{1988}\natexlab{}.
\newblock \showarticletitle{Development of NASA-TLX (Task Load Index): Results of empirical and theoretical research}.
\newblock In \bibinfo{booktitle}{\emph{Advances in psychology}}. Vol.~\bibinfo{volume}{52}. \bibinfo{publisher}{Elsevier}, \bibinfo{pages}{139--183}.
\newblock


\bibitem[Hayes(2013)]%
        {hayes2013new}
\bibfield{author}{\bibinfo{person}{John~R Hayes}.} \bibinfo{year}{2013}\natexlab{}.
\newblock \showarticletitle{A new framework for understanding cognition and affect in writing}.
\newblock In \bibinfo{booktitle}{\emph{The science of writing}}. \bibinfo{publisher}{Routledge}, \bibinfo{pages}{1--27}.
\newblock


\bibitem[Heinonen et~al\mbox{.}(2023)]%
        {heinonen2023synthesizing}
\bibfield{author}{\bibinfo{person}{Ava Heinonen}, \bibinfo{person}{Bettina Lehtel{\"a}}, \bibinfo{person}{Arto Hellas}, {and} \bibinfo{person}{Fabian Fagerholm}.} \bibinfo{year}{2023}\natexlab{}.
\newblock \showarticletitle{Synthesizing research on programmers’ mental models of programs, tasks and concepts—A systematic literature review}.
\newblock \bibinfo{journal}{\emph{Information and Software Technology}} (\bibinfo{year}{2023}), \bibinfo{pages}{107300}.
\newblock


\bibitem[Hoc(1977)]%
        {hoc1977role}
\bibfield{author}{\bibinfo{person}{J-M Hoc}.} \bibinfo{year}{1977}\natexlab{}.
\newblock \showarticletitle{Role of mental representation in learning a programming language}.
\newblock \bibinfo{journal}{\emph{International Journal of Man-Machine Studies}} \bibinfo{volume}{9}, \bibinfo{number}{1} (\bibinfo{year}{1977}), \bibinfo{pages}{87--105}.
\newblock


\bibitem[H{\"o}{\"o}k(2000)]%
        {hook2000steps}
\bibfield{author}{\bibinfo{person}{Kristina H{\"o}{\"o}k}.} \bibinfo{year}{2000}\natexlab{}.
\newblock \showarticletitle{Steps to take before intelligent user interfaces become real}.
\newblock \bibinfo{journal}{\emph{Interacting with computers}} \bibinfo{volume}{12}, \bibinfo{number}{4} (\bibinfo{year}{2000}), \bibinfo{pages}{409--426}.
\newblock


\bibitem[Huang et~al\mbox{.}(2023)]%
        {huang2023anpl}
\bibfield{author}{\bibinfo{person}{Di Huang}, \bibinfo{person}{Ziyuan Nan}, \bibinfo{person}{Xing Hu}, \bibinfo{person}{Pengwei Jin}, \bibinfo{person}{Shaohui Peng}, \bibinfo{person}{Yuanbo Wen}, \bibinfo{person}{Rui Zhang}, \bibinfo{person}{Zidong Du}, \bibinfo{person}{Qi Guo}, \bibinfo{person}{Yewen Pu}, {et~al\mbox{.}}} \bibinfo{year}{2023}\natexlab{}.
\newblock \showarticletitle{ANPL: Compiling Natural Programs with Interactive Decomposition}.
\newblock \bibinfo{journal}{\emph{arXiv preprint arXiv:2305.18498}} (\bibinfo{year}{2023}).
\newblock


\bibitem[Hutchins et~al\mbox{.}(1985)]%
        {hutchins1985direct}
\bibfield{author}{\bibinfo{person}{Edwin~L Hutchins}, \bibinfo{person}{James~D Hollan}, {and} \bibinfo{person}{Donald~A Norman}.} \bibinfo{year}{1985}\natexlab{}.
\newblock \showarticletitle{Direct manipulation interfaces}.
\newblock \bibinfo{journal}{\emph{Human--computer interaction}} \bibinfo{volume}{1}, \bibinfo{number}{4} (\bibinfo{year}{1985}), \bibinfo{pages}{311--338}.
\newblock


\bibitem[Jeffries(1982)]%
        {jeffries1982comparison}
\bibfield{author}{\bibinfo{person}{Robin Jeffries}.} \bibinfo{year}{1982}\natexlab{}.
\newblock \showarticletitle{A comparison of the debugging behavior of expert and novice programmers}. In \bibinfo{booktitle}{\emph{Proceedings of AERA annual meeting}}. \bibinfo{pages}{1--17}.
\newblock


\bibitem[Jiang et~al\mbox{.}(2022)]%
        {jiang2022discovering}
\bibfield{author}{\bibinfo{person}{Ellen Jiang}, \bibinfo{person}{Edwin Toh}, \bibinfo{person}{Alejandra Molina}, \bibinfo{person}{Kristen Olson}, \bibinfo{person}{Claire Kayacik}, \bibinfo{person}{Aaron Donsbach}, \bibinfo{person}{Carrie~J Cai}, {and} \bibinfo{person}{Michael Terry}.} \bibinfo{year}{2022}\natexlab{}.
\newblock \showarticletitle{Discovering the syntax and strategies of natural language programming with generative language models}. In \bibinfo{booktitle}{\emph{Proceedings of the 2022 CHI Conference on Human Factors in Computing Systems}}. \bibinfo{pages}{1--19}.
\newblock


\bibitem[Kery and Myers(2017)]%
        {kery2017exploring}
\bibfield{author}{\bibinfo{person}{Mary~Beth Kery} {and} \bibinfo{person}{Brad~A Myers}.} \bibinfo{year}{2017}\natexlab{}.
\newblock \showarticletitle{Exploring exploratory programming}. In \bibinfo{booktitle}{\emph{2017 IEEE Symposium on Visual Languages and Human-Centric Computing (VL/HCC)}}. IEEE, \bibinfo{pages}{25--29}.
\newblock


\bibitem[Khot et~al\mbox{.}(2022)]%
        {khot2022decomposed}
\bibfield{author}{\bibinfo{person}{Tushar Khot}, \bibinfo{person}{Harsh Trivedi}, \bibinfo{person}{Matthew Finlayson}, \bibinfo{person}{Yao Fu}, \bibinfo{person}{Kyle Richardson}, \bibinfo{person}{Peter Clark}, {and} \bibinfo{person}{Ashish Sabharwal}.} \bibinfo{year}{2022}\natexlab{}.
\newblock \showarticletitle{Decomposed prompting: A modular approach for solving complex tasks}.
\newblock \bibinfo{journal}{\emph{arXiv preprint arXiv:2210.02406}} (\bibinfo{year}{2022}).
\newblock


\bibitem[Kim et~al\mbox{.}(1995)]%
        {kim1995internal}
\bibfield{author}{\bibinfo{person}{Jinwoo Kim}, \bibinfo{person}{F~Javier Lerch}, {and} \bibinfo{person}{Herbert~A Simon}.} \bibinfo{year}{1995}\natexlab{}.
\newblock \showarticletitle{Internal representation and rule development in object-oriented design}.
\newblock \bibinfo{journal}{\emph{ACM Transactions on Computer-Human Interaction (TOCHI)}} \bibinfo{volume}{2}, \bibinfo{number}{4} (\bibinfo{year}{1995}), \bibinfo{pages}{357--390}.
\newblock


\bibitem[LangChain(2023)]%
        {langchain2023}
\bibfield{author}{\bibinfo{person}{LangChain}.} \bibinfo{year}{2023}\natexlab{}.
\newblock
\newblock
\urldef\tempurl%
\url{https://www.langchain.com/}
\showURL{%
\tempurl}


\bibitem[Lewis et~al\mbox{.}(2013)]%
        {lewis2013umuxlite}
\bibfield{author}{\bibinfo{person}{James~R. Lewis}, \bibinfo{person}{Brian~S. Utesch}, {and} \bibinfo{person}{Deborah~E. Maher}.} \bibinfo{year}{2013}\natexlab{}.
\newblock \showarticletitle{UMUX-LITE: When There's No Time for the SUS}. In \bibinfo{booktitle}{\emph{Proceedings of the SIGCHI Conference on Human Factors in Computing Systems}} (Paris, France) \emph{(\bibinfo{series}{CHI '13})}. \bibinfo{publisher}{Association for Computing Machinery}, \bibinfo{address}{New York, NY, USA}, \bibinfo{pages}{2099–2102}.
\newblock
\showISBNx{9781450318990}
\urldef\tempurl%
\url{https://doi.org/10.1145/2470654.2481287}
\showDOI{\tempurl}


\bibitem[Li et~al\mbox{.}(2023)]%
        {li2023chain}
\bibfield{author}{\bibinfo{person}{Chengshu Li}, \bibinfo{person}{Jacky Liang}, \bibinfo{person}{Fei Xia}, \bibinfo{person}{Andy Zeng}, \bibinfo{person}{Sergey Levine}, \bibinfo{person}{Dorsa Sadigh}, \bibinfo{person}{Karol Hausman}, \bibinfo{person}{Xinyun Chen}, \bibinfo{person}{Li Fei-Fei}, {et~al\mbox{.}}} \bibinfo{year}{2023}\natexlab{}.
\newblock \showarticletitle{Chain of Code: Reasoning with a Language Model-Augmented Code Interpreter}. In \bibinfo{booktitle}{\emph{NeurIPS 2023 Foundation Models for Decision Making Workshop}}.
\newblock


\bibitem[Li et~al\mbox{.}(2022)]%
        {li2022competition}
\bibfield{author}{\bibinfo{person}{Yujia Li}, \bibinfo{person}{David Choi}, \bibinfo{person}{Junyoung Chung}, \bibinfo{person}{Nate Kushman}, \bibinfo{person}{Julian Schrittwieser}, \bibinfo{person}{R{\'e}mi Leblond}, \bibinfo{person}{Tom Eccles}, \bibinfo{person}{James Keeling}, \bibinfo{person}{Felix Gimeno}, \bibinfo{person}{Agustin Dal~Lago}, {et~al\mbox{.}}} \bibinfo{year}{2022}\natexlab{}.
\newblock \showarticletitle{Competition-level code generation with alphacode}.
\newblock \bibinfo{journal}{\emph{Science}} \bibinfo{volume}{378}, \bibinfo{number}{6624} (\bibinfo{year}{2022}), \bibinfo{pages}{1092--1097}.
\newblock


\bibitem[Liang et~al\mbox{.}(2023a)]%
        {liang2023large}
\bibfield{author}{\bibinfo{person}{Jenny~T Liang}, \bibinfo{person}{Chenyang Yang}, {and} \bibinfo{person}{Brad~A Myers}.} \bibinfo{year}{2023}\natexlab{a}.
\newblock \showarticletitle{A Large-Scale Survey on the Usability of AI Programming Assistants: Successes and Challenges}. In \bibinfo{booktitle}{\emph{2024 IEEE/ACM 46th International Conference on Software Engineering (ICSE)}}. IEEE Computer Society, \bibinfo{pages}{605--617}.
\newblock


\bibitem[Liang et~al\mbox{.}(2023b)]%
        {liang2023understanding}
\bibfield{author}{\bibinfo{person}{Jenny~T Liang}, \bibinfo{person}{Chenyang Yang}, {and} \bibinfo{person}{Brad~A Myers}.} \bibinfo{year}{2023}\natexlab{b}.
\newblock \showarticletitle{Understanding the Usability of AI Programming Assistants}.
\newblock \bibinfo{journal}{\emph{arXiv preprint arXiv:2303.17125}} (\bibinfo{year}{2023}).
\newblock


\bibitem[Lieberman(2001)]%
        {lieberman2001your}
\bibfield{author}{\bibinfo{person}{Henry Lieberman}.} \bibinfo{year}{2001}\natexlab{}.
\newblock \bibinfo{booktitle}{\emph{Your wish is my command: Programming by example}}.
\newblock \bibinfo{publisher}{Morgan Kaufmann}.
\newblock


\bibitem[Liu et~al\mbox{.}(2023a)]%
        {liu2023wants}
\bibfield{author}{\bibinfo{person}{Michael~Xieyang Liu}, \bibinfo{person}{Advait Sarkar}, \bibinfo{person}{Carina Negreanu}, \bibinfo{person}{Benjamin Zorn}, \bibinfo{person}{Jack Williams}, \bibinfo{person}{Neil Toronto}, {and} \bibinfo{person}{Andrew~D Gordon}.} \bibinfo{year}{2023}\natexlab{a}.
\newblock \showarticletitle{“What It Wants Me To Say”: Bridging the Abstraction Gap Between End-User Programmers and Code-Generating Large Language Models}. In \bibinfo{booktitle}{\emph{Proceedings of the 2023 CHI Conference on Human Factors in Computing Systems}}. \bibinfo{pages}{1--31}.
\newblock


\bibitem[Liu et~al\mbox{.}(2023b)]%
        {liu2023pre}
\bibfield{author}{\bibinfo{person}{Pengfei Liu}, \bibinfo{person}{Weizhe Yuan}, \bibinfo{person}{Jinlan Fu}, \bibinfo{person}{Zhengbao Jiang}, \bibinfo{person}{Hiroaki Hayashi}, {and} \bibinfo{person}{Graham Neubig}.} \bibinfo{year}{2023}\natexlab{b}.
\newblock \showarticletitle{Pre-train, prompt, and predict: A systematic survey of prompting methods in natural language processing}.
\newblock \bibinfo{journal}{\emph{Comput. Surveys}} \bibinfo{volume}{55}, \bibinfo{number}{9} (\bibinfo{year}{2023}), \bibinfo{pages}{1--35}.
\newblock


\bibitem[Luger and Sellen(2016)]%
        {luger2016like}
\bibfield{author}{\bibinfo{person}{Ewa Luger} {and} \bibinfo{person}{Abigail Sellen}.} \bibinfo{year}{2016}\natexlab{}.
\newblock \showarticletitle{" Like Having a Really Bad PA" The Gulf between User Expectation and Experience of Conversational Agents}. In \bibinfo{booktitle}{\emph{Proceedings of the 2016 CHI conference on human factors in computing systems}}. \bibinfo{pages}{5286--5297}.
\newblock


\bibitem[Ma et~al\mbox{.}(2023)]%
        {ma2023demonstration}
\bibfield{author}{\bibinfo{person}{Pingchuan Ma}, \bibinfo{person}{Rui Ding}, \bibinfo{person}{Shuai Wang}, \bibinfo{person}{Shi Han}, {and} \bibinfo{person}{Dongmei Zhang}.} \bibinfo{year}{2023}\natexlab{}.
\newblock \showarticletitle{Demonstration of InsightPilot: An LLM-Empowered Automated Data Exploration System}.
\newblock \bibinfo{journal}{\emph{arXiv preprint arXiv:2304.00477}} (\bibinfo{year}{2023}).
\newblock


\bibitem[Maalej et~al\mbox{.}(2014)]%
        {maalej2014comprehension}
\bibfield{author}{\bibinfo{person}{Walid Maalej}, \bibinfo{person}{Rebecca Tiarks}, \bibinfo{person}{Tobias Roehm}, {and} \bibinfo{person}{Rainer Koschke}.} \bibinfo{year}{2014}\natexlab{}.
\newblock \showarticletitle{On the comprehension of program comprehension}.
\newblock \bibinfo{journal}{\emph{ACM Transactions on Software Engineering and Methodology (TOSEM)}} \bibinfo{volume}{23}, \bibinfo{number}{4} (\bibinfo{year}{2014}), \bibinfo{pages}{1--37}.
\newblock


\bibitem[Microsoft(2023)]%
        {microsoft2023monaco}
\bibfield{author}{\bibinfo{person}{Microsoft}.} \bibinfo{year}{2023}\natexlab{}.
\newblock \bibinfo{title}{The Editor of the Web}.
\newblock
\newblock
\urldef\tempurl%
\url{https://microsoft.github.io/monaco-editor}
\showURL{%
\tempurl}


\bibitem[Mozannar et~al\mbox{.}(2022)]%
        {mozannar2022reading}
\bibfield{author}{\bibinfo{person}{Hussein Mozannar}, \bibinfo{person}{Gagan Bansal}, \bibinfo{person}{Adam Fourney}, {and} \bibinfo{person}{Eric Horvitz}.} \bibinfo{year}{2022}\natexlab{}.
\newblock \showarticletitle{Reading between the lines: Modeling user behaviour and costs in AI-assisted programming}.
\newblock \bibinfo{journal}{\emph{arXiv preprint arXiv:2210.14306}} (\bibinfo{year}{2022}).
\newblock


\bibitem[Navarro-Prieto and Canas(2001)]%
        {navarro2001visual}
\bibfield{author}{\bibinfo{person}{Raquel Navarro-Prieto} {and} \bibinfo{person}{Jose~J Canas}.} \bibinfo{year}{2001}\natexlab{}.
\newblock \showarticletitle{Are visual programming languages better? The role of imagery in program comprehension}.
\newblock \bibinfo{journal}{\emph{International Journal of Human-Computer Studies}} \bibinfo{volume}{54}, \bibinfo{number}{6} (\bibinfo{year}{2001}), \bibinfo{pages}{799--829}.
\newblock


\bibitem[Nguyen and Nadi(2022)]%
        {nguyen2022empirical}
\bibfield{author}{\bibinfo{person}{Nhan Nguyen} {and} \bibinfo{person}{Sarah Nadi}.} \bibinfo{year}{2022}\natexlab{}.
\newblock \showarticletitle{An empirical evaluation of GitHub copilot's code suggestions}. In \bibinfo{booktitle}{\emph{Proceedings of the 19th International Conference on Mining Software Repositories}}. \bibinfo{pages}{1--5}.
\newblock


\bibitem[Norman(1986)]%
        {norman1986cognitive}
\bibfield{author}{\bibinfo{person}{Donald~A Norman}.} \bibinfo{year}{1986}\natexlab{}.
\newblock \showarticletitle{Cognitive engineering}.
\newblock \bibinfo{journal}{\emph{User centered system design}} \bibinfo{volume}{31}, \bibinfo{number}{61} (\bibinfo{year}{1986}), \bibinfo{pages}{2}.
\newblock


\bibitem[Norman(1994)]%
        {norman1994might}
\bibfield{author}{\bibinfo{person}{Donald~A Norman}.} \bibinfo{year}{1994}\natexlab{}.
\newblock \showarticletitle{How might people interact with agents}.
\newblock \bibinfo{journal}{\emph{Commun. ACM}} \bibinfo{volume}{37}, \bibinfo{number}{7} (\bibinfo{year}{1994}), \bibinfo{pages}{68--71}.
\newblock


\bibitem[OpenAI(2023a)]%
        {openai2023chatgpt}
\bibfield{author}{\bibinfo{person}{OpenAI}.} \bibinfo{year}{2023}\natexlab{a}.
\newblock \bibinfo{title}{ChatGPT (Feb 13 version) [Large language model]}.
\newblock
\newblock
\urldef\tempurl%
\url{https://chat.openai.com}
\showURL{%
\tempurl}


\bibitem[OpenAI(2023b)]%
        {openai2023gpt4}
\bibfield{author}{\bibinfo{person}{OpenAI}.} \bibinfo{year}{2023}\natexlab{b}.
\newblock \bibinfo{title}{GPT-4 Technical Report}.
\newblock
\newblock
\showeprint[arxiv]{2303.08774}~[cs.CL]


\bibitem[Pearce et~al\mbox{.}(2023)]%
        {pearce2023examining}
\bibfield{author}{\bibinfo{person}{Hammond Pearce}, \bibinfo{person}{Benjamin Tan}, \bibinfo{person}{Baleegh Ahmad}, \bibinfo{person}{Ramesh Karri}, {and} \bibinfo{person}{Brendan Dolan-Gavitt}.} \bibinfo{year}{2023}\natexlab{}.
\newblock \showarticletitle{Examining zero-shot vulnerability repair with large language models}. In \bibinfo{booktitle}{\emph{2023 IEEE Symposium on Security and Privacy (SP)}}. IEEE, \bibinfo{pages}{2339--2356}.
\newblock


\bibitem[Pudari and Ernst(2023)]%
        {pudari2023copilot}
\bibfield{author}{\bibinfo{person}{Rohith Pudari} {and} \bibinfo{person}{Neil~A Ernst}.} \bibinfo{year}{2023}\natexlab{}.
\newblock \showarticletitle{From Copilot to Pilot: Towards AI Supported Software Development}.
\newblock \bibinfo{journal}{\emph{arXiv preprint arXiv:2303.04142}} (\bibinfo{year}{2023}).
\newblock


\bibitem[Pyodide(2023)]%
        {pyodide2023}
\bibfield{author}{\bibinfo{person}{Pyodide}.} \bibinfo{year}{2023}\natexlab{}.
\newblock \bibinfo{title}{Pyodide is a Python distribution for the browser and Node.js based on WebAssembly.}
\newblock
\newblock
\urldef\tempurl%
\url{https://github.com/pyodide/pyodide}
\showURL{%
\tempurl}


\bibitem[Ritschel et~al\mbox{.}(2022)]%
        {ritschel2022can}
\bibfield{author}{\bibinfo{person}{Nico Ritschel}, \bibinfo{person}{Felipe Fronchetti}, \bibinfo{person}{Reid Holmes}, \bibinfo{person}{Ronald Garcia}, {and} \bibinfo{person}{David~C Shepherd}.} \bibinfo{year}{2022}\natexlab{}.
\newblock \showarticletitle{Can guided decomposition help end-users write larger block-based programs? a mobile robot experiment}.
\newblock \bibinfo{journal}{\emph{Proceedings of the ACM on Programming Languages}} \bibinfo{volume}{6}, \bibinfo{number}{OOPSLA2} (\bibinfo{year}{2022}), \bibinfo{pages}{233--258}.
\newblock


\bibitem[Romero(2001)]%
        {romero2001focal}
\bibfield{author}{\bibinfo{person}{Pablo Romero}.} \bibinfo{year}{2001}\natexlab{}.
\newblock \showarticletitle{Focal structures and information types in Prolog}.
\newblock \bibinfo{journal}{\emph{International Journal of Human-Computer Studies}} \bibinfo{volume}{54}, \bibinfo{number}{2} (\bibinfo{year}{2001}), \bibinfo{pages}{211--236}.
\newblock


\bibitem[Ross et~al\mbox{.}(2023)]%
        {ross2023programmer}
\bibfield{author}{\bibinfo{person}{Steven~I Ross}, \bibinfo{person}{Fernando Martinez}, \bibinfo{person}{Stephanie Houde}, \bibinfo{person}{Michael Muller}, {and} \bibinfo{person}{Justin~D Weisz}.} \bibinfo{year}{2023}\natexlab{}.
\newblock \showarticletitle{The programmer’s assistant: Conversational interaction with a large language model for software development}. In \bibinfo{booktitle}{\emph{Proceedings of the 28th International Conference on Intelligent User Interfaces}}. \bibinfo{pages}{491--514}.
\newblock


\bibitem[Sarkar et~al\mbox{.}(2022)]%
        {sarkar2022like}
\bibfield{author}{\bibinfo{person}{Advait Sarkar}, \bibinfo{person}{Andrew~D Gordon}, \bibinfo{person}{Carina Negreanu}, \bibinfo{person}{Christian Poelitz}, \bibinfo{person}{Sruti~Srinivasa Ragavan}, {and} \bibinfo{person}{Ben Zorn}.} \bibinfo{year}{2022}\natexlab{}.
\newblock \showarticletitle{What is it like to program with artificial intelligence?}
\newblock \bibinfo{journal}{\emph{arXiv preprint arXiv:2208.06213}} (\bibinfo{year}{2022}).
\newblock


\bibitem[Shargabi et~al\mbox{.}(2015)]%
        {shargabi2015program}
\bibfield{author}{\bibinfo{person}{Amal Shargabi}, \bibinfo{person}{Syed~Ahmad Aljunid}, \bibinfo{person}{Muthukkaruppanan Annamalai}, \bibinfo{person}{Shuhaida~Mohamed Shuhidan}, {and} \bibinfo{person}{Abdullah~Mohd Zin}.} \bibinfo{year}{2015}\natexlab{}.
\newblock \showarticletitle{Program comprehension levels of abstraction for novices}. In \bibinfo{booktitle}{\emph{2015 International Conference on Computer, Communications, and Control Technology (I4CT)}}. IEEE, \bibinfo{pages}{211--215}.
\newblock


\bibitem[Snoek et~al\mbox{.}(2012)]%
        {snoek2012practical}
\bibfield{author}{\bibinfo{person}{Jasper Snoek}, \bibinfo{person}{Hugo Larochelle}, {and} \bibinfo{person}{Ryan~P Adams}.} \bibinfo{year}{2012}\natexlab{}.
\newblock \showarticletitle{Practical bayesian optimization of machine learning algorithms}.
\newblock \bibinfo{journal}{\emph{Advances in neural information processing systems}}  \bibinfo{volume}{25} (\bibinfo{year}{2012}).
\newblock


\bibitem[Strobelt et~al\mbox{.}(2022)]%
        {strobelt2022interactive}
\bibfield{author}{\bibinfo{person}{Hendrik Strobelt}, \bibinfo{person}{Albert Webson}, \bibinfo{person}{Victor Sanh}, \bibinfo{person}{Benjamin Hoover}, \bibinfo{person}{Johanna Beyer}, \bibinfo{person}{Hanspeter Pfister}, {and} \bibinfo{person}{Alexander~M Rush}.} \bibinfo{year}{2022}\natexlab{}.
\newblock \showarticletitle{Interactive and visual prompt engineering for ad-hoc task adaptation with large language models}.
\newblock \bibinfo{journal}{\emph{IEEE transactions on visualization and computer graphics}} \bibinfo{volume}{29}, \bibinfo{number}{1} (\bibinfo{year}{2022}), \bibinfo{pages}{1146--1156}.
\newblock


\bibitem[Tang et~al\mbox{.}(2023)]%
        {tang2023empirical}
\bibfield{author}{\bibinfo{person}{Ningzhi Tang}, \bibinfo{person}{Meng Chen}, \bibinfo{person}{Zheng Ning}, \bibinfo{person}{Aakash Bansal}, \bibinfo{person}{Yu Huang}, \bibinfo{person}{Collin McMillan}, {and} \bibinfo{person}{Toby Jia-Jun Li}.} \bibinfo{year}{2023}\natexlab{}.
\newblock \showarticletitle{An Empirical Study of Developer Behaviors for Validating and Repairing AI-Generated Code}. Plateau Workshop.
\newblock


\bibitem[Tukey et~al\mbox{.}(1977)]%
        {tukey1977exploratory}
\bibfield{author}{\bibinfo{person}{John~W Tukey} {et~al\mbox{.}}} \bibinfo{year}{1977}\natexlab{}.
\newblock \bibinfo{booktitle}{\emph{Exploratory data analysis}}. Vol.~\bibinfo{volume}{2}.
\newblock \bibinfo{publisher}{Reading, MA}.
\newblock


\bibitem[Vaithilingam et~al\mbox{.}(2022)]%
        {vaithilingam2022expectation}
\bibfield{author}{\bibinfo{person}{Priyan Vaithilingam}, \bibinfo{person}{Tianyi Zhang}, {and} \bibinfo{person}{Elena~L Glassman}.} \bibinfo{year}{2022}\natexlab{}.
\newblock \showarticletitle{Expectation vs. experience: Evaluating the usability of code generation tools powered by large language models}. In \bibinfo{booktitle}{\emph{Chi conference on human factors in computing systems extended abstracts}}. \bibinfo{pages}{1--7}.
\newblock


\bibitem[Vasconcelos et~al\mbox{.}(2023)]%
        {vasconcelos2023generation}
\bibfield{author}{\bibinfo{person}{Helena Vasconcelos}, \bibinfo{person}{Gagan Bansal}, \bibinfo{person}{Adam Fourney}, \bibinfo{person}{Q~Vera Liao}, {and} \bibinfo{person}{Jennifer~Wortman Vaughan}.} \bibinfo{year}{2023}\natexlab{}.
\newblock \showarticletitle{Generation probabilities are not enough: Exploring the effectiveness of uncertainty highlighting in AI-powered code completions}.
\newblock \bibinfo{journal}{\emph{arXiv preprint arXiv:2302.07248}} (\bibinfo{year}{2023}).
\newblock


\bibitem[von Mayrhauser and Vans(1995)]%
        {von1995industrial}
\bibfield{author}{\bibinfo{person}{Anneliese von Mayrhauser} {and} \bibinfo{person}{A~Marie Vans}.} \bibinfo{year}{1995}\natexlab{}.
\newblock \showarticletitle{Industrial experience with an integrated code comprehension model}.
\newblock \bibinfo{journal}{\emph{Software Engineering Journal}} \bibinfo{volume}{10}, \bibinfo{number}{5} (\bibinfo{year}{1995}), \bibinfo{pages}{171--182}.
\newblock


\bibitem[Von~Mayrhauser and Vans(1995)]%
        {von1995program}
\bibfield{author}{\bibinfo{person}{Anneliese Von~Mayrhauser} {and} \bibinfo{person}{A~Marie Vans}.} \bibinfo{year}{1995}\natexlab{}.
\newblock \showarticletitle{Program comprehension during software maintenance and evolution}.
\newblock \bibinfo{journal}{\emph{Computer}} \bibinfo{volume}{28}, \bibinfo{number}{8} (\bibinfo{year}{1995}), \bibinfo{pages}{44--55}.
\newblock


\bibitem[Wei et~al\mbox{.}(2023)]%
        {wei2023chainofthought}
\bibfield{author}{\bibinfo{person}{Jason Wei}, \bibinfo{person}{Xuezhi Wang}, \bibinfo{person}{Dale Schuurmans}, \bibinfo{person}{Maarten Bosma}, \bibinfo{person}{Brian Ichter}, \bibinfo{person}{Fei Xia}, \bibinfo{person}{Ed Chi}, \bibinfo{person}{Quoc Le}, {and} \bibinfo{person}{Denny Zhou}.} \bibinfo{year}{2023}\natexlab{}.
\newblock \bibinfo{title}{Chain-of-Thought Prompting Elicits Reasoning in Large Language Models}.
\newblock
\newblock
\showeprint[arxiv]{2201.11903}~[cs.CL]


\bibitem[Weisz et~al\mbox{.}(2021)]%
        {weisz2021perfection}
\bibfield{author}{\bibinfo{person}{Justin~D Weisz}, \bibinfo{person}{Michael Muller}, \bibinfo{person}{Stephanie Houde}, \bibinfo{person}{John Richards}, \bibinfo{person}{Steven~I Ross}, \bibinfo{person}{Fernando Martinez}, \bibinfo{person}{Mayank Agarwal}, {and} \bibinfo{person}{Kartik Talamadupula}.} \bibinfo{year}{2021}\natexlab{}.
\newblock \showarticletitle{Perfection not required? Human-AI partnerships in code translation}. In \bibinfo{booktitle}{\emph{26th International Conference on Intelligent User Interfaces}}. \bibinfo{pages}{402--412}.
\newblock


\bibitem[White et~al\mbox{.}(2023)]%
        {white2023prompt}
\bibfield{author}{\bibinfo{person}{Jules White}, \bibinfo{person}{Quchen Fu}, \bibinfo{person}{Sam Hays}, \bibinfo{person}{Michael Sandborn}, \bibinfo{person}{Carlos Olea}, \bibinfo{person}{Henry Gilbert}, \bibinfo{person}{Ashraf Elnashar}, \bibinfo{person}{Jesse Spencer-Smith}, {and} \bibinfo{person}{Douglas~C Schmidt}.} \bibinfo{year}{2023}\natexlab{}.
\newblock \showarticletitle{A prompt pattern catalog to enhance prompt engineering with chatgpt}.
\newblock \bibinfo{journal}{\emph{arXiv preprint arXiv:2302.11382}} (\bibinfo{year}{2023}).
\newblock


\bibitem[Wiedenbeck et~al\mbox{.}(1993)]%
        {wiedenbeck1993characteristics}
\bibfield{author}{\bibinfo{person}{Susan Wiedenbeck}, \bibinfo{person}{Vikki Fix}, {and} \bibinfo{person}{Jean Scholtz}.} \bibinfo{year}{1993}\natexlab{}.
\newblock \showarticletitle{Characteristics of the mental representations of novice and expert programmers: an empirical study}.
\newblock \bibinfo{journal}{\emph{International Journal of Man-Machine Studies}} \bibinfo{volume}{39}, \bibinfo{number}{5} (\bibinfo{year}{1993}), \bibinfo{pages}{793--812}.
\newblock


\bibitem[Wu et~al\mbox{.}(2022a)]%
        {wu2022promptchainer}
\bibfield{author}{\bibinfo{person}{Tongshuang Wu}, \bibinfo{person}{Ellen Jiang}, \bibinfo{person}{Aaron Donsbach}, \bibinfo{person}{Jeff Gray}, \bibinfo{person}{Alejandra Molina}, \bibinfo{person}{Michael Terry}, {and} \bibinfo{person}{Carrie~J Cai}.} \bibinfo{year}{2022}\natexlab{a}.
\newblock \showarticletitle{Promptchainer: Chaining large language model prompts through visual programming}. In \bibinfo{booktitle}{\emph{CHI Conference on Human Factors in Computing Systems Extended Abstracts}}. \bibinfo{pages}{1--10}.
\newblock


\bibitem[Wu et~al\mbox{.}(2022b)]%
        {wu2022ai}
\bibfield{author}{\bibinfo{person}{Tongshuang Wu}, \bibinfo{person}{Michael Terry}, {and} \bibinfo{person}{Carrie~Jun Cai}.} \bibinfo{year}{2022}\natexlab{b}.
\newblock \showarticletitle{Ai chains: Transparent and controllable human-ai interaction by chaining large language model prompts}. In \bibinfo{booktitle}{\emph{Proceedings of the 2022 CHI conference on human factors in computing systems}}. \bibinfo{pages}{1--22}.
\newblock


\bibitem[Xie et~al\mbox{.}(2017)]%
        {xie2017influential}
\bibfield{author}{\bibinfo{person}{Bingjun Xie}, \bibinfo{person}{Jia Zhou}, \bibinfo{person}{Huilin Wang}, {et~al\mbox{.}}} \bibinfo{year}{2017}\natexlab{}.
\newblock \showarticletitle{How influential are mental models on interaction performance? exploring the gap between users’ and designers’ mental models through a new quantitative method}.
\newblock \bibinfo{journal}{\emph{Advances in Human-Computer Interaction}}  \bibinfo{volume}{2017} (\bibinfo{year}{2017}).
\newblock


\bibitem[Xu et~al\mbox{.}(2022)]%
        {xu2022ide}
\bibfield{author}{\bibinfo{person}{Frank~F Xu}, \bibinfo{person}{Bogdan Vasilescu}, {and} \bibinfo{person}{Graham Neubig}.} \bibinfo{year}{2022}\natexlab{}.
\newblock \showarticletitle{In-ide code generation from natural language: Promise and challenges}.
\newblock \bibinfo{journal}{\emph{ACM Transactions on Software Engineering and Methodology (TOSEM)}} \bibinfo{volume}{31}, \bibinfo{number}{2} (\bibinfo{year}{2022}), \bibinfo{pages}{1--47}.
\newblock


\bibitem[Ye and Salvendy(1996)]%
        {ye1996expert}
\bibfield{author}{\bibinfo{person}{Nong Ye} {and} \bibinfo{person}{Gavriel Salvendy}.} \bibinfo{year}{1996}\natexlab{}.
\newblock \showarticletitle{Expert-novice knowledge of computer programming at different levels of abstraction}.
\newblock \bibinfo{journal}{\emph{Ergonomics}} \bibinfo{volume}{39}, \bibinfo{number}{3} (\bibinfo{year}{1996}), \bibinfo{pages}{461--481}.
\newblock


\bibitem[Yoon and Myers(2014)]%
        {yoon2014longitudinal}
\bibfield{author}{\bibinfo{person}{Young~Seok Yoon} {and} \bibinfo{person}{Brad~A Myers}.} \bibinfo{year}{2014}\natexlab{}.
\newblock \showarticletitle{A longitudinal study of programmers' backtracking}. In \bibinfo{booktitle}{\emph{2014 IEEE Symposium on Visual Languages and Human-Centric Computing (VL/HCC)}}. IEEE, \bibinfo{pages}{101--108}.
\newblock


\bibitem[Ziegler et~al\mbox{.}(2022)]%
        {ziegler2022productivity}
\bibfield{author}{\bibinfo{person}{Albert Ziegler}, \bibinfo{person}{Eirini Kalliamvakou}, \bibinfo{person}{X~Alice Li}, \bibinfo{person}{Andrew Rice}, \bibinfo{person}{Devon Rifkin}, \bibinfo{person}{Shawn Simister}, \bibinfo{person}{Ganesh Sittampalam}, {and} \bibinfo{person}{Edward Aftandilian}.} \bibinfo{year}{2022}\natexlab{}.
\newblock \showarticletitle{Productivity assessment of neural code completion}. In \bibinfo{booktitle}{\emph{Proceedings of the 6th ACM SIGPLAN International Symposium on Machine Programming}}. \bibinfo{pages}{21--29}.
\newblock


\end{thebibliography}

\appendix
\onecolumn

\section{Prompt Template}
\label{appendix:prompt}
\begin{figure}[H]
    \centering
    \begin{minipage}[b]{0.47\textwidth}
    \includegraphics[width=\textwidth]{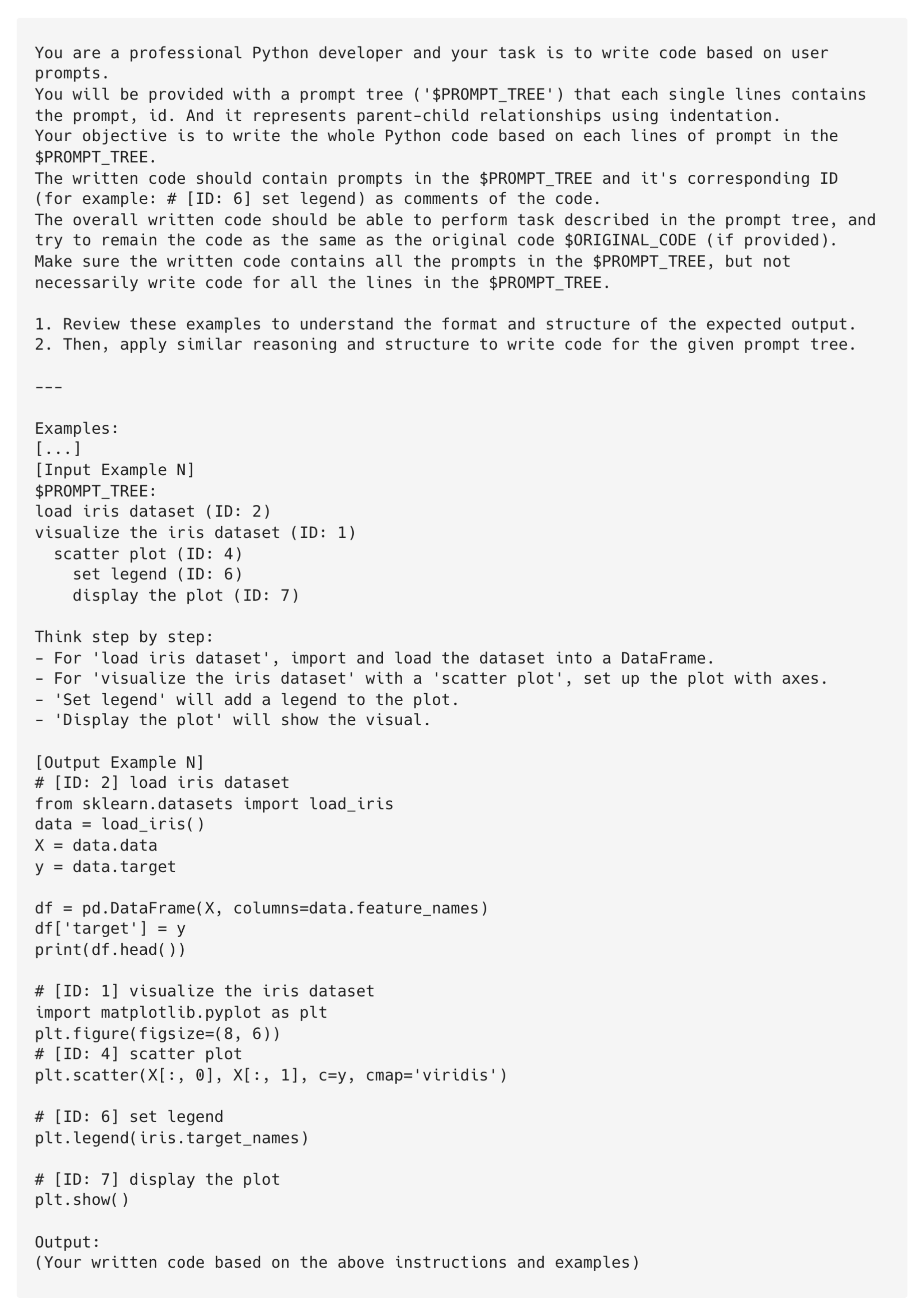}
    \caption{Prompt template for the block's operation [Add]}
    \label{fig:prompt-add}
    \end{minipage}
  \hfill
  \begin{minipage}[b]{0.47\textwidth}
    \includegraphics[width=\textwidth]{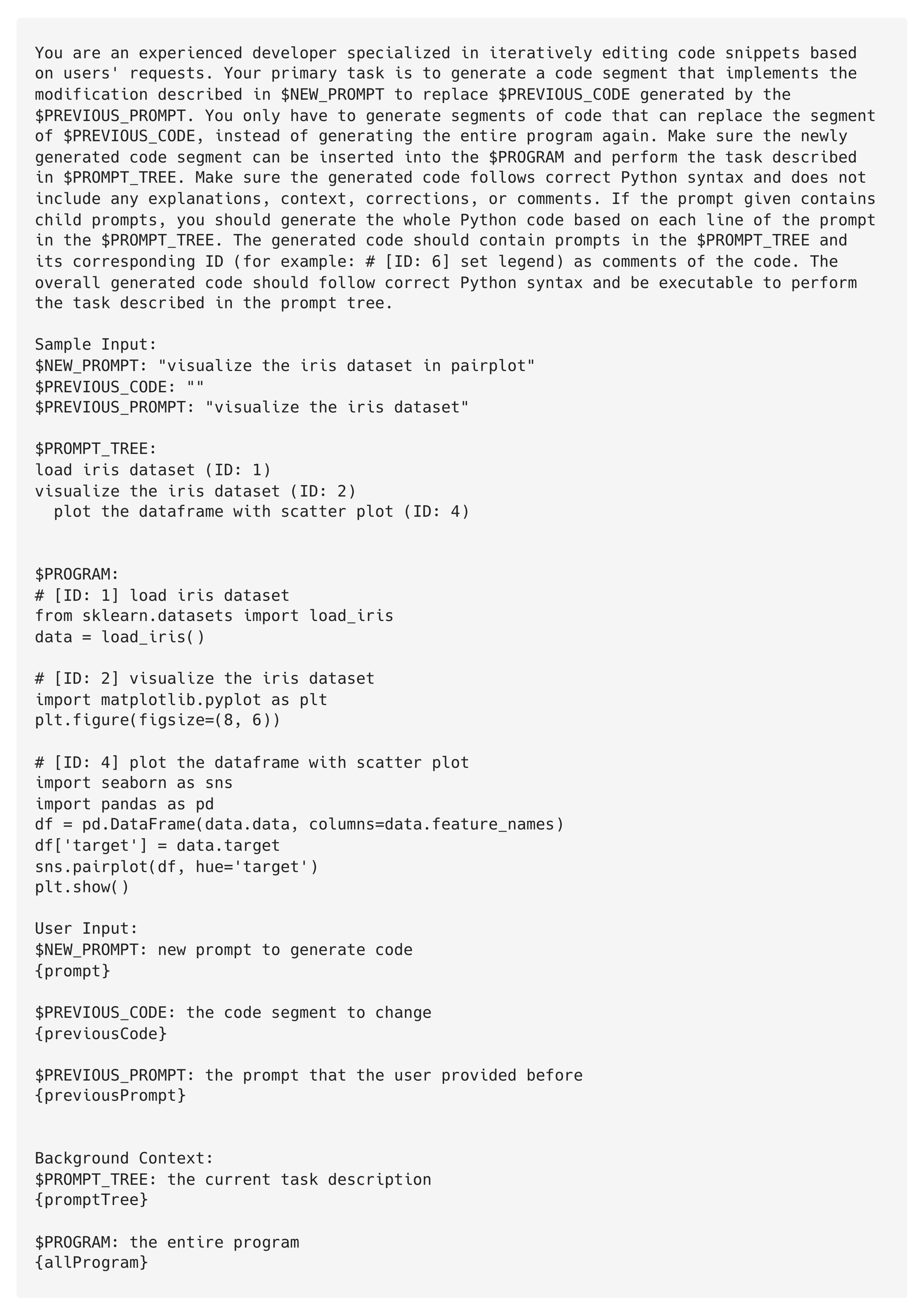}
    \caption{Prompt template for [Edit]}
    \label{fig:prompt-edit}
  \end{minipage}
\end{figure}
\begin{figure}[H]
\centering
    \begin{minipage}[b]{0.65\textwidth}
    \includegraphics[width=\textwidth]{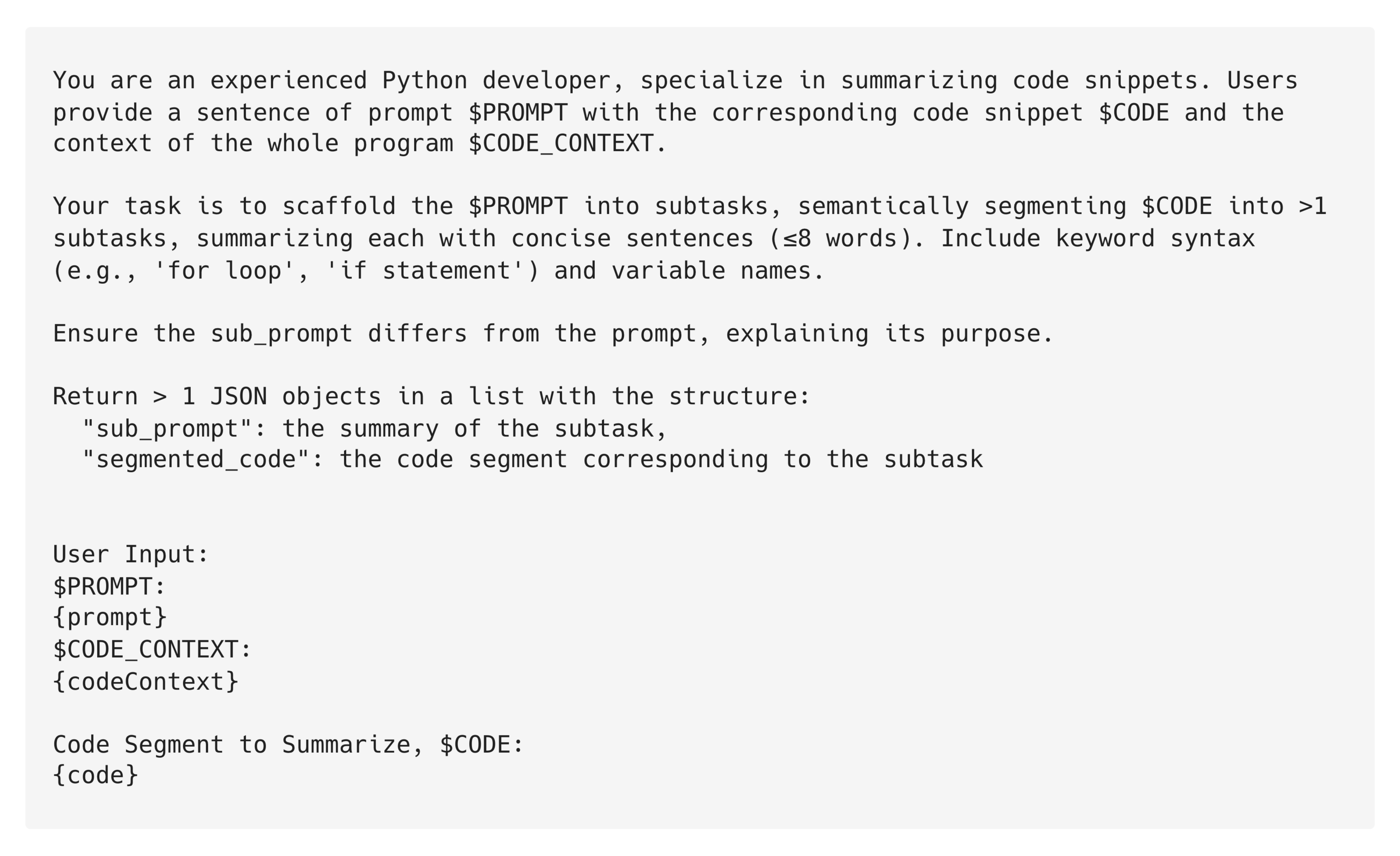}
    \caption{Prompt template for [List Steps]}
    \label{fig:prompt-lists}
    \end{minipage}
  \hfill
  \begin{minipage}[b]{0.65\textwidth}
    \includegraphics[width=\textwidth]{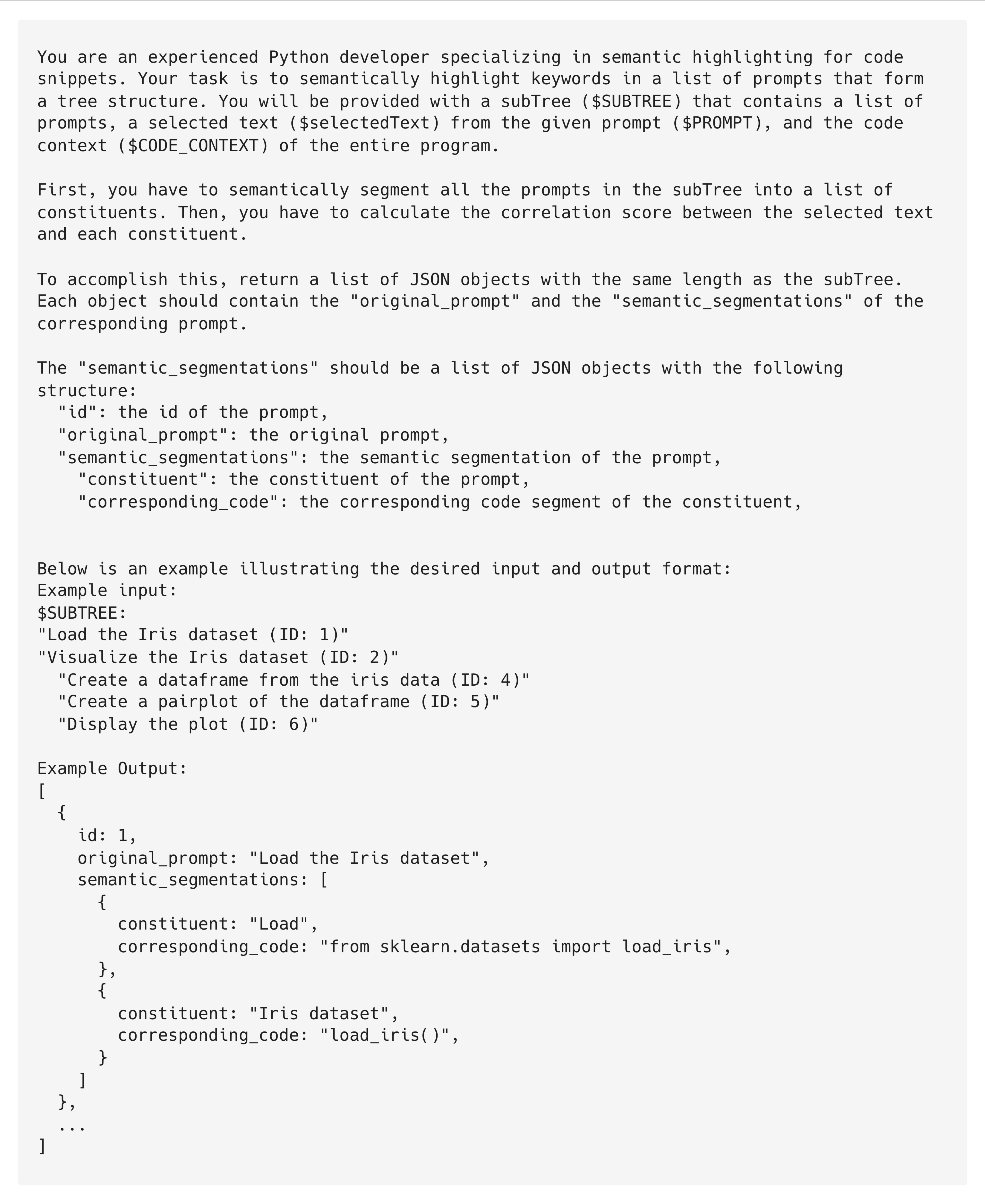}
    \caption{Prompt template for [Semantic Highlighting]}
    \label{fig:prompt-semantic}
  \end{minipage}
\end{figure}

\section{Formative Study Materials}
\label{appendix:formative}
\begin{table}[H]
\centering
\small 
\begin{tabular}{@{}llp{1.5cm}p{1.5cm}p{1.5cm}p{2cm}p{2cm}p{4cm}@{}}
\toprule
\textbf{Sex} & \textbf{Age} & \textbf{Programming Years} & \textbf{AI Familiarity} & \textbf{Programming Familiarity} & \textbf{Usage of LLMs (times/week)} & \textbf{Languages Usage in LLMs} & \textbf{Tasks Usage in LLMs} \\
\midrule
Male         & 26            & 6                          & 5                       & 5                                & >19                              & Python, C\#/C++, JavaScript     & Unfamiliar Code, Algorithm    \\
Female       & 27            & 6                          & 4                       & 4                                & 3–5                              & Python, JavaScript              & Unfamiliar Code, Boilerplate, API Usage \\
Male         & 25            & 5                          & 5                       & 4                                & 5–7                              & Python, Bash                    & Unfamiliar Code, Debugging    \\
Male         & 27            & 10                         & 4                       & 4                                & 7–10                             & JavaScript, Java, C/C++         & Unfamiliar Code, Boilerplate, Code Refactoring \\
Male         & 25            & 7                          & 4                       & 4                                & 3–5                              & Python, Go, Rust                & Debugging, Code Refactoring   \\
Male         & 25            & 6                          & 5                       & 5                                & 3–5                              & Python, C\#/C++, JavaScript     & Unfamiliar Code, Boilerplate, API Usage \\
\bottomrule
\end{tabular}
\caption{Participants in the formative study used various programming languages and accomplished diverse tasks using the LLMs.}
\label{tab:formative-participants}
\end{table}

\section{Evaluation Study Materials}
\subsection{Programming Tasks Categories}
\begin{table}[H]
\centering
\begin{tabular}{ll}
\toprule
\textbf{Category} & \textbf{Tasks} \\
\midrule
Basic Python & T1-1 Randomly generate and sort numbers and characters with dictionary \\
             & T1-2 Date \& time format parsing and calculation with timezone \\
File         & T2-1 Read, manipulate and output CSV files \\
             & T2-2 Text processing about encoding, newline styles, and whitespaces \\
OS           & T3-1 File and directory copying, name editing \\
             & T3-2 File system information aggregation \\
Web Scraping & T4-1 Parse URLs and specific text chunks from web page \\
             & T4-2 Extract table data and images from Wikipedia page \\
Web Server \& Client & T5-1 Implement an HTTP server for querying and validating data \\
                     & T5-2 Implement an HTTP client interacting with given blog post APIs \\
\textbf{Data Analysis \& ML}  & T6-1 Data analysis on automobile data of performance metrics and prices \\
                     & T6-2 Train and evaluate a multi-class logistic regression model given dataset \\
\textbf{Data Visualization}   & T7-1 Produce a scatter plot given specification and dataset \\
                     & T7-2 Draw a figure with three grouped bar chart subplots aggregated from dataset \\
\bottomrule
\end{tabular}
\caption{Overview of 14 programming tasks across 7 categories~\cite{xu2022ide}.}
\label{tab:program-tasks}
\end{table}

\subsection{Programming Tasks}
\label{appendix:tasks}
\begin{figure}[H]
    \centering
    \begin{minipage}[b]{0.49\textwidth}
    \includegraphics[width=\textwidth]{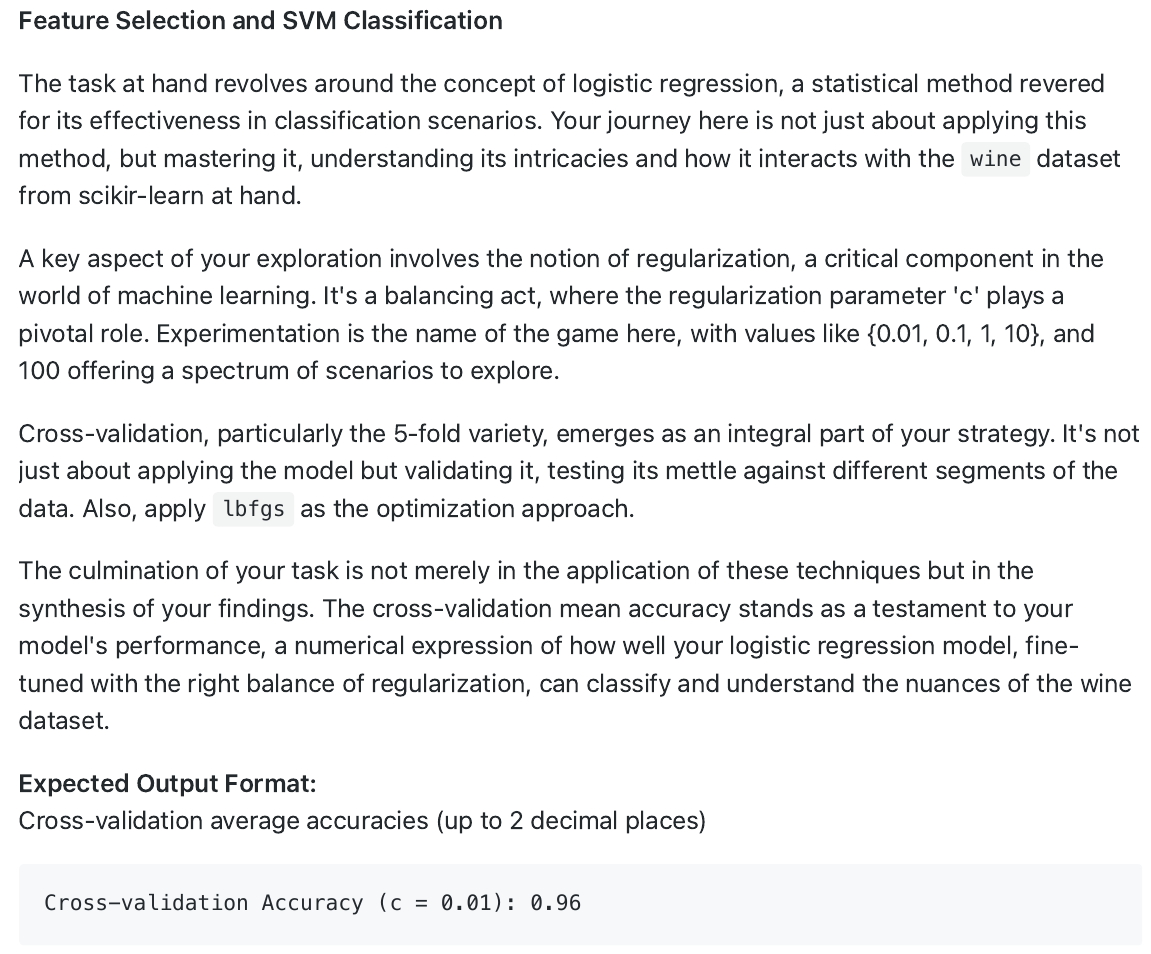}
    \caption{Machine Learning Task 1}
    \label{fig:task-ml1}
    \end{minipage}
  \hfill
  \begin{minipage}[b]{0.49\textwidth}
    \includegraphics[width=\textwidth]{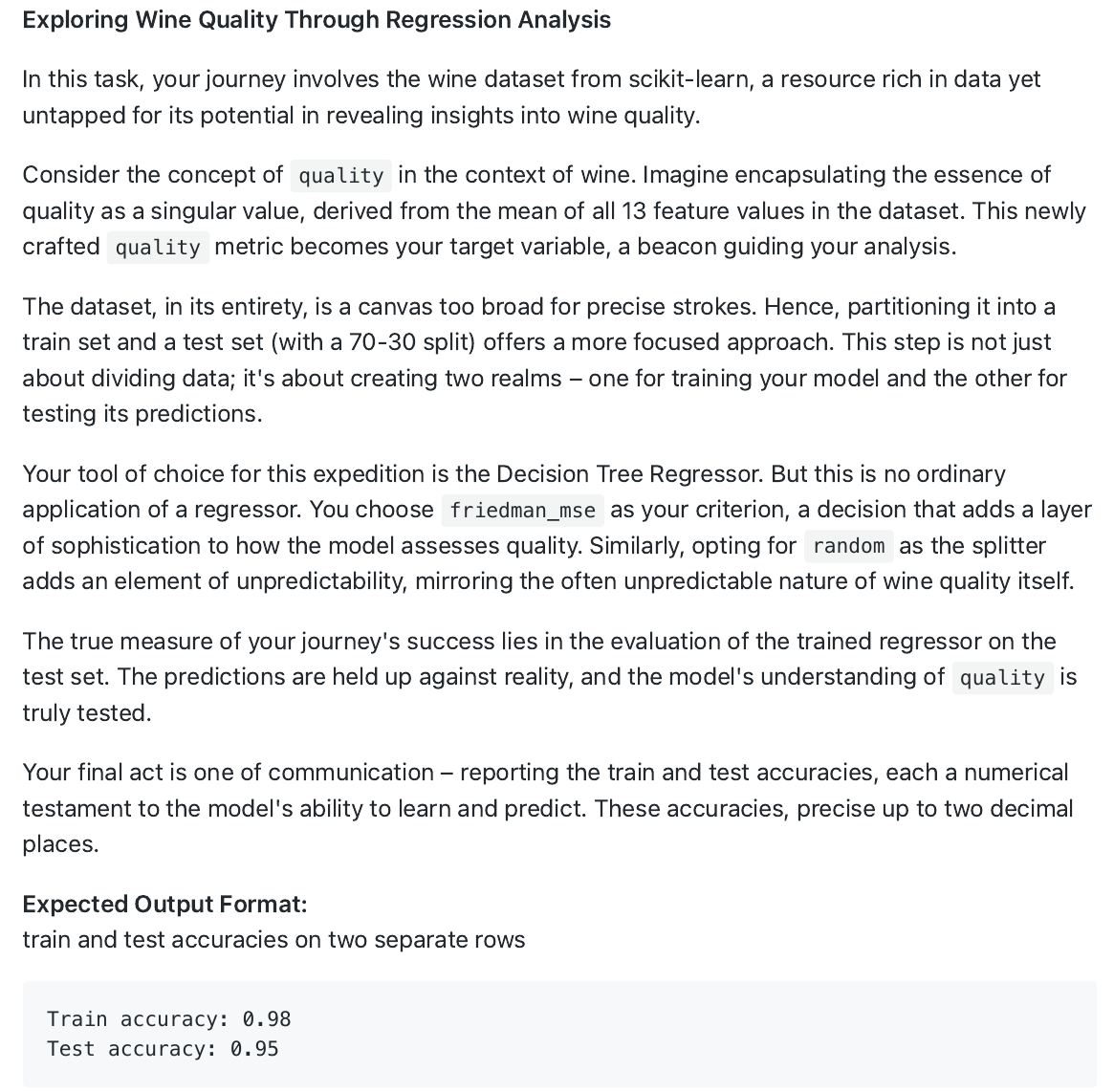}
    \caption{Machine Learning Task 2}
    \label{fig:task-ml2}
  \end{minipage}
\end{figure}

\begin{figure}[H]
    \centering
    \begin{minipage}[b]{0.49\textwidth}
    \includegraphics[width=\textwidth]{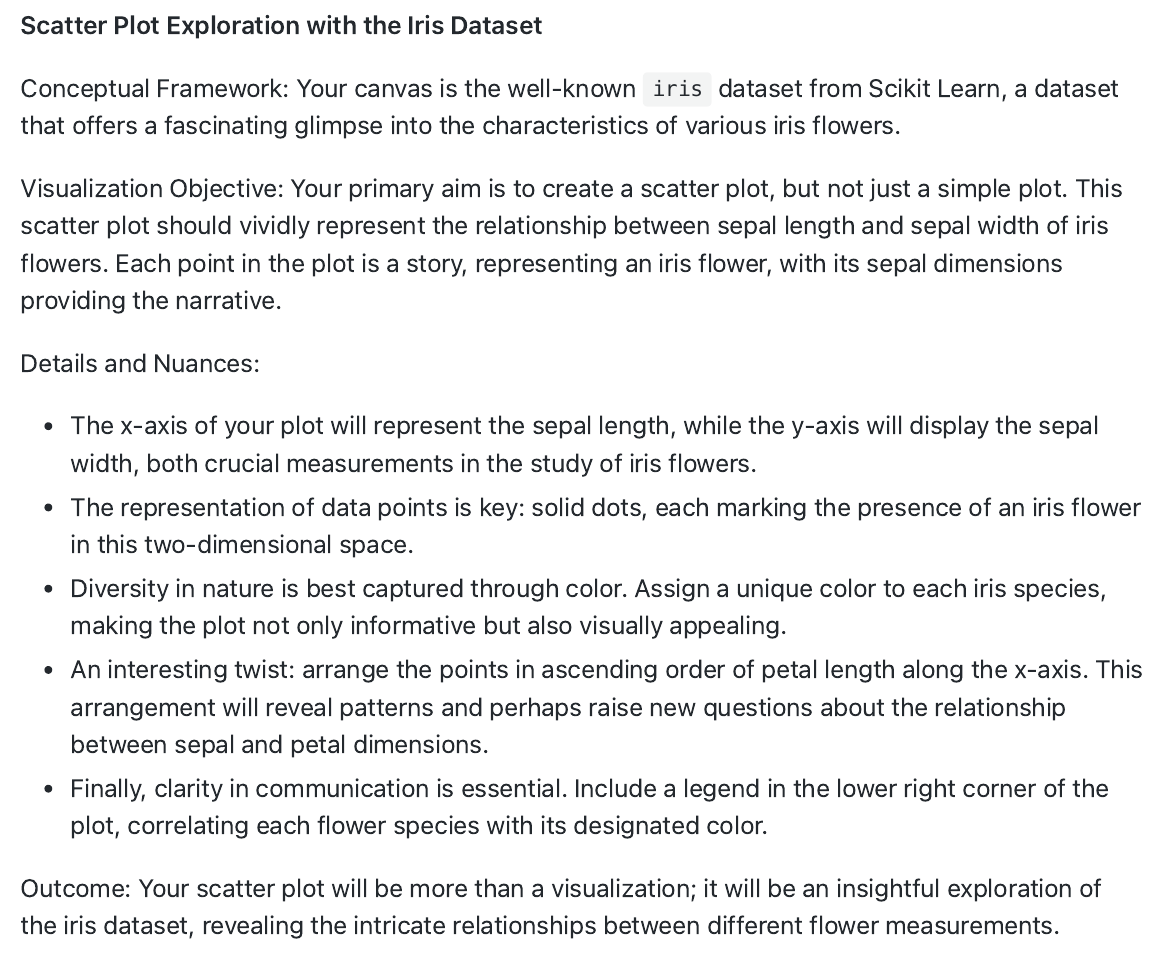}
    \caption{Data Visualization Task 1}
    \label{fig:task-viz1}
    \end{minipage}
  \hfill
  \begin{minipage}[b]{0.49\textwidth}
    \includegraphics[width=\textwidth]{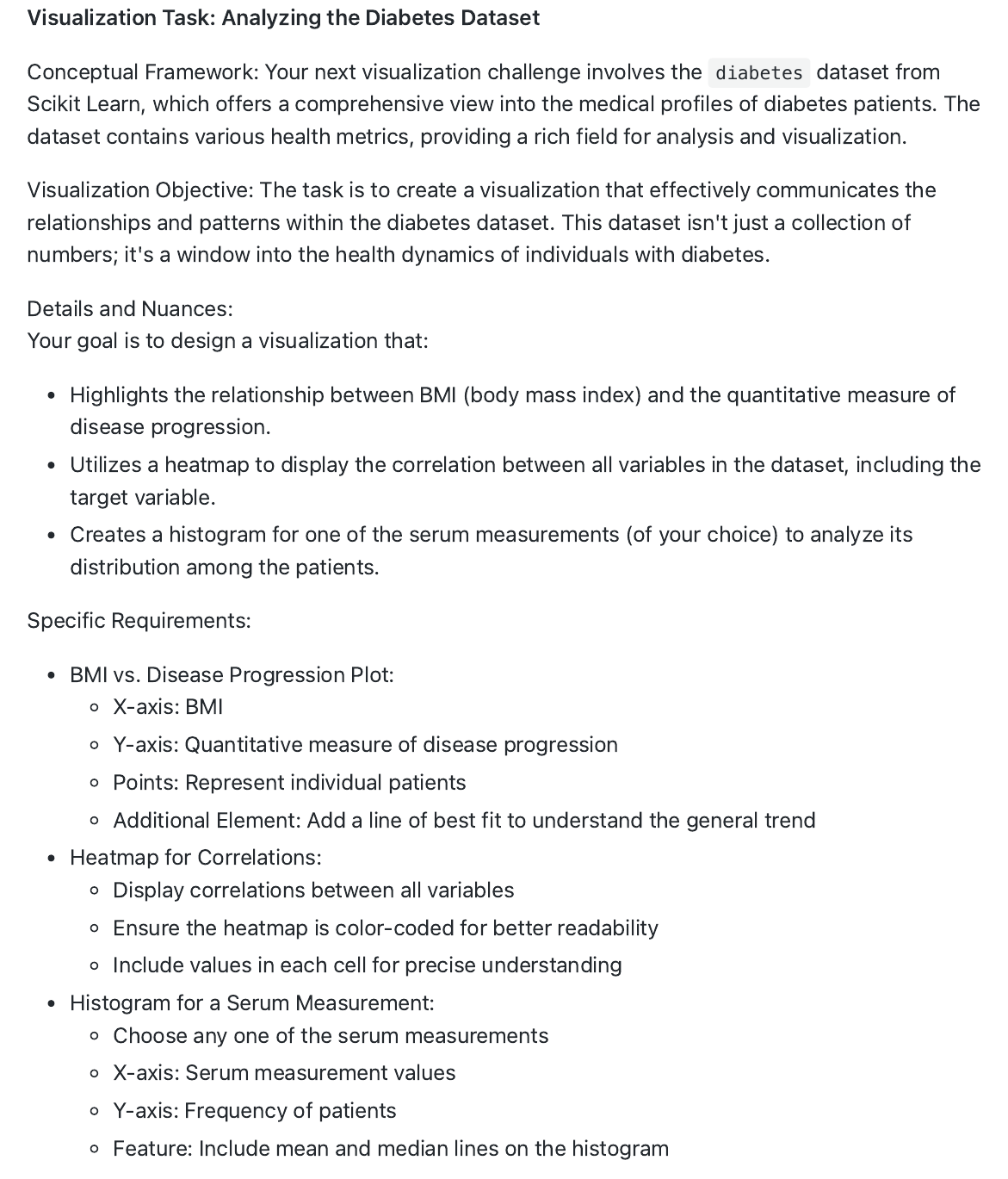}
    \caption{Data Visualization Task 2}
    \label{fig:task-viz2}
  \end{minipage}
\end{figure}

\subsection{Questionnaire}
Below, we list the questions we used in the evaluation study questionnaire.

\subsubsection{UMUX-LITE}
\label{appendix:umux}
\begin{enumerate}
    \item This system's capabilities meet my requirements.
    \item This system is easy to use.
\end{enumerate}

\subsubsection{NASA-TLX}
\label{appendix:nasa}
\begin{enumerate}
    \item How mentally demanding was the task?
    \item How physically demanding was the task?
    \item How hurried or rushed was the pace of the task?
    \item How successful were you in accomplishing what you were asked to do?
    \item How hard did you have to work to accomplish your level of performance?
    \item How insecure, discouraged, irritated, stressed, and annoyed were you?
\end{enumerate}

\subsubsection{Self-Defined Likert Scale Items}
\label{appendix:questionnaire}
\begin{enumerate}
    \item The system reduces the need for cognitive switching between editing and validation.
    \item I had a good understanding of why the system generates such results.
    \item I could steer the system toward the task goal.
    \item The system helps construct a mental model for solving the task.
    \item The system helps scaffold my intents to generate desired code.
    \item I'm satisfied with the overall suggestions from the system.
    \item I am confident that the system generated the correct code.
    \item I understand what my program is about, and how it works.
    \item The system helps me verify the generated results.
\end{enumerate}

\end{document}